\pdfoutput=1 % for arXiv PDFLaTeX submissions
\documentclass[aps,prb,showpacs,twocolumn,floatfix,longbibliography]{revtex4-1}

\usepackage{amsmath,amssymb,amsfonts}
\usepackage{graphicx,epsf}
\usepackage{xspace}
\usepackage{textcomp}
\usepackage{color}
\usepackage[normalem]{ulem}
\usepackage{fancyvrb}

\usepackage{listings}
\usepackage{xcolor}

\usepackage[T1]{fontenc}
\makeatletter
\g@addto@macro\@verbatim\small
\makeatother

\usepackage[htt]{hyphenat}

\usepackage[colorlinks,bookmarks=false,citecolor=blue,linkcolor=blue,urlcolor=blue,hyperfootnotes=true]{hyperref}

\usepackage{enumitem}
\usepackage{algpseudocode}
\usepackage{algorithmicx}
\usepackage{varwidth}

\newcommand{\be}{\begin{equation}}
\newcommand{\ee}{\end{equation}}

\definecolor{codegreen}{rgb}{0,0.6,0}
\definecolor{codegray}{rgb}{0.5,0.5,0.5}
\definecolor{codepurple}{rgb}{0.58,0,0.82}
\definecolor{backcolour}{rgb}{0.95,0.95,0.92}
\definecolor{gray}{HTML}{f4f4f5}
\definecolor{warmgray}{HTML}{888888}
\definecolor{turquis}{HTML}{76b7b2}
\definecolor{orange}{rgb}{255,164,0}

\lstdefinestyle{mystyle}{
    backgroundcolor=\color{gray},
    keywordstyle=\color{black},
    numberstyle=\tiny\color{codegray},
    basicstyle=\ttfamily\scriptsize,
    columns  = fullflexible,
    breakatwhitespace=false,
    breaklines=false,
    captionpos=b,
    keepspaces=true,
    numbers=left,
    numbersep=5pt,
    showspaces=false,
    showstringspaces=false,
    showtabs=false,
    tabsize=2
}

\begin{document}

\title{\textsc{Tkwant}: a software package for time-dependent quantum transport}

\author{Thomas Kloss$^{1}$}
\email{thomas.kloss@cea.fr}
\author{Joseph Weston$^{1,2}$}
\author{Benoit Gaury$^{1}$}
\email{benoit.gaury@asml.com}
\author{Benoit Rossignol$^1$}
\email{benoit-rossignol@hotmail.fr}
\author{Christoph Groth$^1$}
\email{christoph.groth@cea.fr}
\author{Xavier Waintal$^1$}
\email{xavier.waintal@cea.fr}
\address{$^1$Universit\'e Grenoble Alpes, CEA, Grenoble INP, IRIG, PHELIQS, 38000 Grenoble, France}
%\address{$^2$Solid  State  Physics  Laboratory,  ETH  Z{\"u}rich,  8093  Z{\"u}rich,  Switzerland}
\address{$^2$QuTech and Kavli Institute of Nanoscience, Delft University of Technology, 2600 GA Delft, The Netherlands}

\begin{abstract}
\textsc{Tkwant} is a Python package for the simulation of quantum nanoelectronics devices to which external time-dependent perturbations are applied.
\textsc{Tkwant} is an extension of the \textsc{kwant} package (\hyperlink{https://kwant-project.org/}{https://kwant-project.org/}) and can handle the same types of systems: discrete
tight-binding-like models that consist of an arbitrary central region connected to semi-infinite electrodes. The problem is genuinely many-body even 
in the absence of interactions and is treated within the non-equilibrium Keldysh formalism.
Examples of \textsc{tkwant} applications include the propagation of plasmons generated by voltage pulses, propagation of excitations in the quantum Hall regime, spectroscopy of Majorana fermions in semiconducting nanowires, current-induced skyrmion motion in spintronic devices, multiple Andreev reflection, Floquet topological insulators, thermoelectric effects, and more.
The code has been designed to be  easy to use and modular.
\textsc{Tkwant} is free software distributed under a BSD license and can be found at \hyperlink{https://tkwant.kwant-project.org/}{https://tkwant.kwant-project.org/}.
\end{abstract}

\date{February 22, 2021}
% \pacs{..}

\maketitle

%\noindent{\it Keywords\/}: quantum nanoelectronics, non-equilibrium, real-time dynamics, tight binding, numerical simulation

%\tableofcontents

\section{Introduction}
The field of quantum nanoelectronics -- connecting coherent nano- or microscale devices at sub-Kelvin temperatures to macroscopic electronic measuring apparatus -- began in the early eighties and lies at the root of emerging solid-state-based quantum technologies.
A pletorha of new physical effects have been discovered including conductance quantization, electronic interferometry (Aharonov-Bohm effect,\cite{Batelaan09} Mach-Zehnder interferometers\cite{Ji03, *Roulleau08}), interaction effects (Coulomb blockade,\cite{matveev93, *matveev93a} Kondo effect in quantum dots\cite{Inoshita98, Cronenwett98}) hybrid normal-superconducting effects (Andreev reflection\cite{Andreev64}), Klein tunneling (graphene),\cite{Katsnelson06, Stander09} sub-poissonian quantum noise,\cite{Blanter00} and many more. Numerical simulations, featuring increasingly closer connections to experiment, play an important role in the study of these phenomena.

A recent and growing trend in the field is to revisit quantum nanoelectronics at increasingly higher frequencies in the GHz to THz range where one can probe the internal dynamics of a system. While such high-frequency nanoelectronics is still mostly under development, many important milestones have already been reached including the design of coherent single electron sources and their tomography,\cite{Feve08, Dubois13, McNeil11, Fletcher13} the study of the propagation of excitations produced by voltage pulses at zero magnetic field\cite{Roussely2018} and in the quantum Hall regime,\cite{Hashisaka2017} the measurement of photo-assisted shot noise,\cite{Vannucci18} and more. Many phenomena involving superconductors (e.g.\ multiple Andreev reflection) are intrinsically time-dependent due to the appearance of the AC Josephson\cite{Klapwijk82, Averin95, Rokhinson12} effect when a superconducting junction is DC-biased\cite{Sanjose13}.
The recent developments in the manipulation of (semiconducting or superconducting) quantum bits also involve time-resolved dynamics in the GHz range.\cite{Bertoni00, Ionicioiu01, Bautze14, Bauerle18, Glattli20}
There exists, in short, a growing number of experiments that address time-dependent phenomena.

On the othe hand, the theory of time-dependent quantum transport is rather mature.
It involves several formalisms that use either non-equilibrium Green's functions\cite{Caroli71, Croy09} or scattering approaches\cite{Moskalets11}, both being developed either for periodic perturbations (Floquet formalism) or directly in the time domain.
In contrast, numerical simulations, which play an increasingly important role in DC quantum transport, have received limited attention in the time domain.
This is due, in part, to the fact that until recently such simulations were quite computationally intensive, therefore making their application to relevant phenomena rather difficult.
Recent algorithmic progress, however, makes direct time-dependent simulations of relevant quantum devices computationally affordable on a small computing cluster or even on a desktop computer.

This article presents \textsc{tkwant}, a software library that implements state-of-the-art algorithms for the simulation of non-interacting time-dependent quantum transport.\cite{tkwant}
\textsc{Tkwant} (Time-dependent \textsc{kwant}) is an extension of the \textsc{kwant}\cite{groth14} Python library for DC quantum transport.
\textsc{Tkwant} can simulate a wide variety of models for different materials (semiconductors, graphene, topological materials, superconductors, metals, magnets, etc.), different geometries (Hall bars, rings, wires, etc.) in arbitrary dimension (1D, 2D, 3D, \dots), in presence of arbitrary perturbations (voltage pulses, polarized light, static or dynamical disorder, time-dependent electrostatic gates, etc.), and an arbitrary number of connected electrodes. \textsc{Tkwant} has been designed to be easy to learn, to use, and to extend. It is the hope of its authors that it will be useful for many new projects outside of its original range of applications.

The article is organized as follows: Sec.\ \ref{sec:nutshell} introduces \textsc{tkwant} through a simple concrete example: the propagation of a voltage pulse inside an electronic Fabry-Perot cavity. Sec.\ \ref{sec:formalism} provides a brief presentation of the main theoretical objects of time-dependent quantum transport. The different numerical algorithms used in \textsc{tkwant} are discussed in Sec.\ \ref{sec:numerical_algorithms}. Sec.\ \ref{sec:architecture} discusses how the structure of the code is organized to handle one-body and many-body problems. Sec.\ \ref{sec:graphene} illustrates various aspects of \textsc{tkwant} with a full-scale application: propagation of a voltage pulse sent off an electrostatic gate deposited on top of a graphene quatum billard.
Summary and conclusion remarks are given in Sec.\ \ref{sec:conclusion}. Additional technical details on the band structure analysis and calculation of boundary conditions in electrodes are given, respectively, in Appendix \ref{sec:appendix_a} and \ref{sec:appendix_b}. The source code of the the actual simulations that were used to generate the figures of this article is provided as supplementary material.

\section{\textsc{Tkwant} in a nutshell}
\label{sec:nutshell}

This section features a rapid tour of \textsc{tkwant}. We start by formulating the type
of problems that \textsc{tkwant} can solve. Then, we present a simple, yet nontrivial, example calculation for the propagation of an abrupt voltage raise in a one-dimensional Fabry-Perot interferometer. The complete source code for this example is discussed, in order to illustrate the close relation between the short Python code that one writes and the mathematical model than one wants to simulate. Finally, we review various existing applications of \textsc{tkwant}.

\subsection{Problem formulation}

\textsc{Tkwant} can handle general discrete quadratic Hamiltonians of the generic form
\begin{equation}
\label{eq:hamiltonian_gen}
\hat{\mathbf{H}}(t) = \sum_{i,j} \mathbf{H}_{ij}(t) \hat{c}^\dagger_i \hat{c}_{j},
\end{equation}
where the time-dependent matrix $\mathbf{H}_{ij}(t)$ is defined by the user
and $\hat{c}^\dagger_i$ ($\hat{c}_i$) is the fermionic creation (annihilation) operator on site $i$.
A site $i$ may label not only lattice positions, but might also refer to other degrees of freedom, such as spin or orbital numbers. \textsc{Tkwant} inherits from \textsc{kwant}
a comprehensive set of tools for building the Hamiltonian $\hat{\mathbf{H}}(t)$ for devices of arbitrary shapes and dimensions on any lattice (graphene, cubic, amorphous, combinations of those, etc.). Note that even though we consider non-interacting problems, we have defined the above system in terms of a second-quantized Hamiltonian. Indeed, in contrast to DC transport where one can essentially solve the one-body quantum problem at (or close to) Fermi level, here the time-dependent perturbation makes the handling of the Pauli principle nontrivial, even in the non-interacting limit\cite{gaury14}. 

\begin{figure}[tbh]
	\centering
	\includegraphics[width=70mm]{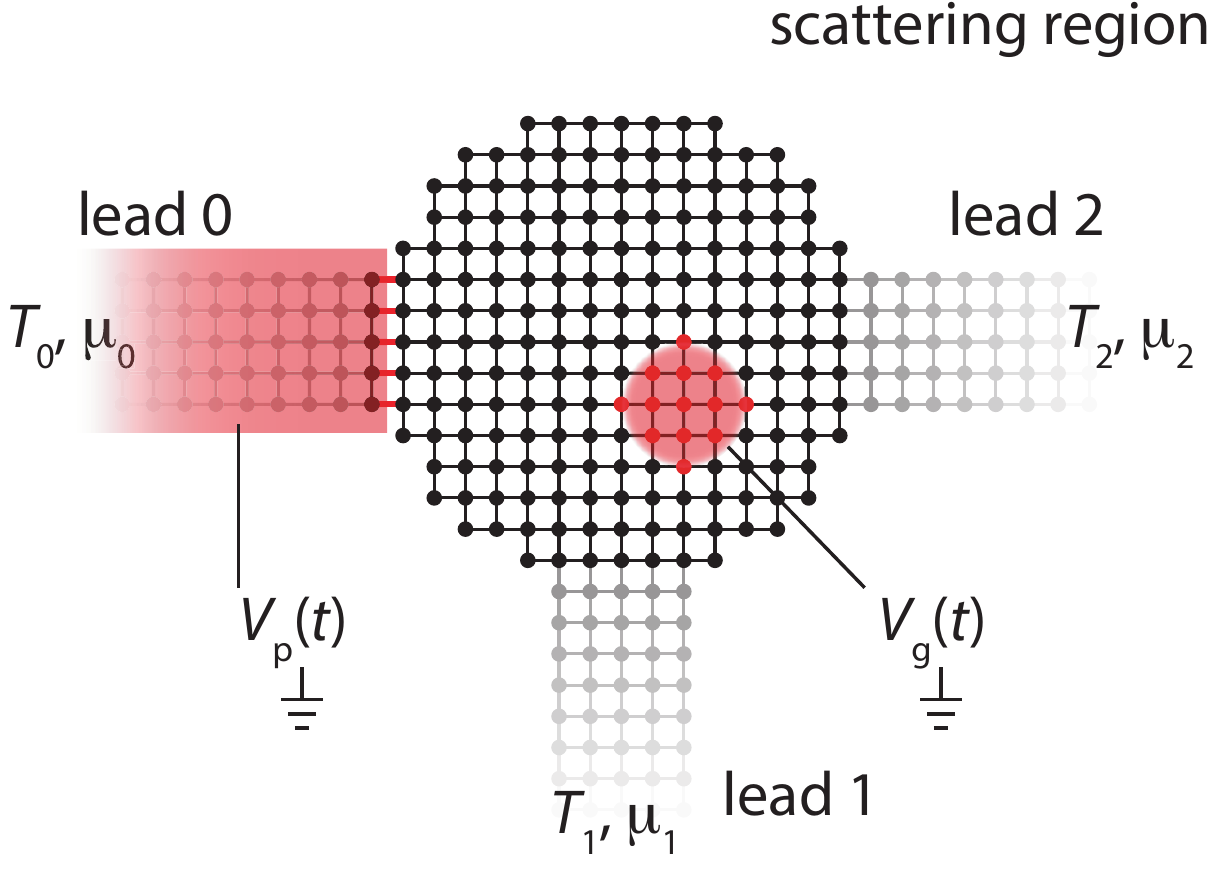}
	\vspace{-1mm}
  \caption{%
Sketch of a typical open quantum system that can be simulated with \textsc{tkwant}. A central scattering region (in black) is connected to several leads (in gray). Each lead represents a translationally invariant, semi-infinite system in thermal equilibrium. Sites and hopping matrix elements are represented by dots and lines.
The regions in red indicate the time-dependent perturbation: in this example a global voltage pulse $V_{\text p}(t)$  on lead 0 and a time-dependent voltage $V_{\text g}(t)$ on a gate inside the scattering region.}
\label{fig:tbsys}
\end{figure}

Although \textsc{kwant} may be used for systems with a finite number of sites, its most common usage is for infinite systems. These systems consist of a finite central region called the scattering region ($s$ with $N_s$ sites) connected to several infinite electrodes called leads ($l$). The leads are semi-infinite and invariant by translation. Such a  Hamiltonian take the form
\begin{equation}
\label{eq:hamltonian_split}
\hat{\mathbf{H}}(t) = \hat{\mathbf{H}}^s(t)
+\sum_l \hat{\mathbf{H}}^l
+  \sum_{l} \hat{\mathbf{H}}^{sl}(t) ,
\end{equation}
where the different terms correspond, respectively, to the
scattering region ($s$), to the leads ($l$) and to the coupling between the  scattering region and the infinite leads ($sl$). A sketch of such a system is shown in Fig.\ \ref{fig:tbsys}. We refer to such infinite systems as open systems. Note that they are different from another class of systems, also refered to as open, that are described by a Lindblad equation and that can be addressed with the software package  \textsc{qutip} \cite{Johansson12a, *Johansson12b} for example.

The Hamiltonian for the scattering region is a general quadratic Hamiltonian,
\begin{equation}
\label{eq:Hs}
\hat{\mathbf{H}}^s(t) = \sum_{i,j } \mathbf{H}^{s}_{ij}(t) \hat{c}^\dagger_i \hat{c}_j .
\end{equation}
Only the finite scattering region $\hat{\mathbf{H}}^s(t)$ and the coupling to the lead $\hat{\mathbf{H}}^{sl}(t)$ contain time-dependent perturbations.
The leads are time-independent with one exception: they may be shifted by a global potential $\mathbf{H}_{ii}^l(t) = V^l(t)$ that is identical on all the sites of a lead.
Indeed in this case, a simple gauge transformation allows one to restore an time-independent lead at the cost of adding a global time-dependent phase
\begin{equation}
\label{eq:phi}
\phi^l(t) = \frac{e}{\hbar} \int_{- \infty}^{t}  V^l(\tau) \, d\tau.
\end{equation}
to the coupling Hamiltonian: $\mathbf{H}^{sl}_{in}(t)\rightarrow e^{-i\phi^l(t)}\mathbf{H}^{sl}_{in}(t) $.
In \textsc{tkwant}, we focus on leads that are invariant by translation: they consist of
unit cells $a$ that are repeated to form a semi-infinite quasi-one dimensional system. Each unit cell contains $N$ sites labeled by indices $n,m$. A site $i$ in the lead is
described by the vector $i=(a,n)$. The Hamiltonian of a lead $l$ is fully characterized by two $N\times N$ matrices $\mathbf{H}^l_0$ and $\mathbf{V}^l$,
\begin{equation}
\hat{\mathbf{H}}^l = \sum_{a=0}^{+\infty}\sum_{n,m}
(\mathbf{H}^{l}_0)_{nm} \hat{c}^\dagger_{a,n} \hat{c}_{a,m} +
\mathbf{V}^{l}_{nm} \hat{c}^\dagger_{a,n} \hat{c}_{a-1,m} +
\textrm{h.c.} .
\end{equation}
The leads are also considered to be in (possibly different) thermal equilibrium characterized by a Fermi function $f^l(E)$ with a time-independent chemical potential $\mu_l$ and temperature $T_l$.
The coupling between the scattering region and the lead is an arbitrary quadratic
Hamiltonian between the scattering region and the first unit cell of the lead,
\begin{equation}
\label{eq:Hsl}
\hat{\mathbf{H}}^{sl}(t)
= \sum_{i,n} \mathbf{H}^{sl}_{in}(t) \hat{c}^\dagger_i \hat{c}_{a=0,n} + \textrm{h.c.} .
\end{equation}
Except for the fact that some matrix elements are time-dependent, the systems considered in \textsc{tkwant} are identical to those in \textsc{kwant}.

The general problem that \textsc{tkwant} adresses is the time evolution of observables such as densities or currents after the system is subject to a time-dependent perturbation for $t>t_0$. The system is initially in a stationary state for $t<t_0$ (in or out of equilibrium). \textsc{tkwant} computes
expectation values such as
\begin{equation}
\langle c^\dagger_ic_j\rangle(t) = {\rm Tr } [c^\dagger_ic_j \hat{\rho}(t) ],
\end{equation}
where $\hat{\rho}(t)$ is the non-equilibrium density matrix of the system. No assumption of adiabaticity or otherwise is made in the calculation and higher-order observables\cite{gaury16} (such as quantum noise) can also be obtained.

\subsection{Diving into \textsc{tkwant} with a simple example}
\label{sec:example_problem}
\begin{figure}[h]
	\centering
	\includegraphics[width=80mm]{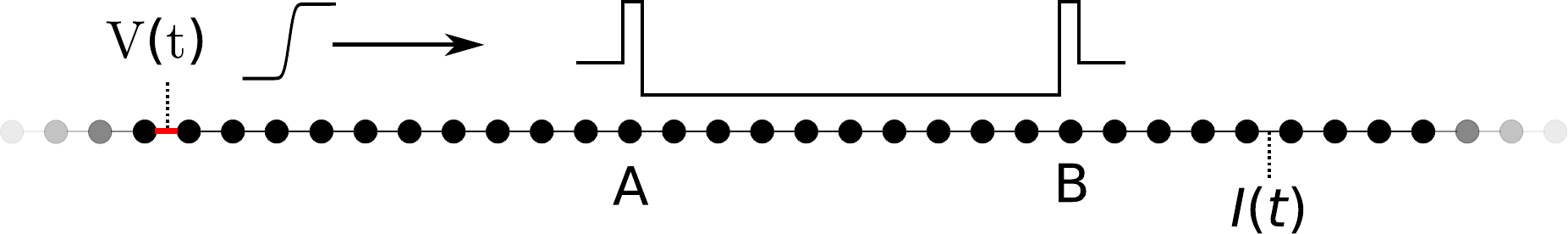}\\
	\vspace{1.2ex}
	\includegraphics[width=75mm]{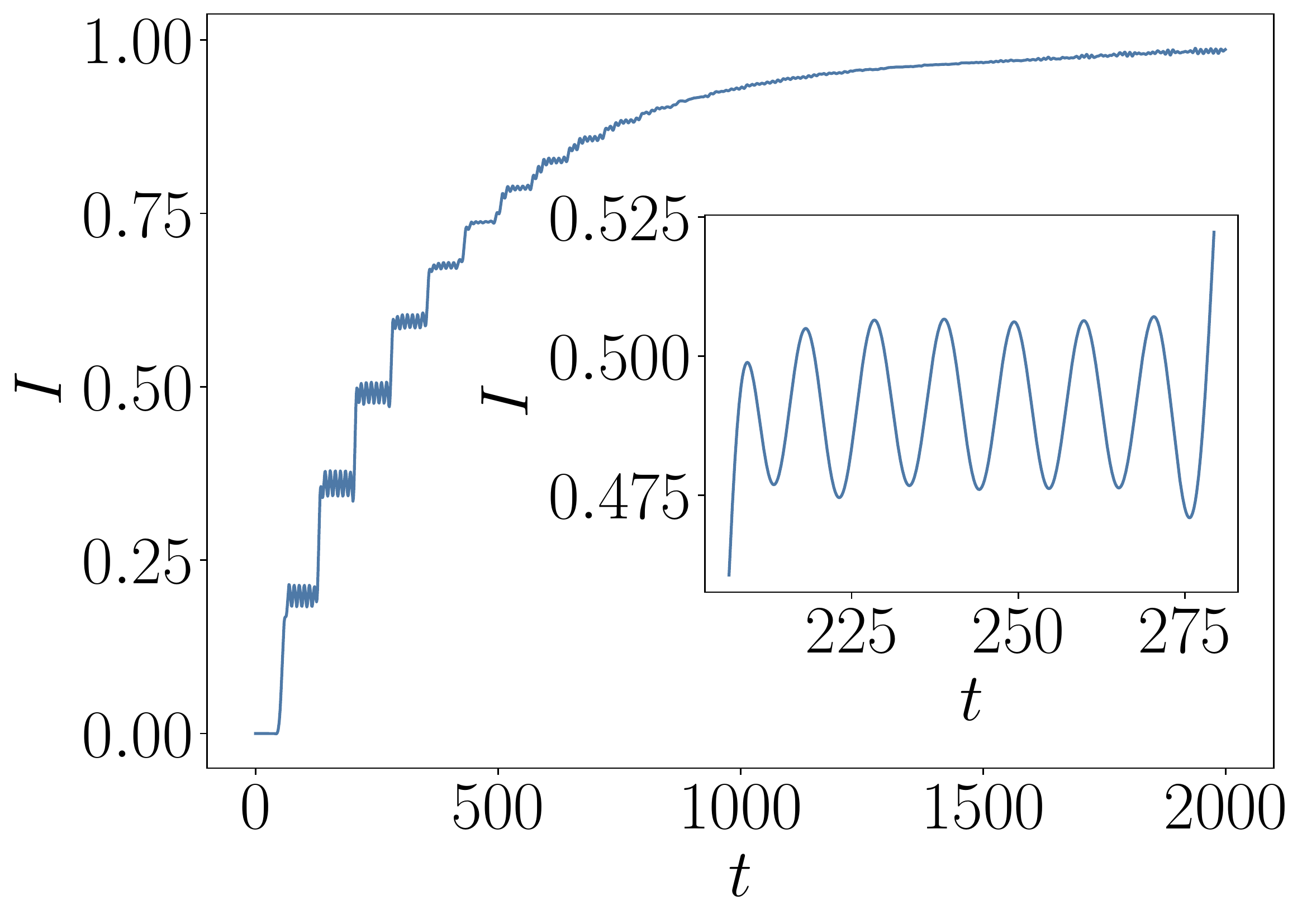}
	\vspace{-2ex}
  \caption{%
Top panel: schematic of the system, a one-dimensional chain with potential barriers on sites A and B that transform the system into a Fabry-Perot cavity.
At $t=0$ one quickly raises the voltage $V(t)$ of the left lead (which induces a phase $\phi(t)$ in the hopping shown in red) from $0$ to $V_{\rm b}$. A similar system has been studied in Ref.~[\onlinecite{gaury15}].
Lower panel: result of the simulation, current $I(t)$ measured on the right of the two barriers A and B.
This plot can be obtained by running the Python code given in the code listing \ref{code:fabry_perot_code}.
Inset: detail of the main figure showing periodic oscillations of the current.}
\label{fig:fabry_perot}
\end{figure}

Below we discuss a numerical experiment for a simple yet nontrivial system.
We consider an infinite one-dimensional chain with nearest-neighbor hoppings. Two potential barriers, A and B, are placed in the system to form a Fabry-Perot cavity. A sketch of the system is shown in the top panel of Fig.\ \ref{fig:fabry_perot}. At $t=0$, the electric potential of the left electrode is suddenly raised from zero to a finite value $V_{\rm b}$ and we want to study the transient regime of the current $I(t)$ before it eventually reaches its stationary value. The lower panel of Fig.\ \ref{fig:fabry_perot} shows the result: the current increases over several plateaus that correspond to the different trajectories through the cavity (direct transmission, reflection at B followed by reflection at A then transmission, etc.). The inset shows an interesting phenomenon: on each plateau, there are small oscillations of the current at a frequency $e V_b/h$.
We refer to Ref.~[\onlinecite{gaury15}] for a detailed discussion of the physics of this system.

The Hamiltonian for this system reads
\begin{equation}
\label{eq:example_hamiltonian}
\hat{\mathbf{H}}(t) =  \sum_{i}^{N_s+1} \epsilon_i \hat{c}^\dagger_i \hat{c}_i - \sum_{-\infty}^{\infty} \hat{c}^\dagger_{i+1} \hat{c}_i -  [e^{i \phi(t)}-1]  \hat{c}^\dagger_{1} \hat{c}_{0} + \text{h.c.},
\end{equation}
where $\epsilon_i$ is a static onsite potential that defines the cavity. The Fermi level is fixed at $E_F=-1$ and the temperature at zero. The time-dependent ramp-up voltage,
\begin{equation}
\label{eq:vp}
       V(t) =
        \begin{cases}
        0, & \text{for } t < 0\\
        \frac{V_{\rm b}}{2} \left ( 1 - \cos\left (\frac{\pi t}{\tau} \right) \right) , & \text{for } 0 \leq t \leq
        \tau \\
        V_{\rm b} , & \text{for } t > \tau
        \end{cases}
\end{equation}
is applied to the left electrode ($i\le 0$). The voltage ramp amounts to adding an extra phase $\phi (t)$ to the hopping from the electrode to the central system, see Eq.\ (\ref{eq:phi}).
The current $I(t)$ takes the form
\begin{equation}
\label{eq:curr}
I(t) = i [ \langle \hat{c}^\dagger_{i_0}\hat{c}_{i_0+1} \rangle (t) - \langle \hat{c}^\dagger_{i_0+1} \hat{c}_{i_0} \rangle (t) ],
\end{equation}
where the site $i_0$ is chosen in the right part of the central region, outside of the Fabry-Perot cavity.
To simulate the system described above with \textsc{tkwant}, it is sufficient to write the short Python
program that can be found below in Listing \ref{code:fabry_perot_code}. Such Python scripts replace the traditional input files of standalone numerical simulation programs while providing more flexibility for defining the system, analyzing the results and combining \textsc{tkwant} with other packages. \textsc{Tkwant} defines objects that represent high-level concepts closely matching the mathematical objects of the underlying formalism. All these objects have a documented application programming interface (API) and are exposed to the user in order to provide as much flexibility as possible.

\begin{lstlisting}[language=Python, label=code:fabry_perot_code, caption={
Python code to simulate the time-dependent current in a Fabry-Perot interferometer
\cite{gaury15} with \textsc{tkwant}. Running this script generates a current vs.\ time plot similar to Fig.\ \ref{fig:fabry_perot}.
Note that most code is related to \textsc{kwant}\cite{groth14} and only a few lines
were added to obtain the time evolution. Running the
code on 48 cores (AMD Opteron 6176 with 2.3 GHz) takes about half an hour.}
]
import tkwant
import kwant
from math import sin, pi
import matplotlib.pyplot as plt


def make_fabry_perot_system():
    # Define an empty tight-binding system on a square lattice.
    lat = kwant.lattice.square(norbs=1)
    syst = kwant.Builder()

    # Central scattering region.
    syst[(lat(x, 0) for x in range(80))] = 0
    syst[lat.neighbors()] = -1
    # Backgate potential.
    syst[(lat(x, 0) for x in range(5, 75))] = -0.0956
    # Barrier potential.
    syst[[lat(4, 0), lat(75, 0)]] = 5.19615

    # Attach lead on the left- and on the right-hand side.
    sym = kwant.TranslationalSymmetry((-1, 0))
    lead = kwant.Builder(sym)
    lead[(lat(0, 0))] = 0
    lead[lat.neighbors()] = -1
    syst.attach_lead(lead)
    syst.attach_lead(lead.reversed())

    return syst, lat


# Phase from the time integrated voltage V(t).
def phi(time):
    vb, tau = 0.6, 30.
    if time > tau:
        return vb * (time - tau / 2.)
    return vb / 2. * (time - tau / pi * sin(pi * time / tau))


times = range(2000)

# Make the system and add voltage V(t) to the left lead (index 0).
syst, lat = make_fabry_perot_system()
tkwant.leads.add_voltage(syst, 0, phi)
syst = syst.finalized()

# Define an operator to measure the current after the barrier.
hoppings = [(lat(78, 0), lat(77, 0))]
current_operator = kwant.operator.Current(syst, where=hoppings)

# Set occupation T = 0 and mu = -1 for both leads.
occup = tkwant.manybody.lead_occupation(chemical_potential=-1)

# Initialize the time-dependent manybody state.
state = tkwant.manybody.State(syst, tmax=max(times),
                              occupations=occup)

# Loop over timesteps and evaluate the current.
currents = []
for time in times:
    state.evolve(time)
    current = state.evaluate(current_operator)
    currents.append(current)

# Plot the normalized current vs. time.
plt.plot(times, currents / currents[-1])
plt.show()
\end{lstlisting}

Understanding this script requires basic familiarity with the Python language.
The first function \texttt{make\_fabry\_perot\_system()} defines a \textsc{kwant} system that implements Eq.\
(\ref{eq:example_hamiltonian}).
The scattering region contains $80$ sites with a barrier $\epsilon_4=\epsilon_{75}=5.19615$ (sites A and B, respectively). An additional gate voltage (here $-0.0956$), applied to all the sites inside the cavity ($4<i<75$), allows to tune the cavity in or out of resonance.
The leads possess an additional translational symmetry so that they are entirely described by a single unit cell (here a single site) and its connection to neighboring unit cells.
We refer to \textsc{kwant}\cite{groth14} documentation for a description of how to define systems.

The other function \texttt{phi(time)} implements Eq.\ (\ref{eq:phi}) with $V(t)$ of Eq.\ (\ref{eq:vp}). Here, the integral of Eq.\ (\ref{eq:phi}) has been calculated analytically, but it can also be calculated numerically for more complex functions $V(t)$.

The main program body calls \texttt{make\_fabry\_perot\_system()} to create the \textsc{kwant} system and ``finalizes'' it.
This operation takes the high-level ``builder'' of \textsc{kwant} and constructs a low-level object better suited for numerical calculations.
Note that from the perspective of the \textsc{kwant} package,  the parameter \texttt{time} is just another parameter of the \textsc{kwant} Hamiltonian as could be e.g.\ a magnetic field or a spin-orbit strength. However, \textsc{tkwant} will recognize a parameter with the name \texttt{time} as the time variable.

The next stage is to define the observables that will be calculated during the simulation.
To this end an instance of \texttt{kwant.operator.Current} is created. This stage is necessary because the internal state of a \textsc{tkwant} simulation can become very large and therefore cannot be recorded for every time step. Most observables must be therefore computed on-the-fly and as such be listed before the begin of the simulation.

The actual \textsc{tkwant} code starts with \texttt{tkwant.manybody.lead\_occupation()}, when one defines a chemical potential of $E_F = -1$ for all leads. The temperature is zero by default. The creation of a \texttt{tkwant.manybody.State} instance initializes the time-dependent many-body state.
This many-body state is evolved according to the many-body Schr\"odinger equation using the \texttt{state.evolve()} method.
The \texttt{state.evaluate()} method is used for the on-the-fly calculation of the current.
Note that the function \texttt{state.evaluate()} returns either scalars or regular Python (NumPy\cite{numpy}) arrays so that any post-processing or plotting of the data can be performed with standard Python tools.

The apparent simplicity of the above script hides a lot of technicalities and default values for certain parameters. \textsc{Tkwant} can be used in this default ``automatic'' mode which is sufficient for many purposes. However, the user can also claim control of all the defaults and other technical aspects as will be explained in the rest of this article.

\subsection{Other examples: a review of \textsc{tkwant} applications}
At the time of writing, \textsc{tkwant} has already been used for various applications. We review them briefly below in order to illustrate some of the problems that can be studied within the \textsc{tkwant} framework. We also review the articles that cover various related technical aspects (algorithm and formalism).

The first \textsc{tkwant} article\cite{gaury14} contains a detailed description of the theoretical framework and in particular shows the equivalence between the non-equilibrium Green's functions formalism and the scattering wave function formalism which is actually used by \textsc{tkwant}.
For computational purposes, we indeed find that the usage of scattering wavefunctions allows to obtain multiple orders of magnitude of speed-up compared to to Green's-function-based approaches.

Ref.~[\onlinecite{weston16b}] contains a simplified presentation of the formalism as well as an application to flying qubits in two-dimensional gases.
Ref.~[\onlinecite{Rossignol18}] extends the study of flying qubits to realistic models.

Ref.~[\onlinecite{weston16a}] improves the algorithms of Ref.~[\onlinecite{gaury14}] to achieve a computational time linear in $t$ (total simulation time) and $N_s$ (number of sites in the scattering region).
\textsc{Tkwant} currently implements its ``source-sink'' algorithm.
This article also discusses the propagation of voltage pulses through Josephson junctions as well as the current-voltage characteristic in presence of multiple Andreev reflection.

Other studies featuring superconductors include a method for performing the spectroscopy of Majorana modes in semiconducting nanowires\cite{weston15} and a mean-field technique to describe the role of electromagnetic environment of Josephson junctions within a microscopic model\cite{rossignol19}.

Refs.~[\onlinecite{gaury14a,gaury15}] discuss the propagation of voltage pulses through
Mach-Zehnder and Fabry-Perot electronic interferometers. Ref.~[\onlinecite{gaury14b}] studies how the propagation of voltage pulses in the quantum Hall regime could be manipulated in real time. Ref.~[\onlinecite{fruchart16}] illustrates how an effective (Floquet) topological insulator could be stabilized with a periodic time-dependent perturbation such as circularly polarized light. Ref.~[\onlinecite{gaury16}] provides the necessary formalism and technicalities to calculate quantum fluctuations (such as current noise) with \textsc{tkwant}. The formalism is illustrated with the calculation of the noise associated with Lorentzian pulses (the so-called Levitons\cite{levitov97, *keeling06, *levitov96})
Ref.~[\onlinecite{Abbout18}] studies the current generated by a moving skyrmion in a magnetic material. Ref.~[\onlinecite{slimane20}] studies time-dependent (electronic) heat transport and thermoelectric effects.
Ref.~[\onlinecite{kloss18}] studies the propagation of plasmons in 1D or quasi-1D geometries and makes contact with the theory of Luttinger liquids.

All the examples discussed above can be simulated with the current version of \textsc{tkwant} with the exception of Refs.~[\onlinecite{kloss18,rossignol19}] which require an extension that is still at prototype level. A few dozen lines of code typically separate one application from another.

\section{Fundamentals of time-dependent quantum transport formalism}
\label{sec:formalism}

In this section, we provide the minimum level of formalism to define the mathematical objects that are calculated in a \textsc{tkwant} simulation. The most popular formalism for time-dependent quantum transport uses the Keldysh formalism of non-equilibrium Green's functions (NEGF).
Starting from the general formalism,\cite{Keldysh64, rammer86, Rammer07} its application to quantum transport in
mesoscopic systems can be found in Refs.~[\onlinecite{Caroli71, Meir92, Wingreen93, Jauho94}]. The alternative -- yet fully equivalent -- approach that \textsc{Tkwant} uses employs scattering wave-functions.  This natural extension of  \textsc{kwant}'s stationary scattering wave-functions is highly advantageous from a computational perspective.  The formalism  is explained in detail in Ref.~[\onlinecite{gaury14}].

\subsection{Definition of Keldysh Green's functions}
The two central objects of NEGF are, respectively, the retarded (R) and lesser ($<$)
Green's functions,
\begin{subequations}
\label{eq:green_func}
\begin{align}
\mathbf{G}_{ij}^R(t, t') &= - i \theta(t - t') \langle \{ \hat{c}^\dagger_i(t) , \hat{c}_j(t') \} \rangle , \label{eq:green_func_r} \\
\mathbf{G}_{ij}^<(t, t') &= i \langle \hat{c}^\dagger_j(t')  \hat{c}_i(t)  \rangle , \label{eq:green_func_g}
\end{align}
\end{subequations}
where $\hat{c}^\dagger_i(t)$ and $\hat{c}_i(t)$ refer to the previously introduced fermionic operators in the Heisenberg picture and the average $\langle\dots\rangle = {\rm Tr } [\dots\rho]$ is taken with respect to a non-equilibrium density matrix that supposes that each lead remains at its own thermodynamic equilibrium while the system is time-independent (in a stationary state) for $t<t_0$ before the time-dependent perturbations are turned on for $t>t_0$. For quadratic models of the form of Eq.\ (\ref{eq:hamltonian_split}), calculating these Green's functions is a two-step procedure,\cite{Jauho94,gaury14} where one first calculates $\mathbf{G}^R$ (solving the quantum mechanical problem) and then $\mathbf{G}^<$ (filling the states according to a non-equilibrium statistical distribution). Calculating the average of a physical observable,
\begin{equation}
\label{eq:observable}
\hat{\mathbf{A}} = \sum_{i,j} \mathbf{A}_{ij} \hat{c}^\dagger_i \hat{c}_j ,
\end{equation}
can be simply done from the knowledge of $\mathbf{G}^<$ at equal times:
\begin{equation}
\langle \hat{\mathbf{A}} \rangle (t)
=-i\ {\rm Tr} [ \mathbf{A} \mathbf{G}^<(t,t) ] .
\end{equation}

As it turns out, the calculation of the retarded and lesser Green's function can be bypassed entirely. Doing so is computationally advantageous in particular when only equal time quantities are needed but also for calculations of quantum noise.\cite{gaury16}

\subsection{Toy model of wavefunction formalism: finite system}

To motivate the scattering wavefunction approach, let us first discuss a simpler situation where the system contains a finite number of sites, a finite number of particles and is initially at equilibrium at zero temperature. For $t<t_0$, the system is described by a time-independent Hamiltonian $\hat{\mathbf{H}}_0$. For $t>t_0$, we write (without loss of generality) the full Hamiltonian as the sum of $\hat{\mathbf{H}}_0$ with whatever time-dependent perturbation $\hat{\mathbf{W}}(t)$ has been added,
\begin{equation}
\label{eq:h_part}
\hat{\mathbf{H}} = \hat{\mathbf{H}}_0 + \hat{\mathbf{W}}(t) .
\end{equation}
Diagonalizing $\hat{\mathbf{H}}_0$ provides the initial wavefunctions $\psi_\alpha$,
where the index $\alpha$ take discrete values. The stationary Schr\"odinger equation is
\begin{equation}
\label{eq:eigenval}
\hat{\mathbf{H}}_0 \psi_\alpha = E_\alpha \psi_\alpha ,
\end{equation}
from which one can build the  Slater determinant that forms the
many-body state of the system at $t<t_0$:
\begin{equation}
\label{eq:slater_det}
| \hat{\psi} \rangle = \prod_{E_\alpha < E_F} \hat{d}^\dagger_\alpha |0 \rangle ,
\end{equation}
with the operators $\hat{d}^\dagger_\alpha$ defined as
\begin{equation}
\label{eq:op_d_c}
\hat{d}^\dagger_\alpha = \sum_i \psi^*_\alpha(i) \ \hat{c}^\dagger_i ,
\end{equation}
and $E_F$ the Fermi level.

Solving the many-body time dependent problem for $t>t_0$ when one switches on the
perturbation is straightforward. It amounts to following the evolution of each of the
wavefunctions according to
\begin{subequations}
\label{eq:schroedinger_eq_global}
\begin{align}
i \partial_t \psi_{\alpha}(t, i) = \sum_{i,j} \mathbf{H}_{ij}(t) \psi_\alpha(t,j) ,
\label{eq:schroedinger_eq} \\
\psi_{\alpha}(t<t_0, i) = \psi_\alpha(i)e^{-i E_\alpha t}.
\label{eq:schroedinger_eq2}
\end{align}
\end{subequations}
One obtains
\begin{equation}
\label{eq:slater_det_t}
| \hat{\psi}(t) \rangle = \prod_{E_\alpha < E_F} \hat{d}^\dagger_\alpha(t) |0 \rangle ,
\end{equation}
with the operators $\hat{d}^\dagger_\alpha$ defined as
\begin{equation}
\label{eq:op_d_c_t}
\hat{d}^\dagger_\alpha (t)= \sum_i \psi^*_\alpha(t,i) \ \hat{c}^\dagger_i.
\end{equation}
The unitary evolution of the wavefunctions $\psi_\alpha(t,i)$ preserves the initial orthonormalization of the
stationary states $\psi_\alpha$ so that one has
\begin{equation}
\sum_i \psi_\alpha^*(t, i) \psi_{\alpha'}(t, i) = \delta_{\alpha \alpha'} \quad \forall t.
\end{equation}
from which the fermionic anticommuation relations of $\hat{d}^\dagger_{\alpha}$ operators follow.

Calculating a physical observable is again straightforward and amounts to calculating the observable for each filled state
\begin{align}
\langle \hat{\mathbf{A}} \rangle (t,i)
&\equiv  \langle \hat \psi(t) |  \hat{\mathbf{A}} |\hat \psi (t) \rangle \nonumber \\
&= \sum_{E_\alpha < E_F} \sum_j \psi_\alpha^*(t,i) \mathbf{A}_{ij} \psi_\alpha(t,j).
\end{align}
Let us emphasize that the sum over occupied states is crucial here as it enforces the Pauli principle\cite{gaury14}.

\subsection{Scattering wavefunction formalism: infinite system}
\label{sec:formalism_inf}

The scattering wavefunction formalism generalizes the previous subsection
to the case of infinite systems that consist of a finite scattering region connected to several leads that remain at their respective thermodynamic equilibrium.
The theory is exact for arbitrary time-dependent perturbations (no adiabatic assumption is necessary).
We partition the Hamiltonian as
\begin{equation}
\label{eq:Tkwant0}
\hat{\mathbf{H}} = \hat{\mathbf{H}}_0 + \hat{\mathbf{W}}(t).
\end{equation}
The scattering wave functions $\psi_{\alpha E}$ at $t<t_0$ are now labeled by the energy, a continuous variable $E$ (the system being infinite, the energy can take any value inside the bandwidth), and a discrete index $\alpha$ that labels all the conducting channels at energy $E$ such that
\begin{equation}
\label{eq:Tkwant1}
\mathbf{H}_0 \psi_{\alpha E} = E \psi_{\alpha E}.
\end{equation}

Note that despite the apparent resemblence of Eq.\ (\ref{eq:Tkwant1}) with Eq.\ (\ref{eq:eigenval}), they are of very different nature. While Eq.\ (\ref{eq:eigenval}) is simply the solution of the eigenvalue problem for a finite matrix, Eq.\ (\ref{eq:Tkwant1}) covers an infinite system with a continuous spectrum. The $\psi_{\alpha E}$ are obtained from wave function matching between the incoming and outgoing modes in the leads.

Very conveniently, $\psi_{\alpha E}$ are direct outputs of the \textsc{kwant} solver.
The scattering wavefunctions of \textsc{kwant} are normalized such that
they correspond to a unit particle current per channel and per energy (i.e.\ before wave matching, the lead
plane waves are normalized to carry unit incoming and outgoing current which guarantees the unitarity of the scattering matrix).
For $t>t_0$, one needs to follow the dynamics of these wavefunctions:
\begin{subequations}
\label{eq:Tkwant23}
\begin{align}
\label{eq:Tkwant2}
i \partial_t \psi_{\alpha E}(t, i) = \sum_j\mathbf{H}_{ij}(t) \psi_{\alpha E}(t,j) ,
 \\
\psi_{\alpha E}(t<t_0,i) = \psi_{\alpha E}(i) e^{-i E t}.
\label{eq:Tkwant3}
\end{align}
\end{subequations}
Unitarity of the time evolution implies
\begin{equation}
\label{eq:wf_normalization}
\sum_{\alpha} \int \frac{dE}{2 \pi}  \psi_{\alpha E}(t, i) \psi^\dagger_{\alpha E}(t, i) = 1, \quad \forall \, t,\, i.
\end{equation}

The observables are then calculated with
\begin{equation}
\label{eq:Tkwant4}
\langle \hat{\mathbf{A}} \rangle (t)
= \sum_{\alpha ij} \int \frac{dE}{2 \pi} f_\alpha(E)  \psi_{\alpha E}^*(t,i) \mathbf{A}_{ij}, \psi_{\alpha E}(t,j)
\end{equation}
where
\begin{equation}
f_\alpha(E) = \frac{1}{e^{(E - \mu_\alpha) / k_{B} T_\alpha} + 1}
\end{equation}
is the Fermi function of the lead to which channel $\alpha$ belongs. In particular, the number $n_i(t)$ of electrons on site $i$ reads
\begin{equation}
n_i(t) \equiv \langle \hat c^\dagger_i \hat c_i \rangle (t)
= \sum_{\alpha} \int \frac{dE}{2 \pi} f_\alpha(E)  |\psi_{\alpha E}(t,i)|^2 ,
\end{equation}
while the particle current $I_{ij}(t)$ from site $i$ to site $j$ reads
\begin{equation}
I_{ij}(t)
= - 2 \text{Im} \sum_{\alpha} \int \frac{dE}{2 \pi} f_\alpha(E)
  \psi_{\alpha E}^*(t,i) \mathbf{H}_{ij} \psi_{\alpha E}(t,j) ,
\end{equation}
with the usual continuity equation
\begin{equation}
\partial_t n_i(t) = \sum_j I_{ji}(t) .
\end{equation}

The above equations suppose that the entire spectrum consists of the continuum of scattering states. It is also possible that some discrete set of bound states $\psi_{b}$ with energy $E_b$ is present\cite{Li07, Dhar06, Khosravi09, Stefanucci07, Khosravi08} with evanescent contributions in the leads. In that case, the formula needs to be modified to account for those:\cite{istas18}
\begin{align}
\label{eq:observable_bs}
\langle \hat{\mathbf{A}} \rangle (t)
&= \sum_{\alpha ij} \int \frac{dE}{2 \pi} f_\alpha(E)  \psi_{\alpha E}^*(t,i) \mathbf{A}_{ij} \psi_{\alpha E}(t,j) &
 \nonumber \\
&+\sum_{b, j} f(E_b)  \psi_{b}^*(t,i) \mathbf{A}_{ij} \psi_{b}(t,j) ,
\end{align}
where the Fermi function $f(E_b)$ refers to the central region.

As announced above, the scattering wavefunction formalism is equivalent to the more standard Keldysh approach. In particular, the retarded and lesser Green's functions can be computed from the scattering wavefunction through simple integrals:\cite{gaury14}
\begin{subequations}
\begin{align}
\label{eq:green_func_wf}
\mathbf{G}_{ij}^R(t, t') &=
- i \theta(t - t') \sum_{\alpha} \int \frac{dE}{2 \pi}  \psi_{\alpha E}(t, i)  \psi^*_{\alpha E}(t', j) ,\\
\mathbf{G}_{ij}^<(t, t') &= i\sum_\alpha \int \frac{d E}{2 \pi} f_\alpha(E) \psi_{\alpha E}(t, i)  \psi^*_{\alpha E}(t', j) .
\end{align}
\end{subequations}
It is also possible to compute the scattering wavefunctions from the knowledge of the retarded Green's function.\cite{gaury14}

Equations (\ref{eq:Tkwant0}),(\ref{eq:Tkwant1}),(\ref{eq:Tkwant23}) and (\ref{eq:Tkwant4}) form the closes set that \textsc{tkwant}
solves.

\section{Numerical approach}
\label{sec:numerical_algorithms}

This section describes the set of algorithms used in \textsc{tkwant} to solve the closed set of time-dependent equations
(\ref{eq:Tkwant0}),(\ref{eq:Tkwant1}), (\ref{eq:Tkwant23}) and (\ref{eq:Tkwant4}).

\subsection{Overview of the different subproblems}

\textsc{Tkwant} consists of algorithms for the following subproblems

\begin{itemize}
\item[(1)] Definition of the model of Eq.\ (\ref{eq:Tkwant0}).  This is done with \textsc{kwant} to which we refer for further information.

\item[(2)] Calculation of the initial scattering states of Eq.\ (\ref{eq:Tkwant1}).  This is also performed using the \textsc{kwant} library.

\item[(3)] Integration of the time-dependent Schr{\"o}dinger equation in an infinite system for each of these states according to Eq.\ (\ref{eq:Tkwant23}). This subproblem is solved using a mapping onto an effective non-hermitian {\it finite} problem which we refer to as the ``source-sink'' algorithm. This finite problem is later integrated using standard schemes for differential equations.

\item[(4)] Calculation of the observables. This amounts to estimating accurately the integral (\ref{eq:Tkwant4}) over the energy using an appropriate quadrature rule. This step is critical in ensuring a proper treatment of the Pauli principle.\cite{gaury14}

\item[(5)] Band structure analysis. The calculation of the integral of subproblem (4) is actually performed in momentum $k$, not in energy $E$. A preliminary step consists in analyzing the band structure of each lead in order to perform the associated change of variable. This subproblem is solved using the package
\textsc{kwantSpectrum}\cite{kwantspectrum} which we also introduce in this article.
\end{itemize}

\subsection{Solving subproblem (3): integration of the time-dependent Schr{\"o}dinger equation for an infinite system}
\label{sec:source_sink}

In this section, we discuss how equation (\ref{eq:Tkwant23}) is integrated. Since the wave functions $\psi_{\alpha E}(t,i)$
are non-zero throughout the infinite system, a direct integration is not possible and one must first map the problem onto a finite problem. This is done
in two steps using the ``source'' and ``sink'' algorithm developed in Ref.~[\onlinecite{weston16a}]. The resulting differential equations are then integrated using
standard integration schemes.

\subsubsection{Source algorithm}
\label{sec:source}
The source algorithm is a simple change of variable where one writes
\begin{equation}
\psi_{\alpha E}(t,i)  = \psi_{\alpha E}(i) e^{-iEt}
+ \tilde\psi_{\alpha E}(t,i)e^{-iEt}.
\end{equation}
The new wavefunction $\tilde\psi_{\alpha E}(t,i)$ encodes the deviation
of the total wavefunction with respect to the stationary one. Inserting the
above definition into Eqs.\ (\ref{eq:Tkwant23}), one arrives at
\begin{subequations}
\begin{align}
\label{eq:Tkwant2_tilde}
i \partial_t \tilde\psi_{\alpha E}(t, i) &= \sum_j(\mathbf{H}_{ij}(t) -E\delta_{ij})
\tilde\psi_{\alpha E}(t,j) + S_{\alpha E}(t,i) ,
 \\
\tilde\psi_{\alpha E}(t<t_0,i) &= 0 ,
\label{eq:Tkwant3_tilde}
\\
S_{\alpha E}(t,i) &= \sum_j \mathbf{W}_{ij}(t) \psi_{\alpha E}(j) .
\end{align}
\end{subequations}
In other words, $\tilde\psi_{\alpha E}(t,i)$ follows a Schr\"odinger equation with an additional source term $S_{\alpha E}(t,i)$ that can be calculated from the scattering state. In return the initial value of the wavefunction is zero everywhere. Since the source term is only localized inside the scattering region, only a finite region of the system needs to be considered. The phase shift $e^{-iEt}$ in the definition of $\tilde\psi_{\alpha E}(t,i)$ is unimportant; it simply absorbs the faster time dependence which allows one to use significantly larger integration steps in the numerical integration.

\subsubsection{Sink algorithm}
\label{sec:sink}
For large simulation times, the wave function $\tilde\psi_{\alpha E}(t,i)$ penetrates deeply into the leads, so that a large finite system must be considered. One can indeed
consider a finite chunk of lead of length $v_{\rm max} t/2$ ($v_{\rm max}$: maximum speed in the system; the factor 2 accounts for the duration of both forward and backward propagation in the lead since the wavepackets get reflected at the lead boundary) to guarantee that no spurious reflection at the end of the finite lead alters the results.
It is important to note that even though one considers a finite system for the time-dependent propagation, the stationary wave-functions $\psi_{\alpha E}(i)$
are still computed for an infinite system, hence the results correspond to an exact solution of the infinite problem (within a given accuracy).
The corresponding algorithm has an overall computational cost
that asymptotically scales as $t^2$ although in many situations the cost is still dominated by the finite scattering region.

The ``sink'' algorithm developed in Ref.~[\onlinecite{weston16a}] allows to overcome this $t^2$ scaling and go down to a computational cost proportional to $t$. Since the leads are invariant by translation, the propagation inside the leads is ballistic: once a wave packet enters a lead, it never comes back to the scattering region and can be ignored.
To take advantage of this fact, one can introduce a ``sink'' in a lead: a purely imaginary potential $i\Sigma(i)$ that absorbs any wavefunction that penetrates into the lead. As a result, the dynamics becomes non-hermitian,
\begin{align}
\label{eq:Tkwant2_source_sink}
i \partial_t \tilde\psi_{\alpha E}(t, i) = \sum_j\left(\mathbf{H}_{ij}(t) -E\delta_{ij}\right)
\tilde\psi_{\alpha E}(t,j) \nonumber \\
+ S_{\alpha E}(t,i) +i\Sigma(i)\tilde\psi_{\alpha E}(t,i) .
\end{align}
The design of the absorbing term $i\Sigma(i)$ must be done with care in order to preserve the original dynamics: the imaginary potential must be increased very smoothly inside the leads as any abrupt variation of $i\Sigma(i)$ creates spurious back-scattering that sends parts of the wavepacket back into the scattering region and spoils the simulation. The concrete procedure to design the absorbing potential is described in details in Appendix \ref{sec:appendix_b}. Eq.\ (\ref{eq:Tkwant2_source_sink}) is the actual equation that is integrated into \textsc{tkwant}.

\subsection{Subproblem (4): Calculation of the physical observables}
\label{sec:momentum_intervals}

In this section, we discuss how \textsc{tkwant} solves the joint problem of performing the summation over conducting channels ($\alpha$) and the integration over energy.
When the time dependent perturbations are sufficiently slow and of small amplitude, 
all the physics happens close to the Fermi level. However, even in this case, 
the integration over the entire filled bands is required to respect Pauli's principle. 
Indeed, Pauli's principle requires the presence of all filled states to guarantee that they do not become 
occupied twice through inelastic processes. 
The unitary evolution of the individual states that was shown in Sec.\ \ref{sec:formalism_inf}
ensures that the antisymmetry of the initial state is preserved along the evolution, see also Sec.\ 1 in Ref.\ [\onlinecite{weston16b}] and Ref.\ [\onlinecite{gaury14}].

\begin{figure}[h]
	\centering
	\includegraphics[width=80mm]{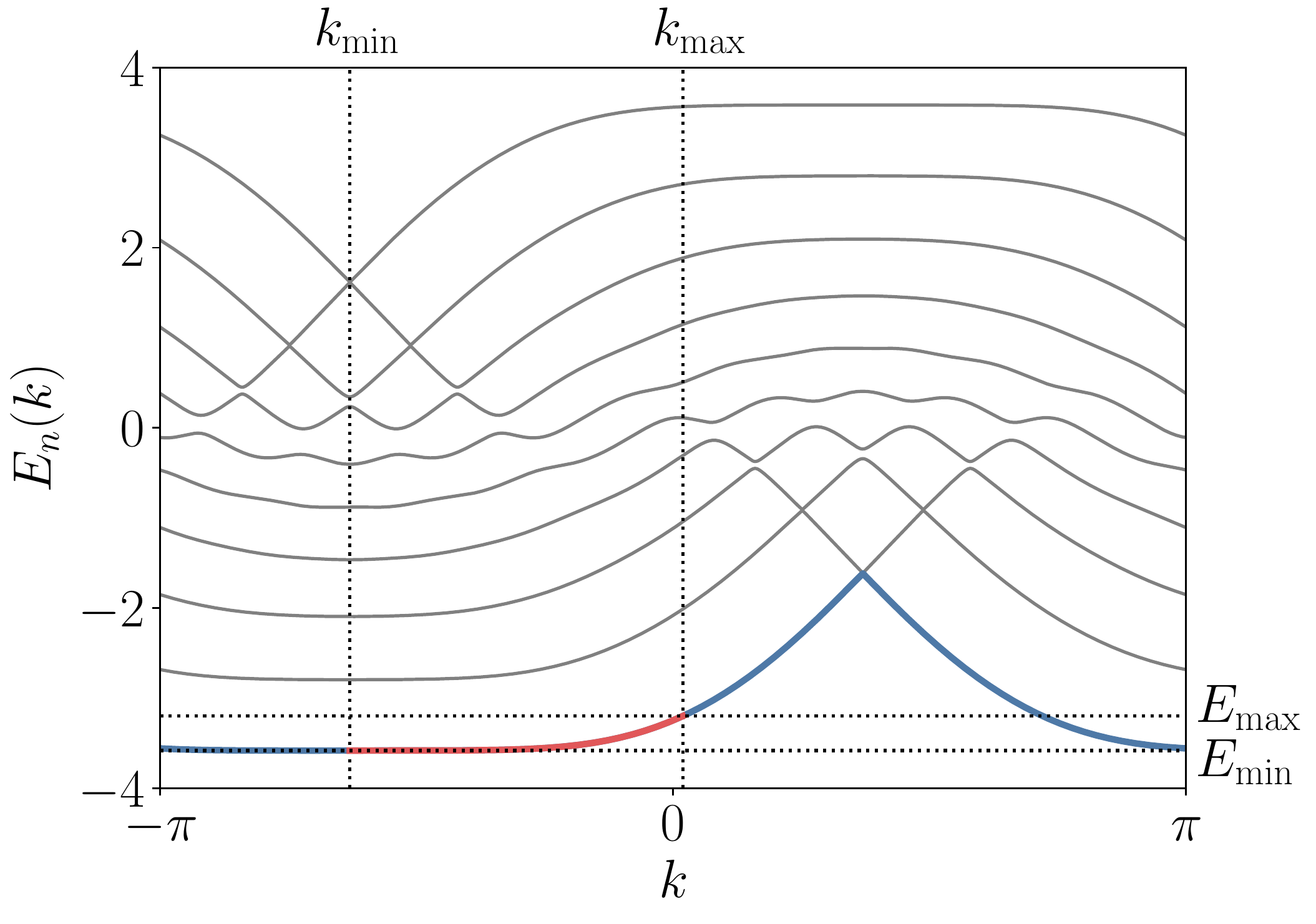}
	\vspace{-2ex}
  \caption{%
Dispersion spectrum $E_n(k)$ in the first Brillouin zone for lead 0 of the Mach-Zehnder interferometer from Ref.\ [\onlinecite{gaury15}]. At zero temperature, only the lowest band with $n = 0$ (blue) has energies below the Fermi energy $E_{\textrm F}$ that will contribute to the many-body state.
The contributing energies with positive velocity $v_0(k) \geq 0$ are highlighted in red.
Energy and momentum boundaries of the contributing area,
$E_{\rm min} = E_0(k_{\rm min})$ and $E_{\rm F} \equiv E_{\rm max} = E_0(k_{\rm max})$, are important to calculate many-body expectation value with Eqs.\ (\ref{eq:Tkwant4}) and (\ref{eq:Tkwant4k}).
}
\label{fig:fermi_filling}
\end{figure}

To understand the strategy in performing the integration and the summation of subproblem (4), it is very illuminating to look at the dispersion relation $E_n(k)$ of the different leads. \textsc{Kwant} provides a direct access to this dispersion relation. The package \textsc{kwantSpectrum} builds on this basic facility to provide a detailed analysis of the $E_n(k)$ curves.

\begin{figure}[tbh]
	\centering
	\includegraphics[width=40mm]{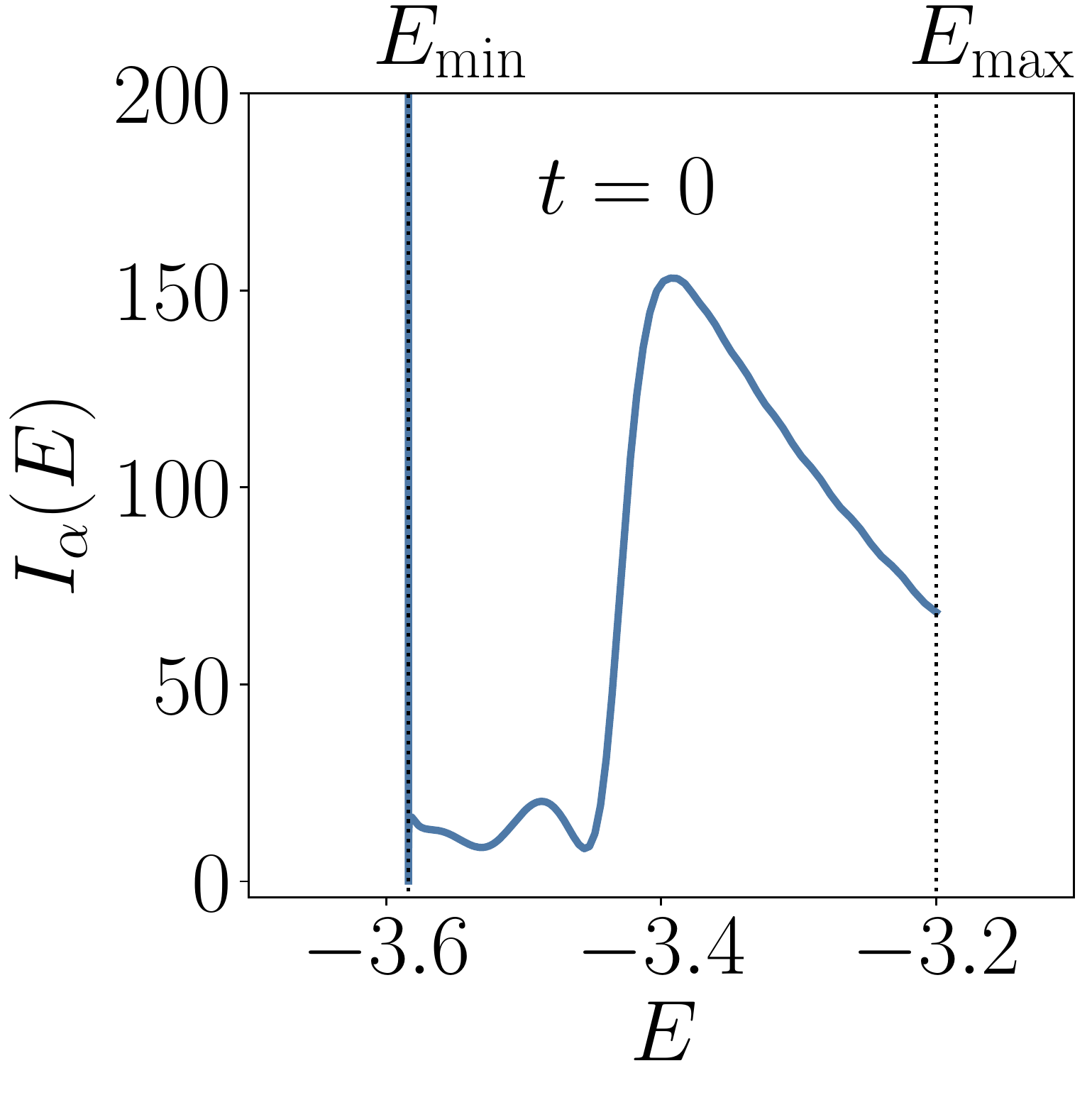}
	\includegraphics[width=41mm]{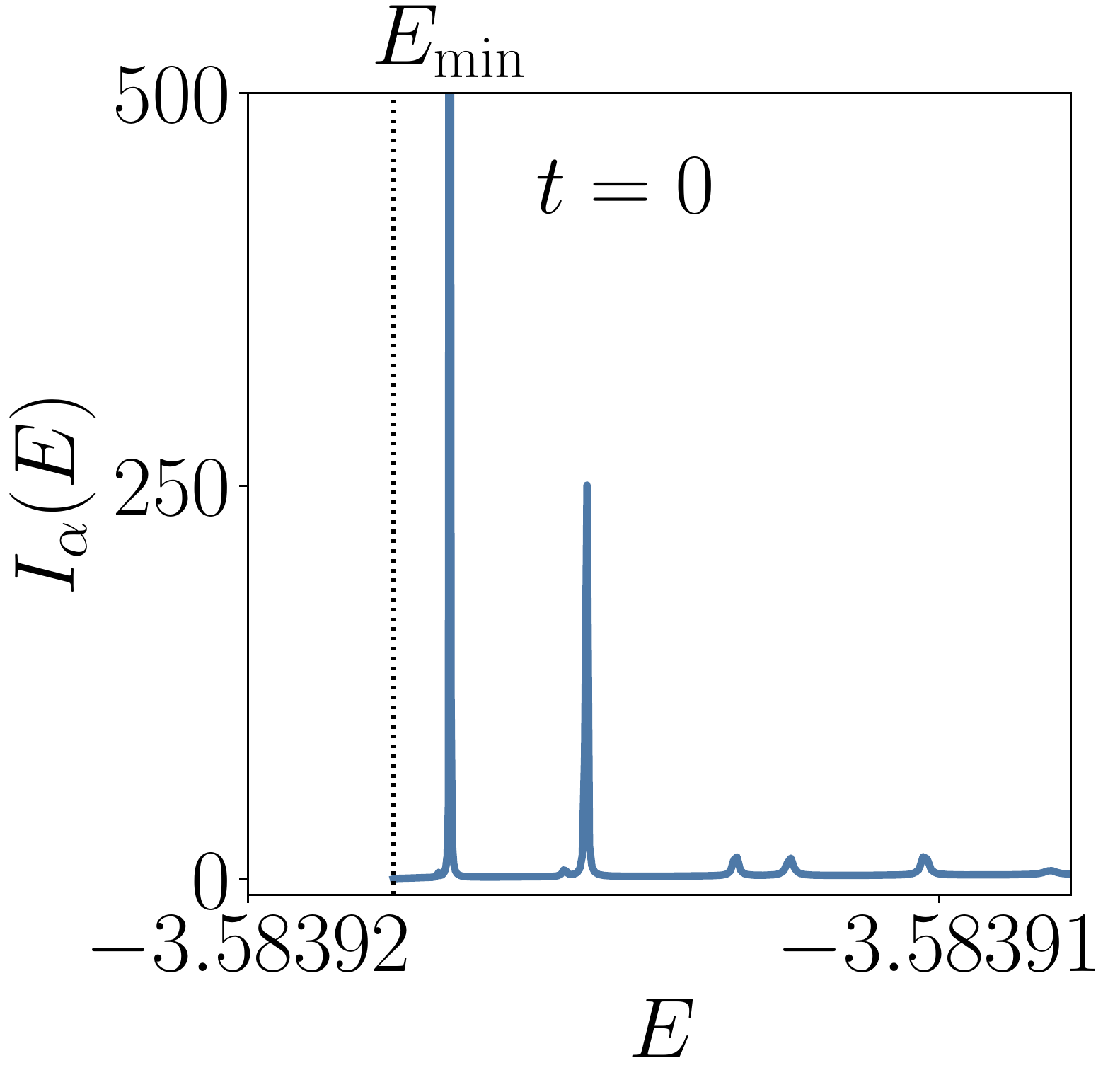}
	\includegraphics[width=40mm]{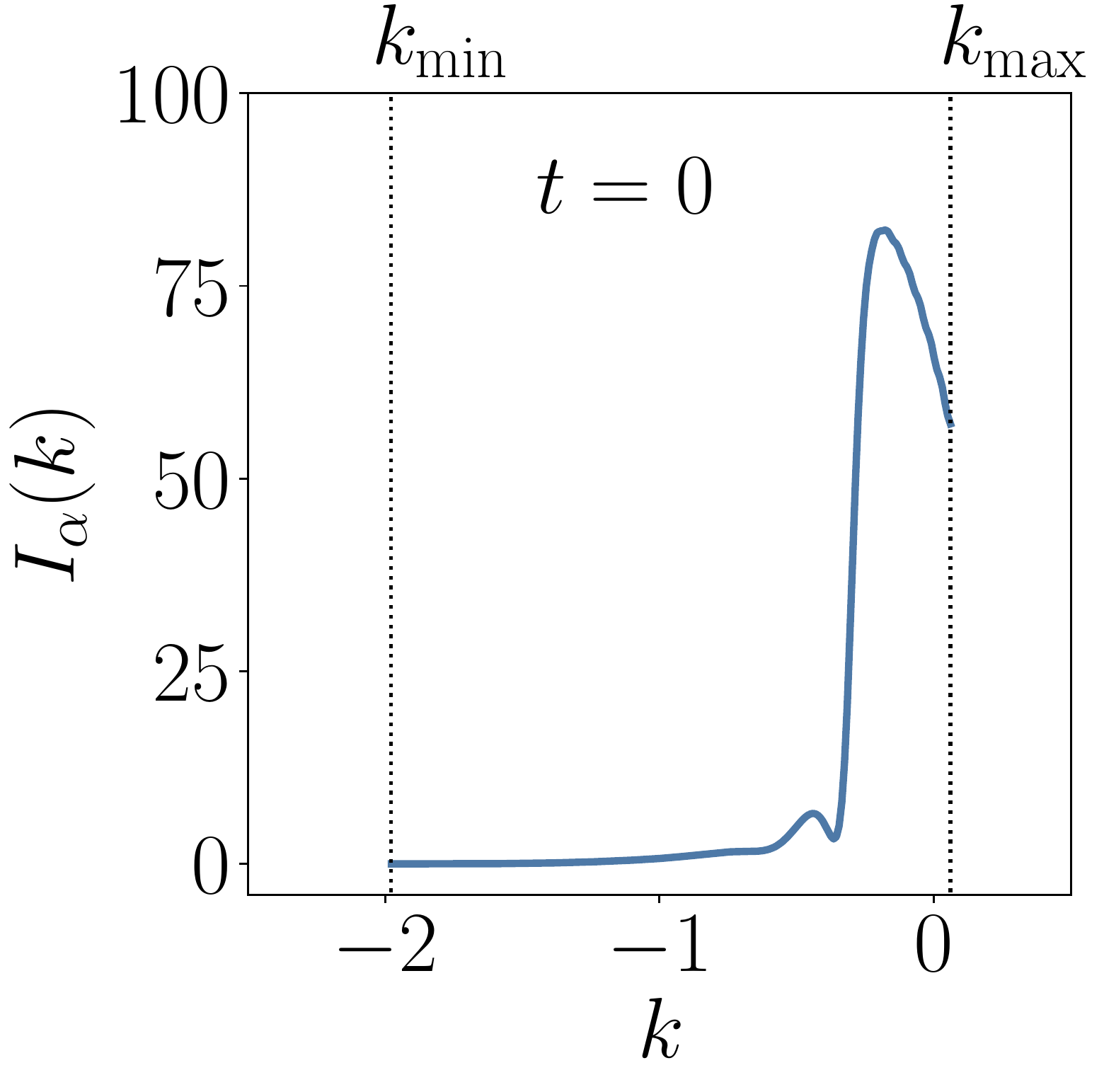}
	\includegraphics[width=40mm]{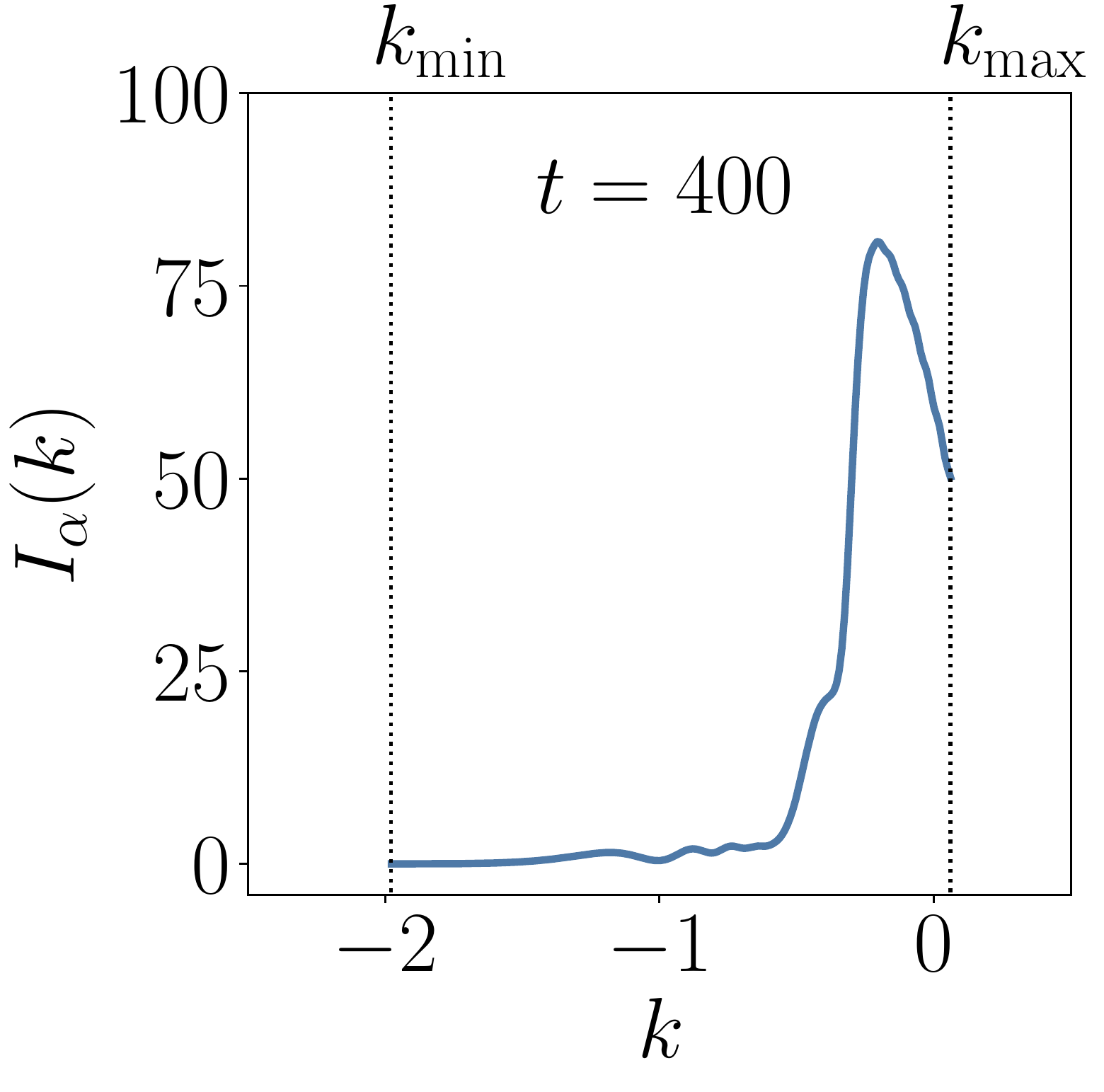}
	\vspace{-2ex}
  \caption{%
Integrand $I_\alpha$ of the many-body observable.
Upper panels: $I_\alpha(E)$ in energy representation Eq.\ (\ref{eq:observable_bs}). The divergence at the lower band gap $E_{\rm min}$ causes numerical inaccuracies, better visible
in the zoom on the right.
Lower panels: $I_\alpha(k)$ in momentum representation Eq.\ (\ref{eq:Tkwant4k}) at two different timesteps.
$I_\alpha(k)$ is a smooth function everywhere inside the integration region.
These integrands correspond to the electronic density in the Mach-Zehnder interferometer from Ref.\ [\onlinecite{gaury15}] that corresponds to lead 0, band $n=0$ contribution, summed over all the sites of the scattering region. Integration bounds correspond to Fig.\ \ref{fig:fermi_filling}.
}
\label{fig:quadrature}
\end{figure}

A typical example of a dispersion relation is shown in Fig.\ \ref{fig:fermi_filling}. This example corresponds to a quasi-one dimensional lead
in presence of a perpendicular magnetic field.
The low-energy bands correspond to the first Landau levels and are therefore very flat. Performing the  summation and the integration over energy amounts to integrating from the bottom of the band to the Fermi level $E_F$ (or up to $E_F$ plus a few time the temperature at finite $T$) and keeping the contributions arising from ``open'' channels. An open channel corresponds to a value of $k$ for which $\exists n, E_n(k)=E$ and the corresponding velocity
\begin{equation}
v_n(k) = \frac{dE_n(k)}{dk}
\end{equation}
is positive. For example in Fig.\ \ref{fig:fermi_filling}, there is a single open channel at $E=E_F$, two at $E=+2.5$ and none at $E=-4$. While such a direct integration over energy is possible, it suffers from serious difficulties. Indeed, close to the bottom of a band, the integrand -- that contains the density of states -- diverges. For a simple quadratic band opening $E\sim k^2$, this results in a $1/\sqrt{E}$ integrable singularity. For the example of Fig.\ \ref{fig:fermi_filling}, the bottom of the band is extremely flat (Landau level) and the associated density of states corresponds to a Dirac function. This is extremely ill-adapted to quadrature methods. An example of the integrand in
energy is shown in the top panels of Fig.\ \ref{fig:quadrature} with a zoom on the right. The very sharp peak associated to the Landau level is very hard to resolve numerically.

In order to avoid these divergences
and more generally to obtain smooth integrands, it is much  more favorable to perform the integral in $k$-space.\cite{weston16b} To do so, one starts by analyzing the band structure $E_n(k)$ in order to extract the intervals of integration $[k_{{\rm min}, \alpha}, k_{{\rm max}, \alpha}]$. Specific algorithms have been developed to perform this analysis (finding the bottom and top of the bands where $v_n(k)=0$, ensuring continuity of the bands at band crossings, etc.).  They correspond to subproblem (5) and are described in Appendix \ref{sec:appendix_a}. In our example of Fig.\ \ref{fig:fermi_filling}, there is a single interval $[k_{{\rm min}, 0}, k_{{\rm max}, 0}]$ (in red) but more intervals would appear as one increases the Fermi energy. Performing the integration in $k$ introduces a Jacobian $|dE_n/dk| = v_n$
that absorbs the divergences of the integrand in energy. The resulting formula for the calculation of an observable reads
\begin{equation}
\label{eq:Tkwant4k}
\langle \hat{\mathbf{A}} \rangle (t)  \!
= \! \sum_{\alpha} \!\!\! \int_{k_{{\rm min}, \alpha}}^{k_{{\rm max}, \alpha}} \!\! \frac{dk}{2 \pi} f_\alpha(E_\alpha(k)) v_\alpha(k) \psi_{\alpha k}^*(t,i) \mathbf{A}_{ij} \psi_{\alpha k}(t,j).
\end{equation}
An example of the corresponding integrand in $k$-space is shown in the lower panels of Fig.\ \ref{fig:quadrature}. These integrands are perfectly smooth, in contrast to their counterparts in $E$-space shown in the upper panels.

The last step, once all the momentum intervals are at hand, it to evaluate the
corresponding integrals using quadrature rules of the form
\begin{equation}
\label{eq:quadrature_rule}
\int_{k_{{\rm min}}}^{k_{{\rm max}}} d k \, g(k) \simeq \sum_i w_i \, g(k_i).
\end{equation}
\textsc{Tkwant} uses two kinds of quadrature rules with either a fixed number of points
(Gauss-Legendre rules) or an adaptive number of points (Gauss-Kronrod rules\cite{quadpack, numerical_recipes}). Both quadratures have the additional advantage that the integrand is not evaluated at the boundaries of the interval where band opening leads to ill-defined behavior of the integrand ($1/\sqrt{E}$ singularities for the integration in energy domain).

\section{Software architecture and main concepts}
\label{sec:architecture}

In this section, we describe how \textsc{tkwant} is organized. \textsc{tkwant} implements several concepts that provide a clean separation between the different subproblems and
allow the package to be easily modified or extended. For instance, although \textsc{tkwant}'s main focus is time-dependent nanoelectronic problems, it can also be used for simpler problems such as the propagation of a single-particle wave packet in an infinite or even finite system.

\textsc{Tkwant} has separate APIs for one-body problems and many-body problems.
For each of these, it proposes a low-level interface that exposes all the mathematical objects used in the algorithms and a high-level interface that provides additional functionality as well as heuristics to propose robust values
of the simulation parameters (such as the imaginary potential or the number quadrature points in the calculation of the integrals). The low-level API of both
one-body and many-body problems has been designed to be compatible but independent from \textsc{kwant} while the high-level interface relies on \textsc{kwant} more heavily.

\subsection{Solving one-body problems}

To illustrate the one-body solvers, let us consider the simple problem of the propagation of a wavepacket in one dimension.  This means we want to integrate
\begin{equation}
i \hbar \partial_t \psi(t, x) = - \frac{\hbar^2}{2m}\partial_x^2 \psi(t, x)
\end{equation}
with some initial condition, for example
\begin{equation}
\psi(t=0, x) = \psi_0(x) = - \frac{1}{\sqrt{\pi}} e^{-\frac{x^2}{2} + ikx}.
\end{equation}
The first step for such a simulation is to discretize the spatial variable $x$.
This can be done automatically\footnote{\textsc{Kwant} provides a discretizer to translate continuum into tight-binding models.}
or manually by approximating the $\partial^2_x$ operator with a three-point rule on an equidistant grid $x_i = i a$ where $a$ is the discretization lattice constant.
One arrives at a tight binding model of the form of Eq.\ (\ref{eq:Tkwant2})
with the Hamiltonian matrix
\begin{equation}
H = \left( \begin{array}{cccc}2 & -1 & & 0\\-1 & \ddots & \ddots &  \\& \ddots & \ddots &  -1 \\0 &  & -1 & 2  \end{array}  \right),
\label{fig:hamil_mat_expl}
\end{equation}
with energies [times] measured in units of $\hbar^2 / (2 m a^2)$ [$2 m a^2/\hbar$]. \textsc{Tkwant} provides solvers for the above equation, possibly in presence of a time-dependent and spatially-dependent potential for both finite and infinite systems.

\begin{figure}[h]
	\centering
	\includegraphics[width=80mm]{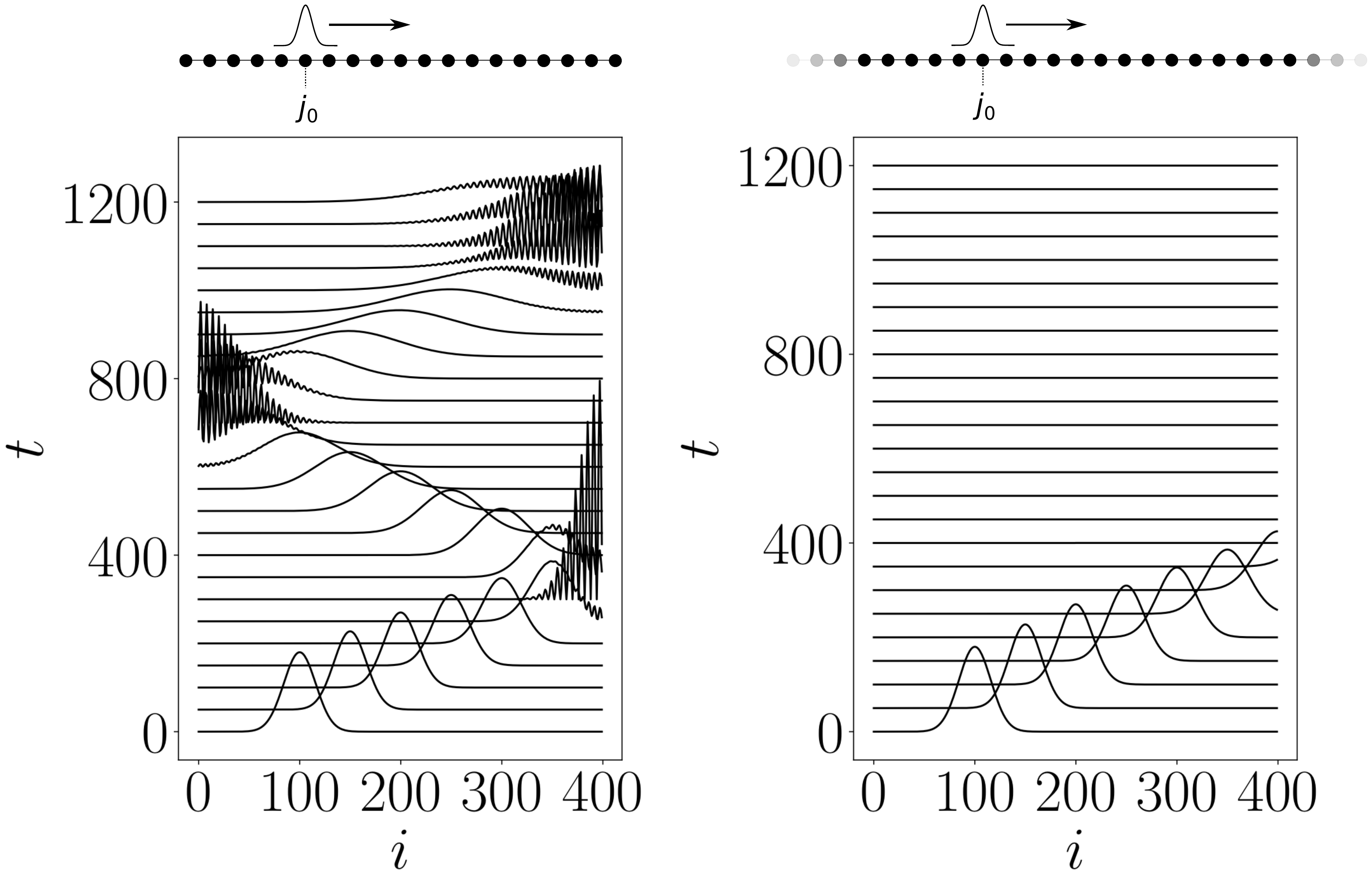}
	\vspace{-2ex}
  \caption{%
Time evolution of the probability density $|\psi(t,i)|^2$ on a one-dimensional chain. On the left panel, the chain has a finite size, so that
the pulse gets  reflected successively on the left and the right boundary.
The right panel shows the same simulation for an infinite chain, where the
the pulse continues its propagation without reflection by leaving the central scattering region.
Initial condition
$\psi(t = 0, j) = e^{- b (j - j_0)^2 + i k j}$,
$b = 0.001$, $j_0 = 100$, $k = \pi / 6$, $N_s = 400$ (central scattering region).
The simulations were performed with code listings \ref{lst:closed_simple}, or \ref{lst:closed_kwant} and \ref{lst:open_kwant}. The plots show traces of density versus space at different times. Each trace is offset by a constant proportional to time in order to make the propagation apparent.
}
\label{fig:onebody_pulse}
\end{figure}

\subsubsection{Finite systems}

The dynamics of the probability density $|\psi(t, x)|^2$ for a finite system of $N_s=400$ sites is shown in the left panel of Fig.\ \ref{fig:onebody_pulse}. The initial condition is a Gaussian wave packet centered at $j_{\text 0}=100$ with a momentum $k=\pi /6$. As the dispersion relation of the infinite chain is $E(k) = 2 - 2\cos k$,  the wavepacket has initial group velocity $v(k) = \partial_k E = 1$ (in units of lattice spacing $a$ per time unit) towards the right of the system. As the system is finite, the wave packet gets reflected at the boundaries and displays a ping-pong like dynamics while at the same time the wavepacket spreads.

Listing \ref{lst:closed_simple} uses the \textsc{tkwant} low-level interface, namely the class \texttt{onebody.WaveFunction}, to obtain the data of the left panel of Fig.\ \ref{fig:onebody_pulse}.
After defining the temporal and spatial grids (\texttt{time} and \texttt{xi}), the  Hamiltonian matrix \texttt{H0} of  Eq.\ (\ref{fig:hamil_mat_expl}) is constructed with standard Python tools. The one-body Schr\"odinger equation is finally solved in lines 21--25. The \texttt{evolve()} method in line 25 propagates the one-body state forward in time. \textsc{Tkwant} currently employs an explicit Runge-Kutta method of order (4)5 with adaptive Dormand and Prince stepsize control\cite{hairer93} for this task.

\begin{lstlisting}[language=Python, label=lst:closed_simple,
caption={
Python code to calculate the time-evolution of the probability density
$|\psi(t,i)|^2 $ on a finite one-dimensional chain (left panel of Fig.\ \ref{fig:onebody_pulse}).
The Hamiltonian matrix Eq.\ (\ref{fig:hamil_mat_expl}) is defined explicitly using sparse matrices, \texttt{psi0} is the initial condition and the one-body Schr\"odinger equation is solved using \textsc{tkwant}.  Running this script takes only a few seconds on a desktop computer.
}
]
from tkwant import onebody
import numpy as np
import scipy
import matplotlib.pyplot as plt

# Define spacial and temporal grids.
xi = np.arange(400)
times = np.arange(0, 1201, 50)

# Initial condition.
k = np.pi / 6
psi0 = np.exp(- 0.001 * (xi - 100)**2 + 1j * k * xi)

# Hamiltonian matrix.
diag = 2 * np.ones(len(xi))
offdiag = - np.ones(len(xi) - 1)
H0 = scipy.sparse.diags([diag, offdiag, offdiag], [0, 1, -1],
                        dtype=complex)

# Initialize the onebody wavefunction solver.
wave_func = onebody.WaveFunction(H0, W=None, psi_init=psi0)

# Loop over the timesteps and plot the result.
for time in times:
    wave_func.evolve(time)
    psi = wave_func.psi()
    density = np.real(psi * psi.conjugate())
    # Prefactor and shift for representation purpose.
    plt.plot(xi, 180 * density + time)
\end{lstlisting}

Listing \ref{lst:closed_kwant} performs the same task as Listing \ref{lst:closed_simple} but uses \textsc{kwant}\cite{groth14} for the construction of the Hamiltonian matrix and for the calculation of the density. For such a simple example, using \textsc{kwant} is superfluous. However in more complex situations (time-dependent systems of various shapes, with different lattices or topologies, etc.) it becomes very handy.
The method \texttt{evaluate()} calculates the expectation value of an operator.  The class \texttt{onebody.WaveFunction} interprets any argument with name \texttt{time} automatically as the time argument and attributes the corresponding Hamiltonian elements to the $W(t)$ matrix.
\begin{lstlisting}[language=Python, label=lst:closed_kwant,
caption={
This Python code is similar to Listing \ref{lst:closed_simple} except that the Hamiltonian matrix and the density operator are defined using \textsc{kwant}\cite{groth14}. The object \texttt{syst} is the Kwant object that represents the finite system. It contains the Hamiltonian matrix and can be used by \textsc{tkwant}'s one-body solver \texttt{onebody.WaveFunction}. Running this script takes only a few seconds on a desktop computer.
}
]
from tkwant import onebody
import kwant
import numpy as np
import matplotlib.pyplot as plt

def make_system(L):

    # Define an empty tight-binding system on a square lattice.
    lat = kwant.lattice.square(a=1, norbs=1)
    syst = kwant.Builder()

    # Central scattering region.
    syst[(lat(x, 0) for x in range(L))] = 2
    syst[lat.neighbors()] = -1

    return syst

# Build the system using Kwant.
syst = make_system(400).finalized()

# Get lattice positions and define temporal grid.
xi = np.array([site.pos[0] for site in syst.sites])
times = np.arange(0, 1201, 50)

# Define observables using Kwant.
density_operator = kwant.operator.Density(syst)

# Initial condition.
k = np.pi / 6
psi0 = np.exp(- 0.001 * (xi - 100)**2 + 1j * k * xi)

# Initialize the onebody wavefunction solver.
wave_func = onebody.WaveFunction.from_kwant(syst, psi0)

# Loop over the timesteps and plot the result.
for time in times:
    wave_func.evolve(time)
    density = wave_func.evaluate(density_operator)
    # Prefactor and shift for representation purpose.
    plt.plot(xi, 180 * density + time)
\end{lstlisting}

\subsubsection{Infinite systems}

The dynamics of the probability density $|\psi(t, x)|^2$ for an {\it infinite} system is shown in the right panel of Fig.\ \ref{fig:onebody_pulse}. In contrast to the previous example, the wavepacket is not reflected on the boundary of the system but continues its propagation indefinitely. The finite system here only corresponds to the window that we are monitoring but the physical system is strictly infinite and translationally invariant.

The corresponding code is shown in Listing \ref{lst:open_kwant}. The differences with Listing \ref{lst:closed_kwant} are highlighted in blue. The chain is extended to positive and negative infinity by attaching semi-infinite leads to the \textsc{kwant} system.  The Hamiltonian matrix of the infinite (or open) system has a block structure similar to that of Eq.\ (\ref{eq:hamltonian_split}). Note that we have to provide special boundary conditions (imaginary potential, cf.\ Sec.\ \ref{sec:appendix_b}) to the \textsc{tkwant} solver (line 41) to deal with infinite systems.

\begin{lstlisting}[language=Python, label=lst:open_kwant,
caption={
Python code to calculate the time-evolution of the probability density
$|\psi(t,i)|^2 $ for an infinite one-dimensional chain (right panel of Fig.\ \ref{fig:onebody_pulse}). The object \texttt{syst} is the \textsc{kwant} system that represents the infinite system. It has leads on both sides of the finite scattering region that extend the chain to $\pm \infty$. Additional boundary conditions must be provided to the \texttt{onebody.WaveFunction} solver for a system with leads.
New lines of code (in blue) and comments (in gray) are highlighted to show the difference to Listing \ref{lst:closed_kwant}. Running this script takes only a few seconds on a desktop computer.
}
]
from tkwant import onebody<@\textcolor{blue}{, leads}@>
import kwant
import numpy as np
import matplotlib.pyplot as plt

def make_system(L):

    # Define an empty tight-binding system on a square lattice.
    lat = kwant.lattice.square(a=1, norbs=1)
    syst = kwant.Builder()

    # Central scattering region.
    syst[(lat(x, 0) for x in range(L))] = 2
    syst[lat.neighbors()] = -1

    <@\textcolor{warmgray}{\# Attach lead on the left- and on the right-hand side.}@>
    <@\textcolor{blue}{sym = kwant.TranslationalSymmetry((-1, 0))}@>
    <@\textcolor{blue}{lead\_left = kwant.Builder(sym)}@>
    <@\textcolor{blue}{lead\_left[lat(0, 0)] = 2}@>
    <@\textcolor{blue}{lead\_left[lat.neighbors()] = -1}@>
    <@\textcolor{blue}{syst.attach\_lead(lead\_left)}@>
    <@\textcolor{blue}{syst.attach\_lead(lead\_left.reversed())}@>

    return syst

# Build the system using kwant.
syst = make_system(400).finalized()

# Get lattice positions and define temporal grid.
xi = np.array([site.pos[0] for site in syst.sites])
times = np.arange(0, 1201, 50)

# Define observables using Kwant.
density_operator = kwant.operator.Density(syst)

# Initial condition.
k = np.pi / 6
psi0 = np.exp(- 0.001 * (xi - 100)**2 + 1j * k * xi)

<@\textcolor{warmgray}{\# make boundary conditions for the system with leads}@>
<@\textcolor{blue}{boundaries = leads.automatic\_boundary(syst.leads,\,
                                      tmax=max(times))}@>

# Initialize the onebody wavefunction solver.
wave_func = onebody.WaveFunction.from_kwant(syst,<@\,@>psi0<@\textcolor{blue}{,\,boundaries}@>)

# Loop over timesteps and plot the result.
for time in times:
    wave_func.evolve(time)
    density = wave_func.evaluate(density_operator)
    # Prefactor and shift for representation purpose.
    plt.plot(xi, 180 * density + time)
\end{lstlisting}

\subsubsection{Infinite systems with initial scattering states}

In \textsc{tkwant} special support exists for the simulation of infinite systems whose initial state is a scattering state of the system.
The scattering states are obtained from the numerical solution of Eq.\ (\ref{eq:Tkwant1}), and this step is conveniently performed with \textsc{kwant}.
For the one-dimensional chain, the scattering states have the simple form
\begin{equation}
\psi_\alpha(t, x) = \frac{1}{\sqrt{v(k)}} e^{i (k x - E t) }.
\end{equation}

In presence of a time-dependent perturbartion, scattering states  immediately become more complex as a reflected wave must be added in the left lead while the wave on the right gets multiplied by a transmission amplitude and the wave in the scattering region loses its plane-wave structure.
Scattering state initial conditions are somewhat special in two ways. First, scattering states are eigenstates of $\hat{\mathbf{H}}_0$ hence a time-dependent perturbation is needed to observe a nontrivial time evolution. Second, these initial conditions are defined everywhere in the infinite system (hence the appearance of the source terms in the Schr\"odinger equation, see Section \ref{sec:source_sink}) as opposed to just inside the scattering region as is the case for a simple wave packet.

The high-level class \texttt{onebody.ScatteringStates} handles the calculation of these initial conditions, of the associated source terms, and provides robust automatic heuristics for setting up proper boundary conditions in the leads (imaginary potential).
Listing \ref{lst:open_scatt} shows an example of the API of \texttt{onebody.ScatteringStates}. An instance of \texttt{onebody.ScatteringStates} is an iterable object that returns \texttt{onebody.WaveFunction} objects upon iteration.

\begin{lstlisting}[language=Python, label=lst:open_scatt,
caption={Python code snippet to set up a one-body solver for the time-dependent Schr\"odinger equation for an infinite system that starts in an initial scattering state.
To use this snippet, one should replace lines 38--45 in listing \ref{lst:open_kwant} by the above lines.
\texttt{syst} is a \textsc{kwant} system with leads, \texttt{energy} the energy of the state and \texttt{lead=0} refers to the left lead.  Note that the boundary conditions are built automatically on the fly.
}
]
wave_func = onebody.ScatteringStates(syst, energy=0., lead=0,
                                     tmax=1200)[0]
\end{lstlisting}

\subsection{Solving the many-body problem}
\label{sec:tkwant_manybody}
Let us now turn to the many-body solver of \textsc{tkwant}. Solving the many-body Schr\"odinger equation with \textsc{tkwant} requires several steps as described in Sec.\  \ref{sec:numerical_algorithms}. \textsc{Tkwant} provides two interfaces for solving the many-body problem.

The first, class \texttt{manybody.WaveFunction}, provides a low-level interface for the problem. Its main task is to handle the evolution of multiple scattering states (in parallel for multi-core computers) and perform the integration over energy using a static number of scattering states. When using \texttt{manybody.WaveFunction} the different preprocessing steps must be handled manually. They consist of
\begin{itemize}
\item the calculation of the dispersion relation $E_n(k)$ for all leads,
\item the analysis of $E_n(k)$ to obtain the $k$-intervals for the integration,
\item the calculation of the imaginary potential in the leads,
\item and the calculation of the initial scattering states at $t = 0$.
\end{itemize}

The other class, \texttt{manybody.State}, provides a high-level interface that offers additional functionality: it uses heuristics to automatically handle the preprocessing steps; it implements an adaptive integration scheme that allows one to refine the integration by adding new points on the fly. Note that in what follows, we concentrate on the treatment of the energy/momentum integration on the continuum part of the spectrum. Bound states, if present, must also be accounted for.  We refer to \textsc{tkwant} documentation for a description of the corresponding API.\cite{tkwant}

\subsubsection{Low-level API}
Listing \ref{lst:manybody_wf} showcases usage of the low-level interface, supposing that a \textsc{kwant} system \texttt{syst} has already been constructed. Line 8 calculates and analyzes the dispersion relations of the different leads.
Line 12 sets up the Fermi functions of the different electrodes. Line 13
calculates the maximum energy $E_{\rm max}$ of the energy integration (energy above which the Fermi functions are effectively zero). Line 16 sets up an imaginary potential in the leads adapted to their actual spectrum. Lines 19--21
set up the ``quadrature intervals'' that will be used for the integration. A quadrature interval is an interval in $k$ to which a quadrature rule (here Gauss-Legendre) is associated
along with the order in which this rule will be used (here 20, meaning that 20 points will be used per interval). The function \texttt{split\_intervals} allows to split one interval into several subintervals in order to obtain a higher accuracy of the integration. Line 24 sets up the different ``tasks'', i.e.\ the different one-body problems that must be integrated. Line 25 calculates the initial condition for each task. All this information is gathered (line 28) by the \texttt{manybody.WaveFunction} instance that is in charge of integrating the different one-body problems and performing the integration. Note that at this level, the integration is performed on a fixed number of predefined points.

The core routines of \texttt{manybody.WaveFunction} handle the different tasks in parallel using the Message Passing Interface (MPI) \cite{mpi32_standard} framework. As the problem is embarrassingly parallel it easily scales to thousands of cores.
In addition to saving computing time, the distribution of tasks in a parallel execution also lowers the memory footprint per core,
so that \textsc{tkwant} simulation are usually not limited by the amount of memory available.
The time-resolved simulation of a system whose static \textsc{kwant} simulation runs on a single core, typically requires around one hundred cores or more if comparable computation times are desired.

\begin{lstlisting}[language=Python, label=lst:manybody_wf,
caption={Python code snippet to build up the many-body wavefunction manually.
The different steps reflect the numerical algorithm and the comments follow the preprocessing step described in Sec.\ \ref{sec:tkwant_manybody}. Note that the number of interval splits (\texttt{number\_subintervals=10} in the example) is highly system-dependent.
}
]
from tkwant import leads, manybody as mb
import kwantspectrum
import functools.partials as part

dens_op = kwant.operator.Density(syst)

# Calculate the spectrum E(k) for all leads.
spectra = kwantspectrum.spectra(syst.leads)

# Estimate the cutoff energy Ecut from T, \mu and f(E).
# All states are effectively empty above E_cut
occupations = mb.lead_occupation(chemical_potential=0, temperature=0)
emin, emax = mb.calc_energy_cutoffs(occupations)

# Define boundary conditions.
bdr = leads.automatic_boundary(spectra, tmax, emin=emin, emax=emax)

# Calculate the k intervals for the quadrature.
intvl_type = part(mb.Interval, order=20, quadrature='gausslegendre')
intervals = mb.calc_intervals(spectra, occupations, interval_type)
intervals = mb.split_intervals(intervals, number_subintervals=10)

# Calculate all one-body scattering states at time t = 0.
tasks = mb.calc_tasks(intervals, spectra, occupations)
psi_init = mb.calc_initial_state(syst, tasks, bdr)

# Set up the manybody wave function.
wave_function = mb.WaveFunction(psi_init, tasks)

for time in range(tmax):  # Loop over timesteps.
    wave_function.evolve(time)
    density = wave_function.evaluate(dens_op)
\end{lstlisting}

The low-level interface has been designed to be very modular so that it can be adapted or extended  to new situations easily. The convergence of the integral of Eq.\ (\ref{eq:Tkwant4k}) must be checked manually. Increasing its accuracy is possible by using quadrature rules of higher order and by splitting the initial intervals (such as the one shown in Fig.\ \ref{fig:fermi_filling}) into subintervals. We have found empirically that using 10--20 points per sub-interval is usually optimal while using higher orders often brings little benefit. The number of sub-intervals must then be increased until the result converges.  However this number is dependent very much on the paricular system under study. The main advantage of the high-level interface described below is adaptative refinement of the integral.

\subsubsection{High-level API}
The class \texttt{manybody.State} forms the high-level interface for the many-body problem. It takes care of all the preprocessing steps automatically so that setting up a simulation becomes as simple as

\begin{lstlisting}[language=Python, label=lst:manybody_state,
caption={Python code snippet to compute a many-body wavefunction in an automatic way.
This code should give results similar to the code of listing \ref{lst:manybody_wf} but
an additional adaptive quadrature helps to assure the numerical accuracy.
}]
from tkwant import manybody as mb

dens_op = kwant.operator.Density(syst)

occupations = mb.lead_occupation(chemical_potential=0, temperature=0)

state = mb.State(syst, tmax, occupations, error_op=dens_op)

for time in range(tmax):  # Loop over timesteps.
    state.evolve(time)
    state.refine_intervals(atol=1e-05, rtol=1e-05)
    density = state.evaluate(dens_op)
    error = state.estimate_error()

\end{lstlisting}
While being slightly less flexible than the low-level approach, it is more convenient and sufficient in most cases.

The main additional facility provided by \texttt{manybody.State} is the ability to dynamically adapt the number of points used to perform the energy/momentum integral.
The function \texttt{refine\_interval()} on line 11 of Listing \ref{lst:manybody_state} estimates the error in the integration and then proceeds to split the integration interval into subintervals if necessary.
Line 13 shows the corresponding estimate of the integration error using the \texttt{state.estimate\_error()} method.
A global adaptive strategy, based on Quadpack's algorithm \cite{quadpack} is used for the refinement cycle and the error estimate.

The adaptive calculation of the integral is a non-trivial and computationally intensive problem. Indeed, here the integrand {\it depends on time}. The regions in $k$-space that dominate the integral at a given time might be different from the regions that dominate at a later time. Furthermore, anytime the algorithm decides that more points are necessary in a certain part of $k$-space to achieve a given accuracy, these new points must be evolved all the way from $t=0$ to the current time of the simulation. From a computational perspective this is suboptimal as it interferes with parallelization (computing cores must wait until the new tasks ``catch up''). To minimize this effect, we found empirically that it is best to perform the refinement early in the simulation with a slightly smaller error tolerance than the ultimately targeted one.

\subsection{Overall architecture and code design}

\begin{figure}[h]
	\centering
	\includegraphics[width=80mm]{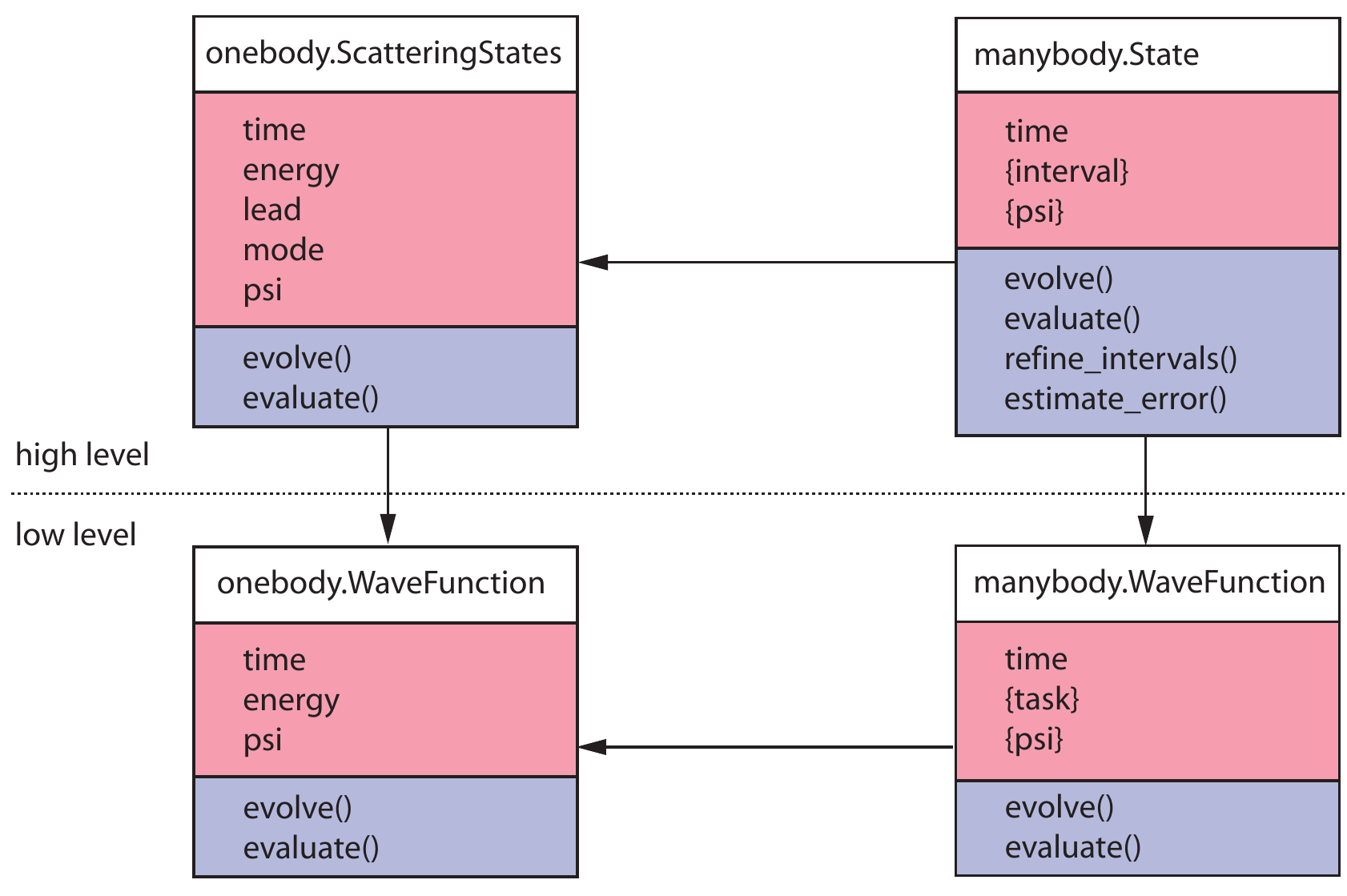}
	\vspace{-2ex}
  \caption{%
Relation diagram of the four solver classes implemented in \textsc{tkwant}. For two classes A and B, the notation A $\leftarrow$ B indicates a reliance of B on A's interface specification, with A being totally unaware on B, or in other words: ``A is used by B''. 
Representative attributes and methods are listed inside the red and respectively blue boxes and curly brackets \{ \} depict sets of objects.
Note that implementation details might differ from above representation.
}
\label{fig:structure_diagram_solver}
\end{figure}

The design of \textsc{Tkwant} is centered around the four classes that have already been introduced above.
They implement, respectively, the one-body/many-body states of the system at low/high level of abstraction.
Functions exist to help with the various pre-calculations that arise at the beginning of a simulation.
Fig.\ \ref{fig:structure_diagram_solver} shows the relation between the main \textsc{tkwant} classes.

The four solver classes provide at least two methods: an \texttt{evolve()} method to evolve the wavefunction(s) forward in time and an \texttt{evaluate()} method, to calculate expectation values of an operator.  Extending the functionality of the solvers can be achieved by providing classes with a similar interface (``duck typing'').
We find this approach preferable to inheritance mechanisms.

\begin{figure}[h]
	\centering
	\includegraphics[width=80mm]{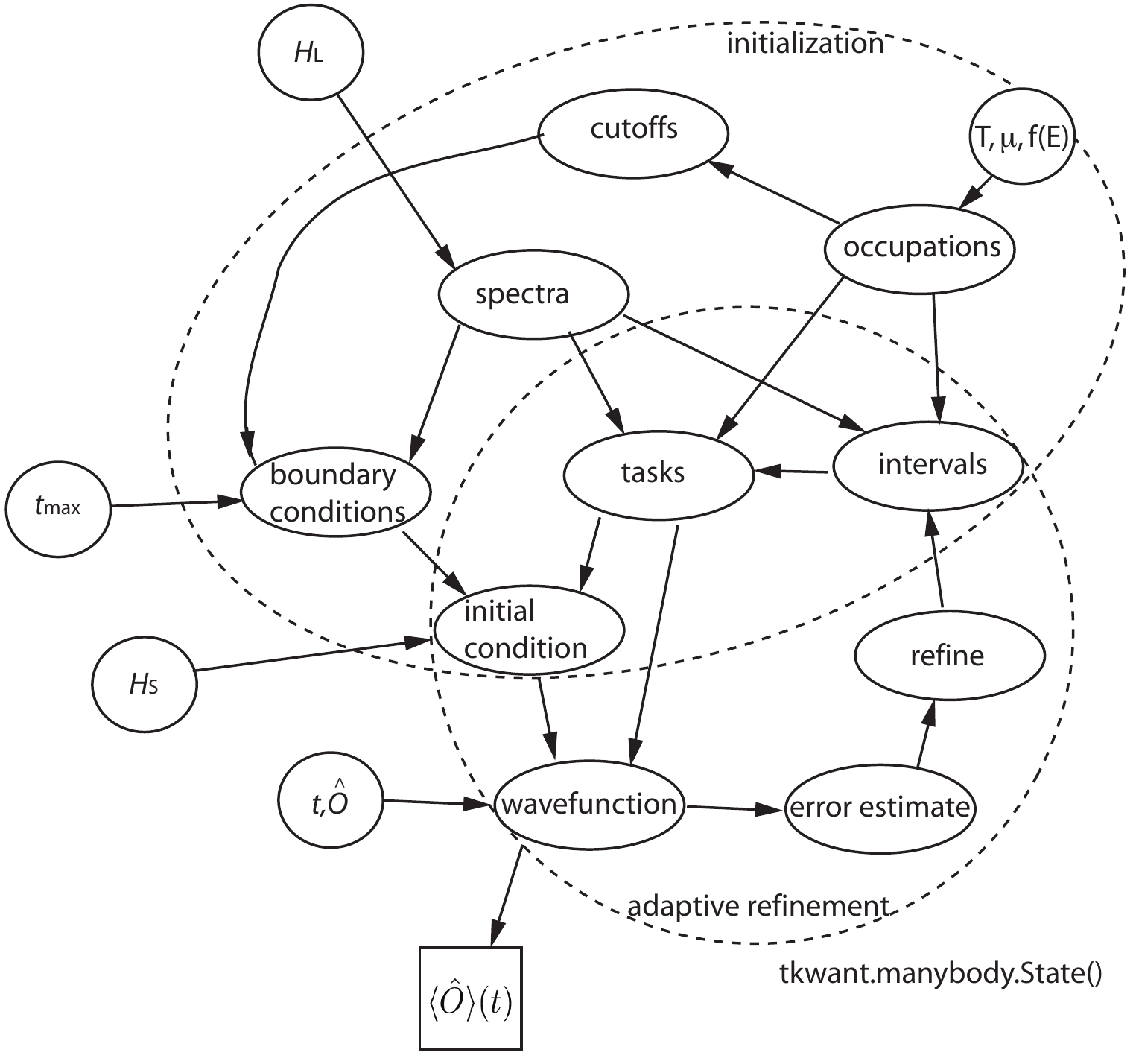}
	\vspace{-2ex}
  \caption{%
Data flow diagram of the high-level adaptive manybody solver \texttt{manybody.State}. Arrows point in the direction of the data flow between
input values (round circles), methods/functions (oval circles) and the result (rectangle).
Note the similarity in the dashed upper initialization circle and the data transfer between the functions in (low-level) code listing
\ref{lst:manybody_wf}.
The lower dashed circle mark methods involved in the adaptive refinement cycle.
}
\label{fig:data_flow_manybody}
\end{figure}

Additional methods, like for instance adaptive refinement, are present in ``high-level'' classes which are  more specialized.
The public attributes follow a similar logic. While all solver classes have at least one time attribute which holds the current time of the state, additional attributes such as lead or mode index are already a specialization to a specific usecase. The overall data flow diagram of the high-level solver \texttt{manybody.States}
is shown in Fig.\ \ref{fig:data_flow_manybody} with the various steps of preprocessing, evolution, and on-the-fly refinement of the integral.

Array-valued numerical data, especially for performance-critical parts, are usually represented in form of NumPy\cite{numpy}-arrays within \textsc{tkwant}.
For more complex and heterogeneous data, such as the sequence of quadrature intervals, \textsc{tkwant} uses flat lists of data classes. By data class, we mean a class without methods, which is only used to store data as attributes.  This is practical because the data is easily readable by humans and can be manipulated without having to care about side effects from stateful objects.

\section{A real-life application: Pulse propagation in a graphene quantum billiard}
\label{sec:graphene}

We end this article with a real-life example of \textsc{tkwant} usage. The device is a small
graphene sample of chaotic shape, connected to a semi-infinite graphene ribbon.
An electrostatic gate deposited on top of the system is pulsed and one follows the associated ripple of density that propagates inside the sample.

\begin{figure*}
	\centering
	\includegraphics[width=\textwidth]{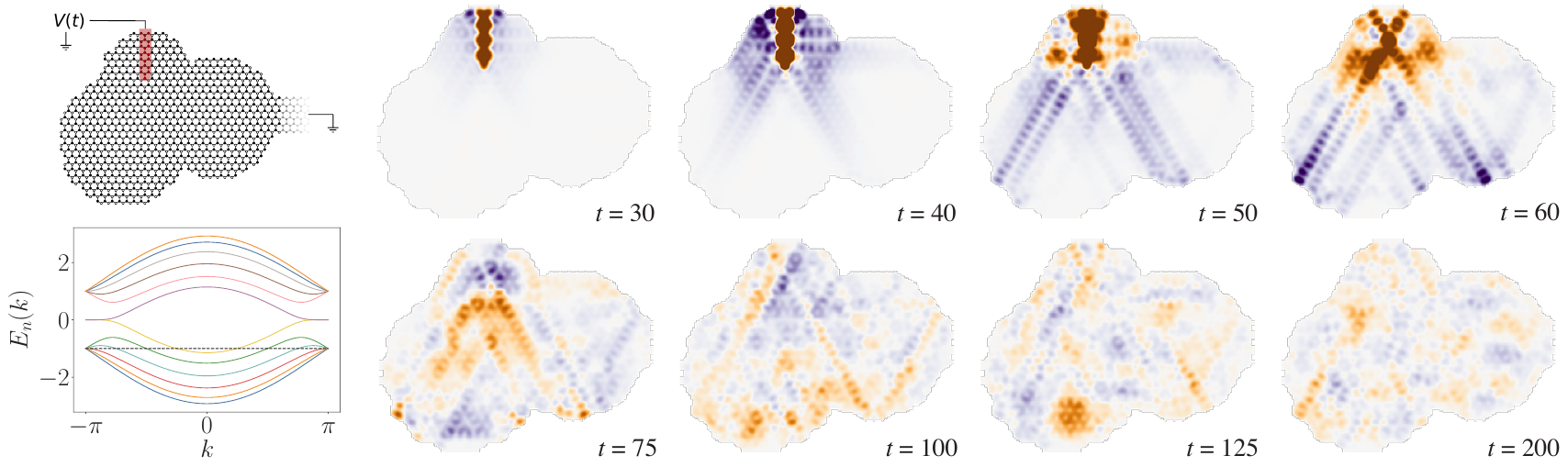}
	\vspace{-1mm}
  \caption{%
  Time evolution of the electron density $\langle c^\dagger_i(t) c_i(t) \rangle$ in an irregularly shaped graphene dot after a Gaussian pulse. The snapshots visualize positive (blue) and negative (yellow) density fluctuations around the equilibrium density. Upper left panel: Schematic sketch of the graphene dot with a lead attached from the right and the electrode position to perturb the system with a local onsite potential $V(t)$.
 The plots can be obtained by running the Python code given in code listing  \ref{lst:graphene}.
}
\label{fig:graphene}
\end{figure*}

Snapshots of the electron density $\langle c^\dagger_i (t) c_i(t) \rangle$ are shown in Fig.\ \ref{fig:graphene} along with a sketch of the system (leftmost panel). One observes first a clear ballistic propagation of the ripple, followed by a more complex speckle like interference pattern as the waves get reflected by the boundaries of the billiards. Eventually, at very long time the ripple leaves the sample entirely through the semi-infinite ribbon.

The typical workflow of a \textsc{tkwant} project starts with an analysis of the static properties, such as the dispersion relation of the leads or (the energy dependance of) the conductance matrix of the system. This static analysis allows one to estimate and tune the relevant timescales of the system and can be done for example with \textsc{kwant}. Here, we skip this part for brevity and focus on the time-dependent simulations.

The complete Python script to perform this numerical simulation and to plot the result is given in code Listing \ref{lst:graphene}.
The structure of the script is quite similar to the first example in code Listing \ref{code:fabry_perot_code} and most of the lines are again related to the construction of the system with \textsc{kwant}.
The code in Listing \ref{lst:graphene} can be optionally run in parallel on several cores to speed up the computation.
A few additional lines (related to the user-defined function \texttt{am\_master()}) are needed to redirect all output to the master MPI process ``rank zero'' responsible for plotting the data.

\begin{lstlisting}[language=Python, label=lst:graphene,
caption={Python code to simulate the electron density of the graphene dot after perturbation with a pulse. Running the code generates the density snapshots shown in Fig.\ \ref{fig:graphene}. Note that the code can be run in parallel using MPI.
Running the code on 48 cores (AMD Opteron 6176 with 2.3 GHz) takes about 2 hours.}
]
import tkwant
import kwant
import numpy as np
import functools as ft

def am_master():
    # returns true if the MPI rank is the master
    return tkwant.mpi.get_communicator().rank == 0

def make_system():

    def onsite_potential(site, time):
        return 0.001 * np.exp(- 0.01 * (time - 40)**2)

    def circle(pos, x0, y0, r):
        x, y = pos
        return (x - x0)**2 + (y - y0)**2 < r**2

    def electrode_shape(pos):
        x, y = pos
        upper_arc = circle(pos, -2.7, 4.8, 6.8)
        return (-4 < x < -2) and (5 < y < 15) and upper_arc

    def lead_shape(site):
        x, y = site.pos
        return -2.5 < y < 3.0

    # Define the graphene lattice.
    lat = kwant.lattice.honeycomb(a=1, norbs=1)
    a, b = lat.sublattices

    # Create graphene model.
    model = kwant.Builder(kwant.TranslationalSymmetry(
        lat.vec((1, 0)), lat.vec((0, 1))))
    model[[a(0, 0), b(0, 0)]] = 0
    model[lat.neighbors()] = -1

    # Central scattering region.
    funs = [ft.partial(circle, x0=7, y0=0, r=8.3),
            ft.partial(circle, x0=-2.7, y0=4.8, r=6.8),
            ft.partial(circle, x0=-5.9, y0=-3, r=9)]
    syst = kwant.Builder()
    syst.fill(model, lambda site: any(f(site.pos)
                                      for f in funs), a(0, 0))
    syst.eradicate_dangling()
    syst[lat.shape(electrode_shape, (-3, 10))] = onsite_potential

    # Define leads using a trick to avoid ugly diag. interfaces.
    sym = kwant.TranslationalSymmetry(lat.vec((1, 0)))
    sym.add_site_family(a, other_vectors=[(-1, 2)])
    sym.add_site_family(b, other_vectors=[(-1, 2)])

    lead = kwant.Builder(sym)
    lead.fill(model, lead_shape, a(0, 0))
    syst.attach_lead(lead)

    return syst

times = np.arange(0, 201, 5)

# Build the system using Kwant.
syst = make_system().finalized()

# Define observables using Kwant.
density_operator = kwant.operator.Density(syst)

# Set a non-zero chemical potential, temperature is T = 0.
occup = tkwant.manybody.lead_occupation(chemical_potential=-1)

# Initialize the time-dependent manybody state.
# Lower numerical accuracy for refinement to speed-up simulation.
state = tkwant.manybody.State(syst, max(times), occup,
                              refine=False)
state.refine_intervals(rtol=1E-3, atol=1E-3)

density0 = state.evaluate(density_operator)

# Loop over timesteps and evaluate and plot the density.
for time in times:
    state.evolve(time)
    if time <= 100:  # Adaptive refinement only for early times.
        state.refine_intervals(rtol=1E-3, atol=1E-3)
    density = state.evaluate(density_operator)
    if am_master:
    		kwant.plotter.density(syst, density - density0)
\end{lstlisting}

\section{Conclusion and outlook}
\label{sec:conclusion}

Recent years have seen a radical shift in the way with which the scientific community approaches numerical simulations. First, open source software -- a necessary condition for an efficient distribution of both old and novel algorithms -- has become increasingly popular. Second, the monolithic approach to scientific programming is progressively yielding to the advent of versatile libraries, often in high-level languages such as Python, that facilitate extensions and combining of different packages. Scientific projects involving computer simulations are increasingly expected to promote transparency and reproducibility by publishing the code that was used to produce the data.

The authors of this work also subscribe to an approach that could be described as ``computer-assisted theory'', where algorithms follow closely the theoretical approach that one would use in an analytical calculations. In particular \textsc{tkwant} exposes all the mathematical objects of the relevant theory (e.g.\ Green's functions, wave functions, dispersion relations, etc.) and explicitly solves a given mathematical problem. The application to specific physical problems is left to the end user. This is in contrast to the ``numerical experiments'' approach where the modeling and associated stream of approximations is often partly implicit.

In this article, we have presented the package \textsc{tkwant} for time-dependent quantum transport. The design of \textsc{tkwant} itself aims at lowering the entrance cost to new users as far as possible.
Exhaustive documentation is available online including a tutorial, additional examples, and complete reference documentation.\cite{tkwant}
The authors hope that \textsc{tkwant} will be used with success by many research groups.

Extensions to \textsc{Tkwant} exist that are not yet included in the official release. One of them extends the non-interacting model to a time-dependent mean field model which already goes beyond the random phase approximation. This extension has been used in Ref.\ [\onlinecite{kloss18}] to describe how charge excitations get renormalized into plasmons in presence of electron-electron interaction (Luttinger liquids). It is used in Ref.\ [\onlinecite{rossignol19}] to study the effect of an electromagnetic environment on the properties of superconducting Josephson junctions. More extensions could be envisioned such as the inclusion of Lindblad-like terms in the dynamics or a treatment of correlations beyond mean field using e.g.\ the novel quantum quasi Monte-Carlo technique\cite{macek20, bertrand19}. It would also be very interesting to combine \textsc{tkwant} with a proper treatment of electrostatics such as the one performed in Ref.\  [\onlinecite{armagnat19}].

\section*{ACKNOWLEDGMENTS}
This project has received funding from the European Union's H2020 research and innovation programme under grant agreement No 86268.
X.W. acknowledges funding from FET open ``UltraFastNano'', ANR Flagera ``Gransport'', French-Japan ANR QCONTROL Project ANR-18-JSTQ-0001 and French-USA ANR PIRE. Early work on \textsc{tkwant} was funded by the ERC consolidator MesoQMC. 
T.K. likes to thank Vera Kontos for help with the figures and acknowledges support from the HLRS Stuttgart.
We thank early external users of \textsc{tkwant} for useful feedback including Michel Fruchard, Pierre Delplace, David Carpentier, Adel Abbout, Aur\'{e}lien Manchon, Genevi\`{e}ve Fleury, Adel Kara Slimane, Phillipp Reck, Matthieu Santin, Manuel Houzet, Tatiane Pereira dos Santos, Pac\^{o}me Armagnat, Baptiste Anselme Martin.

\section*{SUPPLEMENTARY}

The Python codes to generate the plots in this article are given as supplementary material.
The calculations for this article have been performed using \textsc{tkwant} v.\ $1.0.0$, \textsc{kwant} v.\ $1.4.1$ and \textsc{kwantSpectrum} v.\ $0.1.0$.

\begin{appendix}

\renewcommand{\theequation}{A\arabic{equation}}
\section{Smooth dispersion relation reconstruction}
\label{sec:appendix_a}
\setcounter{equation}{0}

In \textsc{kwant} and \textsc{tkwant}, the leads are semi-infinite systems that are invariant by translations. They are described by a unit cell containing $N$ sites. This unit cell is repeated up to infinity.
A lead is characterized by two $N\times N$ matrices: The Hamiltonian matrix inside a unit cell $\mathbf{H}_0$ and the hopping matrix $\mathbf{V}$ that connects one unit cell to the next.
These two matrices can directly be retrieved with \textsc{kwant}\cite{groth14}. In this appendix we discuss the underlying principles of a small package \textsc{kwantSpectrum}\cite{kwantspectrum} that calculates and analyzes the lead dispersion relation.

\begin{figure}[h]
	\centering
	\includegraphics[height=55mm]{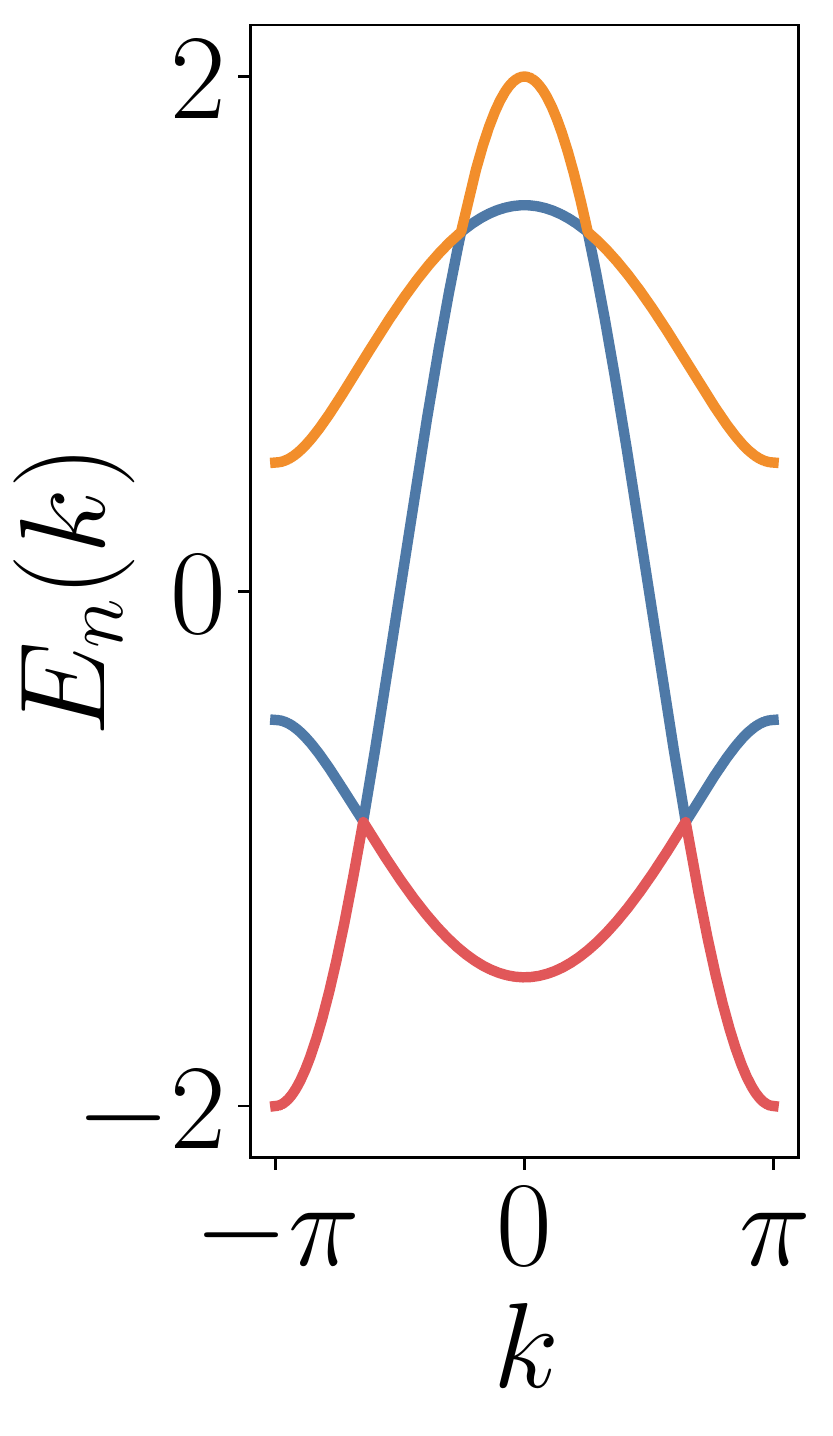}
	\includegraphics[height=55mm]{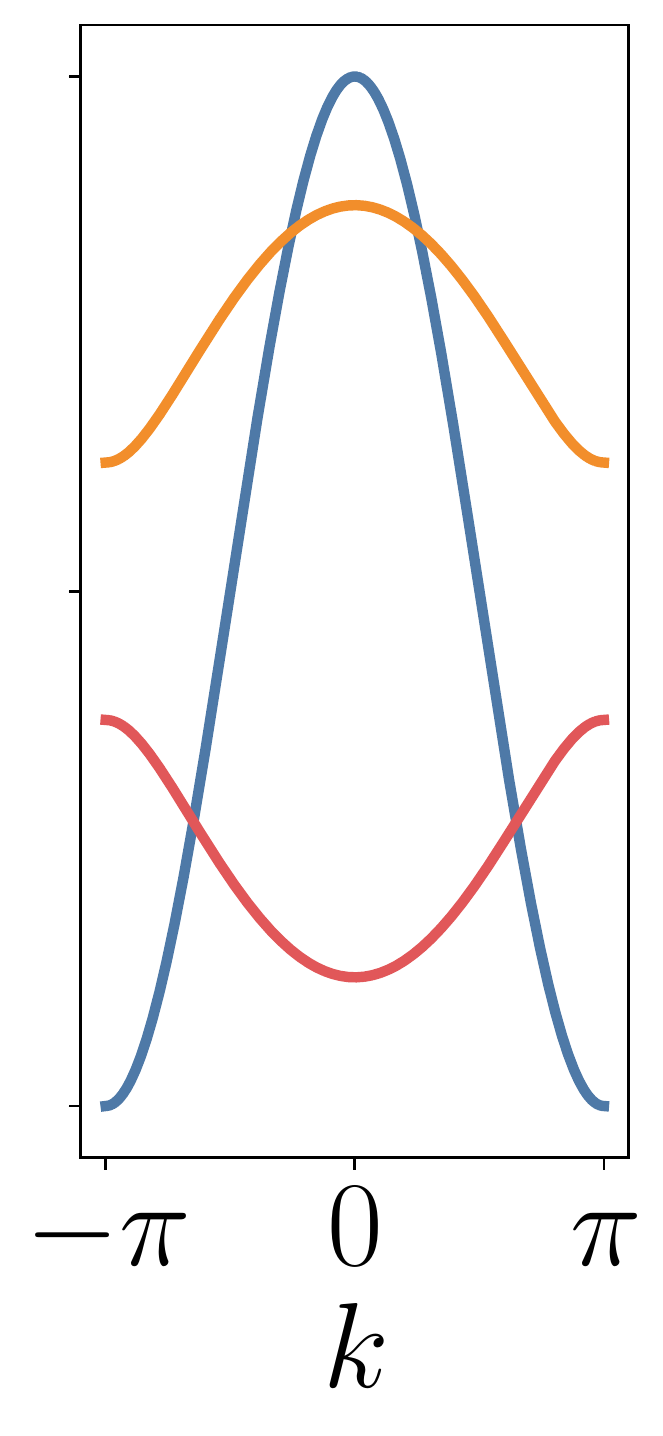}
	\includegraphics[height=55mm]{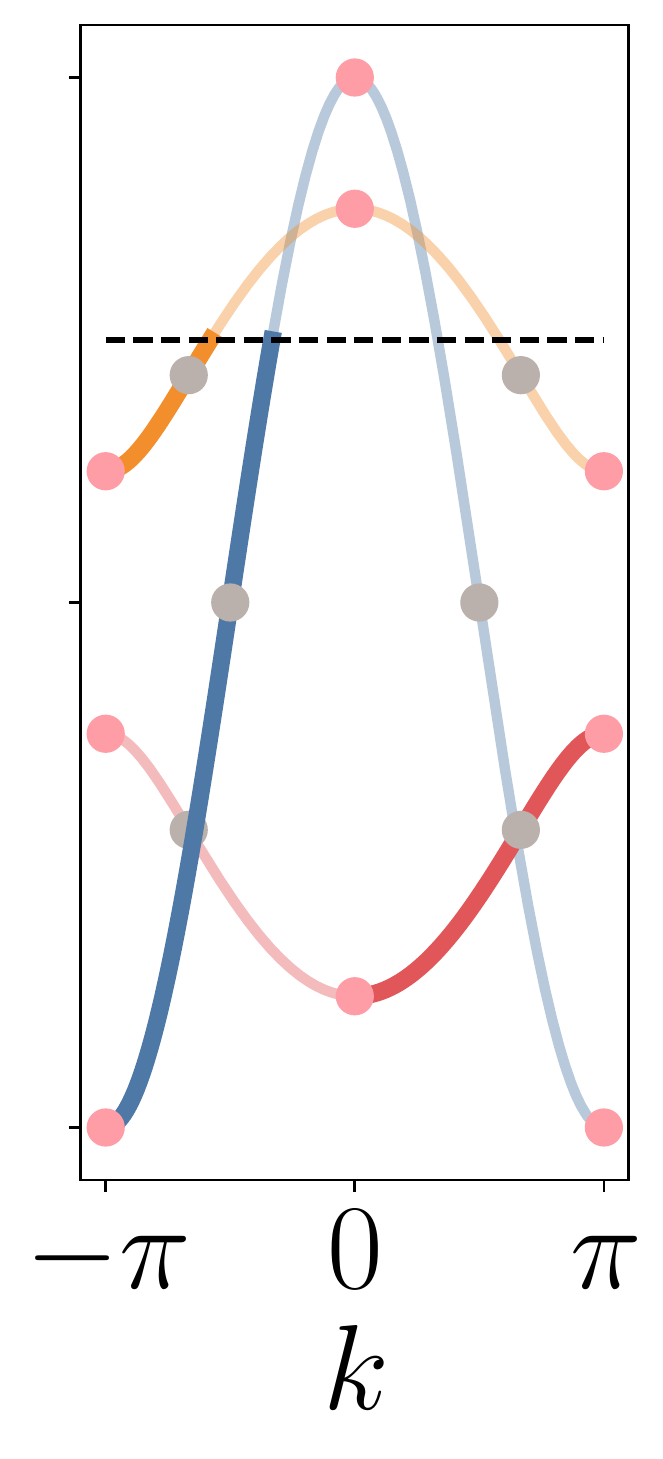}
	\vspace{-2ex}
  \caption{%
Dispersion relation of a simple 3 bands model.
Left panel: Direct diagonalization of  Eq.\ (\ref{eq:lead_diag2}). For each value of $k$ the numerical routine returns the three bands ordered from smallest to largest $E_1(k)\le E_2(k) \le E_3(k)$. The resulting plots reflects this ordering with abrupt change of the
derivative of the curves at the crossing point.
Middle panel: Dispersion relation after reconstruction by \textsc{kwantSpectrum}. The different bands now have continuous derivative.
Right panel: Extraction of important special points by \textsc{kwantSpectrum}: extremas (pink), inflection points (gray) and integration regions (bold) with positive velocity and below a certain energy threshold (dashed).
}
\label{fig:matching}
\end{figure}

\subsection{Problem formulation}
Introducing the matrix
\begin{equation}
\label{eq:lead_diag}
 \mathbf{H}(k) = \mathbf{H}_0 + e^{-i k} \mathbf{V} + e^{i k} \mathbf{V}^\dagger,
\end{equation}
the dispersion relation of the lead is simply given by diagonalizing $\mathbf{H}(k)$:
\begin{equation}
\label{eq:lead_diag2}
 \mathbf{H}(k)  \psi_{\alpha k} = E_\alpha(k) \psi_{\alpha k}.
\end{equation}
While diagonalizing such a matrix for a set of values of $k$ is straightforward numerically, such a direct approach has an important drawback. The problem is best shown on a simple example. The left panel of Fig.~\ref{fig:matching} shows a plot of the dispersion relations for a simple three band models. While the three bands are smooth functions of $k$ the numerical diagonalization make different calculations of the different bands for different values of $k$.  Energies for a given value of $k$ are typically returned ordered from smallest to highest value, so that the smooth bands are only known up to a permutation.
This is apparent from the wrong coloring of the bands in the left panel of Fig.\ \ref{fig:matching}.

Quadrature techniques for integration rely, however, on smooth integrands. The task of \textsc{kwantSpectrum} is to perform a ``smooth dispersion relation reconstruction'', i.e.\ for each value of $k$, finds the permutation that goes from the left panel of
Fig.\ \ref{fig:matching} to its middle panel.
\textsc{KwantSpectrum} returns a precise interpolant of the smooth bands that can be used to analyze the dispersion relations and define the proper integration intervals in $k$-space.
The resulting plot is shown in the middle panel of Fig.\ \ref{fig:matching}.

$k$-integration in \textsc{tkwant} is performed on bands and values of $k$
that satisfy $E_\alpha(k) \le E_F$ and positive velocity $\partial E_\alpha(k)/\partial k \ge 0$. Calculating the corresponding intervals of integration (shown in bold in the right panel of Fig.\ \ref{fig:matching}) requires the knowledge of various special points. The interpolation of \textsc{kwantSpectrum} provides direct access to these special points: maximum and minimum of each bands, inflection points (where the velocity is maximum), solutions of  $E_\alpha(k) = E_F$, see the right panel of Fig.\ \ref{fig:matching}. Another application of \textsc{kwantSpectrum} is the unfolding of the spectrum from the first Brillouin zone to a larger zone in $k$-space.

The rest of this appendix briefly describes \textsc{kwantSpectrum} API and then proceeds to describe the algorithm used for the smooth dispersion relation reconstruction.

\subsection{\textsc{KwantSpectrum} package}
Listing \ref{lst:kwantspectrum} shows the code used to generate the right panel
of Fig.\ \ref{fig:matching}. Lines 6--14 define a lead using \textsc{kwant}. \textsc{Kwant} automatically handles the translational symmetry, i.e.\ it automatically constructs the two matrices
$\mathbf{H}_0$ and $\mathbf{V}$ that are needed for the calculation. The actual computation of the spectrum (matching algorithm and interpolation) is performed in line 18.
The function \texttt{kwantspectrum.spectrum()} computes the interpolant of the different bands. It returns an object that provides various methods for calculating the intervals of integrations and special points that are used in the rest of the script.

\begin{lstlisting}[language=Python, label=lst:kwantspectrum,
caption={Python code snippet to calculate the right panel in Fig.\ \ref{fig:matching}
of the model dispersion using the \textsc{kwantSpectrum}\cite{kwantspectrum} package.
Running the script needs seconds on a standard desktop computer.
}
]
import kwantspectrum as ks
import kwant
import numpy as np
import matplotlib.pyplot as plt

def make_lead_with_crossing_bands():
    lat = kwant.lattice.square(a=1, norbs=1)
    sym = kwant.TranslationalSymmetry((-2, 0))
    H = kwant.Builder(sym)
    H[[lat(0, 0), lat(0, 1), lat(1, 0)]] = 0
    H[lat(1, 0), lat(0, 0)] = 1
    H[lat(2, 1), lat(0, 1)] = 1
    H[lat(2, 0), lat(1, 0)] = 0.5
    return H.finalized()

# Build the system and perform the reconstruction.
lead = make_lead_with_crossing_bands()
spec = ks.spectrum(lead)

momenta = np.linspace(-np.pi, np.pi, 500)
upper_energy = 1

for band in range(spec.nbands):

    # Plot the dispersion of band with index `band`.
    plt.plot(momenta, spec(momenta, band))

    # Intervals with E(k) <= upper_energy and v(k) >= 0.
    eint = spec.intervals(band, upper=upper_energy)
    vint = spec.intervals(band, lower=0, derivative_order=1)
    intervals = ks.intersect_intervals(eint, vint)

    for interval in intervals:
        # Interval is a tuple (kmin, kmax).
        k = np.linspace(*interval)
        plt.plot(k, spec(k, band), linewidth=5.0)

    # Find the special points.
    vel_zeros = spec.intersect(0, band, derivative_order=1)
    plt.plot(vel_zeros, spec(vel_zeros, band), 'o')

    curv_zeros = spec.intersect(0, band, derivative_order=2)
    plt.plot(curv_zeros, spec(curv_zeros, band), 'o')

plt.plot([-np.pi, np.pi], [upper_energy] * 2, '--k')
plt.show()
\end{lstlisting}

\subsection{Overview of the reconstruction algorithm}

Let us start by describing the building block of the algorithm we use for reconstructing the smooth dispersion relations. They are as follows:

\subsubsection{Matching}
Considering an interval $[k_l,k_r]$.

\begin{itemize}

\item First we calculate the dispersion relation $E_{\alpha,l} = E_\alpha(k_l)$ ($E_{\alpha,r} = E_\alpha(k_r)$) at momentum $k_l$ ($k_r$) by diagonalizing Eq.\ (\ref{eq:lead_diag2}). We also obtain (as explained below) the velocities $v_{\alpha,l} = \partial E_\alpha(k_l)/\partial k$
and $v_{\alpha,r} = \partial E_\alpha(k_r)/\partial k$ at the same points.

\item Second, we construct a cost matrix $M_{\alpha\beta}$ that measures how likely is band $\beta$ at point $k_r$ to be assigned to band $\alpha$ at point $k_l$.  The underlying idea for the construction of the cost matrix is straightforward: Given $E_{\alpha,l}$ and its derivative $v_{\alpha,l}$, we make a linear extrapolation of the band $\alpha$ at point $k_r$. The resulting value $E_{\alpha,l} + (k_r - k_l) v_{\alpha,l}$ is compared to the value of the different bands $\beta$ at $k_r$. A possible choice for the cost matrix is therefore $M_{\alpha\beta} = [E_{\alpha,l} + (k_r - k_l) v_{\alpha,l} - E_{\beta,r}]^2$. The actual form of $M_{\alpha\beta}$ that we use is more robust as it also takes advantage of our knowledge of $v_{\alpha,r}$ and is fully symmetric. The detailed form of the cost matrix will be given below in Eq.\ (\ref{eq:cost_matrix}).
For a perfect match of $\alpha$ and $\beta$ the corresponding element $M_{\alpha\beta}$
vanishes.

\item Third, once the cost matrix has been constructed we are back to a standard ``linear assignment problem'': one must find the permutation $P$ of the index $\beta$ that brings the vanishing elements of the cost matrix onto the diagonal, i.e.\
we look for the permutation $P$ that minimizes
\begin{equation}
\sum_\alpha M_{\alpha, P(\alpha)} .
\end{equation}
For this problem, we use the ``Hungarian method'' \cite{Kuhn55} as implemented in the SciPy\cite{scipy20} package.
\end{itemize}

\subsubsection{Interpolating.}
\begin{itemize}

\item Once the matching has been done, we construct a cubic interpolation
$E_\alpha^{lr}(k)$ of the different smooth bands inside $[k_l,k_r]$. The function
$E_\alpha^{lr}(k)$ is a polynomial of degree three that satisfies $E_\alpha^{lr}(k_l)=E_{\alpha,l}$, $E_\alpha^{lr}(k_r)=E_{P(\alpha),r}$, $\partial E_\alpha^{lr}(k_l)/\partial k = v_{\alpha,l}$ and   $\partial E_\alpha^{lr}(k_r)/\partial k = v_{P(\alpha),r}$. The precise form of the interpolant is given below in Eq.\
(\ref{eq:2-point-cubic}).

\item An important part of the algorithm is the evaluation of the quality of the interpolant and of the validity of the matching. We introduce the error $\delta$ of the interpolant.
To estimate $\delta$ we first split the interval $[k_l,k_r]$ in two and perform the matching and interpolation on the two subintervals $[k_l,k_c]$ and $[k_c,k_r]$ where
$k_c = (k_l+k_r)/2$. $\delta$ measures the difference between the interpolant $E_\alpha^{lr}(k)$ and the two subinterpolants $E_\alpha^{lc}(k)$ and $E_\alpha^{cr}(k)$.
Its precise definition is given below in Eq.\ (\ref{eq:error_estimate}).

\end{itemize}

\subsubsection{Overall adapting algorithm.}

The overall algorithm works as follows. We start with $k_l=-\pi$ and $k_r=+\pi$ and apply the matching algorithm and interpolation on the interval $[-\pi,+\pi]$. The interval is then split in two for the evaluation of the error $\delta$. If $\delta$ is smaller than a preset tolerance level $\epsilon$, the algorithm stops. Otherwise the same procedure is applied to the two subintervals $[-\pi,0]$ and $[0,+\pi]$. One proceeds recursively by dividing each sub-interval for which the error $\delta$ lies above the tolerance threshold $\epsilon$. When $\delta<\epsilon$ for each interval the recursive splitting stops. Note that through the quality of the interpolant, the tolerance $\epsilon$ also controls the validity of the matching. Indeed, the cubic interpolant does not converge if
the underlying function has discontinuous derivatives.

\begin{figure}[h]
	\centering
	\includegraphics[width=40mm]{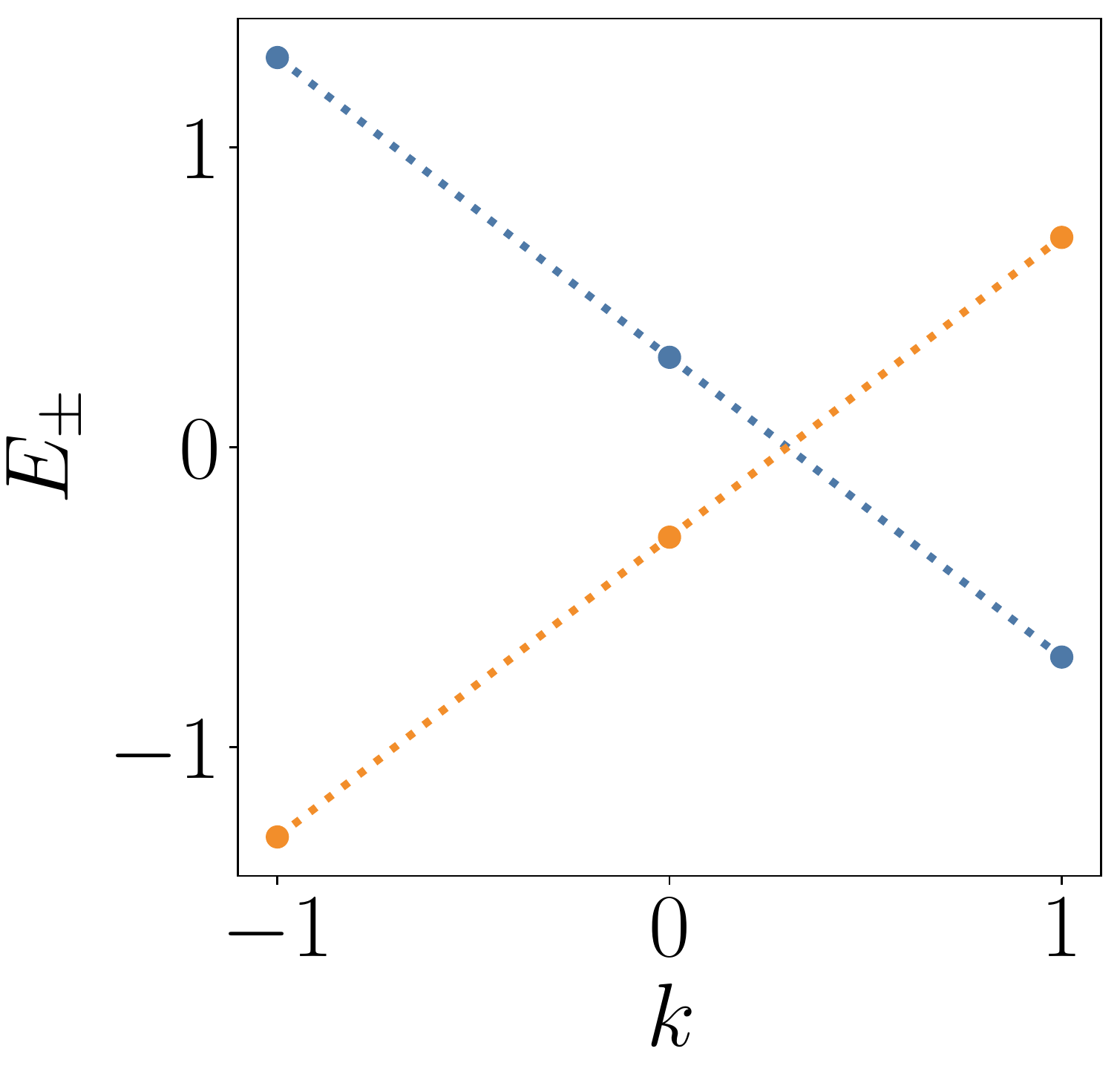}
	\includegraphics[width=40mm]{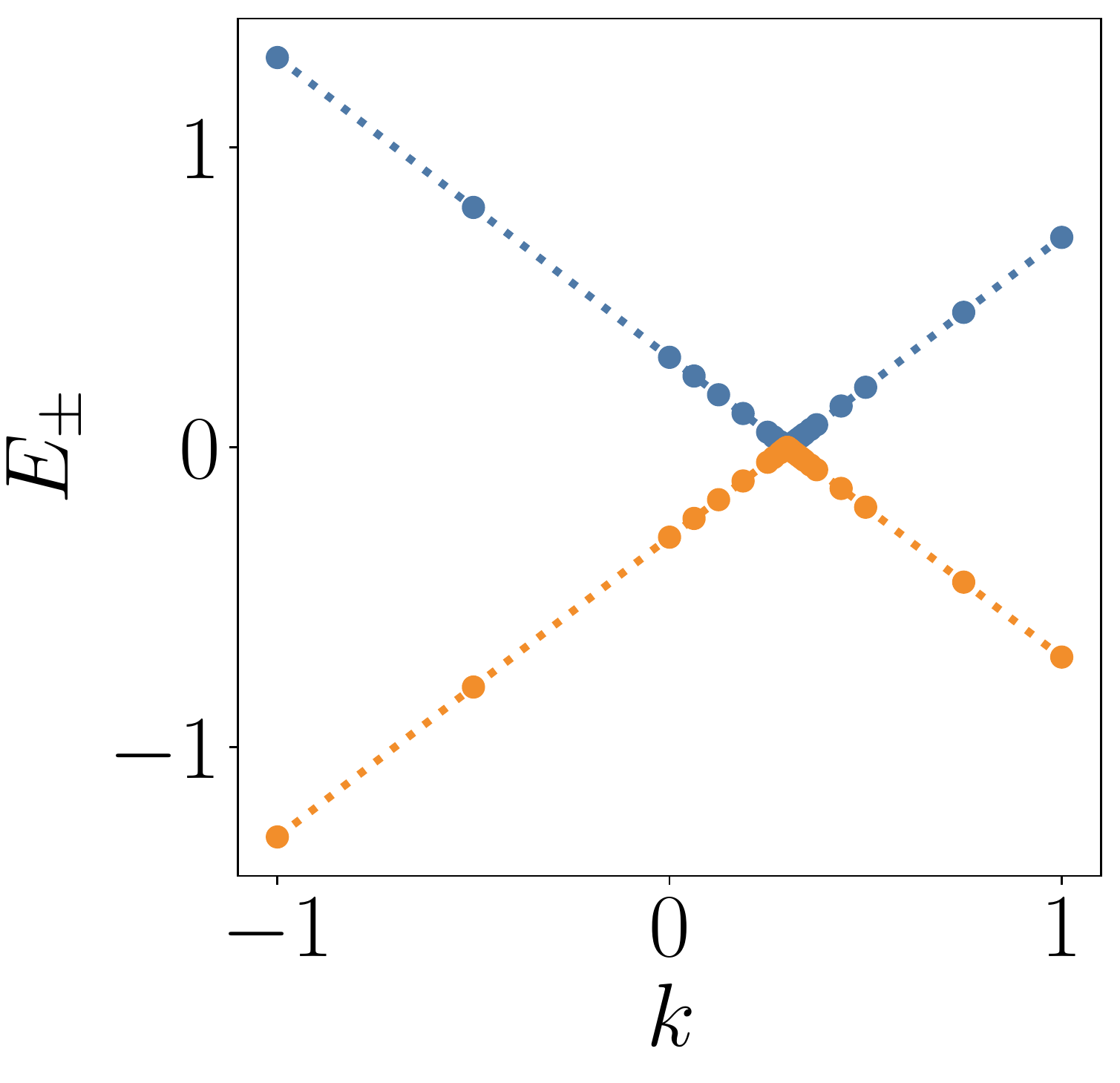}
	\vspace{-2ex}
  \caption{%
Result of the matching algorithm on the model spectrum $E_{\pm}(k) = \pm \sqrt{(k - k_0) + \Delta^2}$.
Left panel: the accuracy is too low ($\epsilon \geq \Delta$) so that the algorithm is not able to detect the small gap $\Delta$ (not visible on this scale). Right panel: the accuracy is high enough ($\epsilon \ll \Delta$), so that the gap is correctly found. The data point correspond to the values of $k$ where the dispersion relation has actually been computed by the algorithm. Parameters: $k_0 = 0.3, \, \Delta = 0.001,\, \epsilon = 0.001\, \text{(left)},\, \epsilon = 10^{-5}\, \text{(right)}$.
}
\label{fig:matching-practice}
\end{figure}

To understand the role of the tolerance parameter $\epsilon$, let us consider an extreme (yet perfectly physical) scenario where two bands almost cross but there is a small
avoided crossing $\Delta \ll 1$ between the two bands. The Hamiltonian $\mathbf{H}(k)$ reads
\begin{equation}
\mathbf{H}(k) =
\begin{pmatrix}
k-k_0 & \Delta \\
\Delta & k_0-k
\end{pmatrix} ,
\end{equation}
so that the two bands are $E_\pm (k) = \pm \sqrt{(k-k_0)^2 +\Delta^2}$. An example of the matching algorithm for this model is shown in Fig.\ \ref{fig:matching-practice} for two values of the tolerance $\epsilon$. When $\epsilon > \Delta$, the algorithm will ignore the small avoided crossing (left panel). When $\epsilon < \Delta$, the algorithm is sensitive to the avoided crossing and labels the band accordingly (right panel)

\subsection{Cost matrix}

To evaluate the cost matrix in an interval $[k_l,k_r]$, we use two different linear extrapolations of the spectrum starting from the left and right points respectively.
The linear interpolation from the left is
\begin{equation}
E_\alpha(k) \approx E_{\alpha,l} + v_{\alpha,l} (k - k_l) ,
\end{equation}
while the interpolation from the right is
\begin{equation}
E_\beta(k) \approx E_{\beta,r} + v_{\beta,r} (k - k_r) .
\end{equation}
Using these two approximations, the cost matrix is simply defined as the average of the
square of the difference between the two approximations:
\begin{eqnarray}
 M_{\alpha\beta} = \frac{1}{k_r-k_l} \int_{k_l}^{k_r} dk
 \left[ E_{\alpha,l} + v_{\alpha,l} (k - k_l) \right.\nonumber \\
 \left. - E_{\beta,r} - v_{\beta,r} (k - k_r)\right]^2 .
\end{eqnarray}
Performing the integration, we arrive at,
\begin{equation}
\label{eq:cost_matrix}
 M_{\alpha\beta} = (A_{\alpha\beta})^2
 +  \frac{(B_{\alpha\beta})^2}{12}(k_r-k_l)^2 ,
\end{equation}
with
\begin{subequations}
\begin{align}
A_{\alpha\beta} &= E_{\alpha,l} - E_{\beta,r} + \left(v_{\beta,r} - v_{\alpha,l}\right)
\frac{k_r-k_l}{2} ,\\
B_{\alpha\beta} &= v_{\beta,r} - v_{\alpha,l}  .
\end{align}
\end{subequations}

\subsection{Cubic Interpolation}

We use a piecewise cubic Hermite interpolation in each of the intervals $[k_l,k_r]$.
Piecewise cubic Hermite interpolation has the advantage that the function and the first derivative  of the interpolation function are exact on the boundaries $k_l$ and $k_r$.
Moreover, these interpolations provide a local (in contrast to a global, which would be the case for Splines) error estimate in each interval.

The interpolation takes the form
\begin{equation}
\label{eq:2-point-cubic}
E_\alpha^{lr}(k) = E_{\alpha,l} (1-t) + E_{\alpha,r} t + t(1-t)[a (1-t) + bt ] ,
\end{equation}
with
\begin{subequations}
\begin{align}
t &= \frac{k - k_l}{k_{r} - k_l} ,\\
a &= v_{\alpha,l} (k_{r} - k_l) - (E_{\alpha,r} - E_{\alpha,l}) , \\
b &= -v_{\alpha,r}(k_{r} - k_l) + (E_{\alpha,r} - E_{\alpha,l}) .
\end{align}
\end{subequations}

To estimate the error of the interpolation $E_\alpha^{lr}(k)$
in the interval $[k_l,k_r]$, we construct the two interpolants
$E_\alpha^{lc}(k)$ and $E_\alpha^{cr}(k)$ with $k_c = (k_l+k_r)/2$
and compute the average of the square of the differences between the two interpolant:
\begin{equation}
 \delta_i =  \sqrt{\frac{105}{2}  I} ,
  \label{eq:interpol_error}
\end{equation}
(the factor $105$ is there purely for convenience) with
\begin{eqnarray}
I = \frac{1}{k_c-k_l}\int_{k_l}^{k_c} dk \, \left[E_\alpha^{lr}(k) - E_\alpha^{lc}(k)\right]^2 , \nonumber \\
 + \frac{1}{k_r-k_c}
 \int_{k_c}^{k_r} dk \, \left[E_\alpha^{lr}(k) - E_\alpha^{cr}(k)\right]^2 .
 \label{eq:interpol_integral}
\end{eqnarray}

To perform each of these integrals, let us remark that they amount respectively to calculating the variance of a cubic interpolant with zero value and derivative on the left (right) while the right  (left) values of the interpolant are given by
$\Delta_\alpha \equiv E_\alpha^{lr}(k_c) -E_{\alpha, c}$ with
$V_\alpha \equiv \partial E_\alpha^{lr}/\partial k (k_c) -v_{\alpha, c}$
for the corresponding derivative. With
 \begin{equation}
E_\alpha^{lr}(k_c) =
\frac{E_{\alpha,l} + E_{\alpha,r}}{2}
 + \frac{v_{\alpha,l} - v_{\alpha,r}}{8}(k_r-k_l) ,
\end{equation}
and
\begin{equation}
\frac{\partial E_\alpha^{lr}}{\partial k}(k_c) =
\frac{3}{2}\frac{E_{\alpha,r} - E_{\alpha,l}}{k_r-k_l}
 - \frac{v_{\alpha,l} + v_{\alpha,r}}{4} .
\end{equation}
Performing the integral, we arrive at
\begin{equation}
I = V_\alpha^2 (k_c - k_l)^2 \frac{2}{105}
+ \Delta_\alpha^2 \frac{26}{35} ,
\end{equation}
so that,
\begin{equation}
\label{eq:error_estimate}
\delta = \sqrt{V_\alpha^2 (k_c - k_l)^2 + 39\Delta_\alpha^2 } .
\end{equation}
It is important to notice that this error, which consists of a weighted sum of
the deviation of the value {\it and its derivative} at the middle point $k_c$
is much more robust than an estimate that would include only one of this two quantities would be. Such error kind of estimates have been used in the context of quadrature
rules\cite{Gonnet10, Gonnet12}.

\subsection{Derivatives of the energy spectrum}
We end this appendix by summarizing the basic results of perturbation theories that we use to calculate the derivative and second derivative of $E_\alpha (k)$ with respect to $k$. Although only the first derivative has been used in the matching algorithm (attempts to use the second derivative have been found to be less robust), the second derivative will be used in the next appendix for the calculation of the effective mass needed for setting the imaginary potential.

Introducing
\begin{equation}
\label{eq:dHdk}
\mathbf{H'}(k)\equiv \frac{d}{dk}\mathbf{H}(k) = i (e^{i k} \mathbf{V}^\dagger -e^{-i k} \mathbf{V}),
\end{equation}
and
\begin{equation}
\label{eq:d2Hdk2}
\mathbf{H''}(k) \equiv \frac{d^2}{dk^2}\mathbf{H}(k) = -(e^{i k} \mathbf{V}^\dagger + e^{-i k} \mathbf{V}),
\end{equation}
one has
\begin{equation}
\label{eq:dEdk}
 v_\alpha (k) \equiv \frac{d}{dk}E_\alpha(k) = \psi_{\alpha k}^\dagger \mathbf{H'}(k) \psi_{\alpha k} ,
 \end{equation}
and
\begin{equation}
\label{eq:d2Edk2}
 \frac{d^2}{dk^2}E_\alpha(k) = \frac{1}{2}
 \psi_{\alpha k}^\dagger \mathbf{H''}(k) \psi_{\alpha k}
 + \sum_{\beta\ne\alpha} \frac{\left|\psi_{\alpha k}^\dagger \mathbf{H'}(k) \psi_{\beta k}\right|^2}{E_\alpha(k) -E_\beta(k)} .
 \end{equation}

\renewcommand{\theequation}{B\arabic{equation}}
\section{Heuristic for setting the absorbing imaginary potential}
\setcounter{equation}{0}
\label{sec:appendix_b}

\subsection{Problem formulation}

Since we consider leads that are invariant by translation, any wave packet that enters the lead will propagate ballistically inside the lead towards infinity and therefore never come back to the scattering region. In \textsc{tkwant}, we use an imaginary potential
$\Sigma^l (a)$ inside the lead $l$ to absorb these wave packets to that they do not create spurious signal in the simulations. This imaginary potential depends on the cell $a$ insides the lead. The addition of the imaginary potential amounts to the change
\begin{equation}
\label{eq:hamltonian_lead}
\hat{\mathbf{H}}^l \rightarrow  \hat{\mathbf{H}}^l +
i \sum_{a \geq 0,n,m} \mathbf{\Sigma}^{l}(a)\hat{c}^\dagger_{a,n} \hat{c}_{a,m}
\end{equation}
in the lead Hamiltonian. The corresponding ``sink'' algorithm was discussed in Ref.~[\onlinecite{weston16a}]. The present discussion expands on the original algorithm and adds simple heuristics for the choice of the
$\Sigma^l(a)$ function.

The choice of $\Sigma^l(a)$ is an optimization problem where one seeks to minimize the amount of signal reflected into the scattering region. Two effects work in opposite direction: on one hand one wants a large imaginary potential so that the waves get absorbed before they reach the end of the system. On the other hand any abrupt increase of $\Sigma^l(a)$ creates backscattering that sends spurious waves back into the scattering region. Hence, one aims to construct an imaginary potential that rises very smoothly to avoid these reflections.

\begin{figure}[h]
	\centering
	\includegraphics[width=80mm]{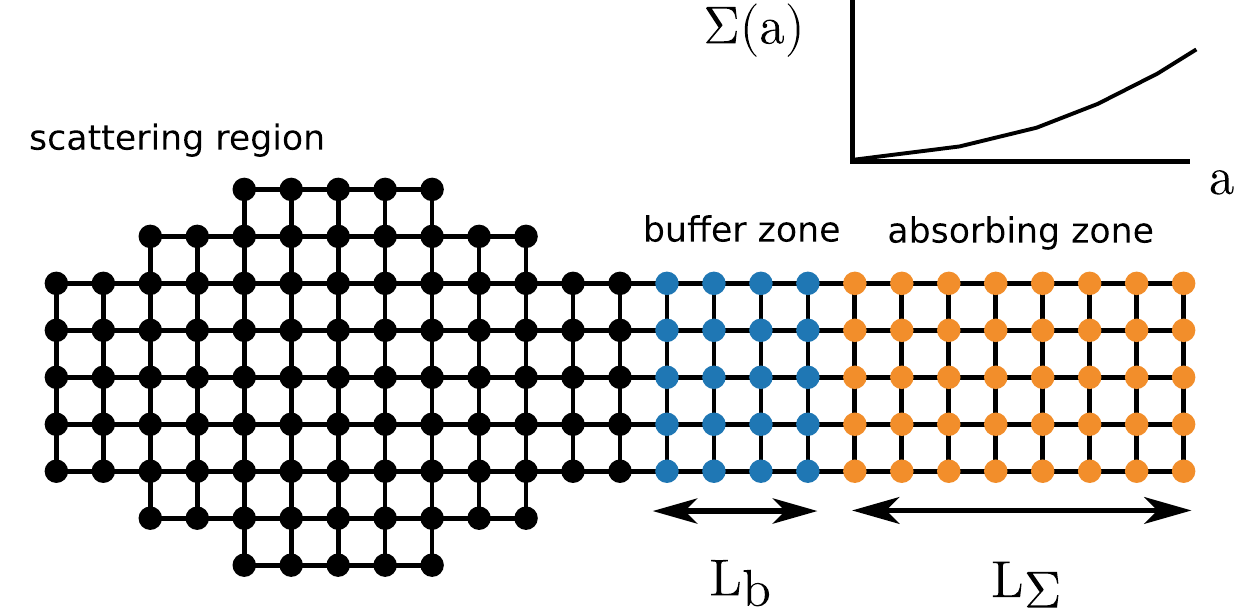}
	\vspace{-2ex}
  \caption{%
Sketch of the imaginary potential used in the leads. A finite portion of the lead is included in the time-dependent simulation (the initial calculation of the scattering states is done with infinite leads). This finite portion is split into a buffer zone (blue) and an absorbing zone (yellow) where the imaginary potential is slowly raised.
}
\label{fig:boundary-cells}
\end{figure}

In practice, the spurious reflection due to the variation of $\Sigma^l(a)$ is dominated by the
long wave length part of the spectrum. Indeed,
when the wave length $\lambda = 2\pi/k$ of the wave
is large, any variation of $\Sigma^l(a)$ looks abrupt. On the other hand, the corresponding wave are typically very slow. Hence, if a spurious reflection is created, it typically takes a long time to reach the scattering region.
The strategy used in \textsc{tkwant} is to split the lead region into two sub-regions: the absorbing zone where the imaginary potential is applied and a buffer zone, see Fig.\ \ref{fig:boundary-cells} for a sketch.
The size of the respective sub-regions are optimized to guarantee -- for a given level of precision -- that the spurious reflections do not have time to reach the scattering region in the duration of the simulation. This is a very conservative ``safe'' mode of \textsc{tkwant}. Experienced users can use less stringent conditions but need to check the accuracy of the results manually.

Following Ref.\ [\onlinecite{weston16a}], we use
a polynomial shape of the imaginary potential,
\begin{equation}
 \Sigma(x) = (n + 1) A x^{n}.
 \label{eq:aborbing_pot}
\end{equation}
The length of the buffer and of the absorbing zone are denoted as $L_b$ and $L_\Sigma$ unit cells respectively.

To define the spurious reflection, we consider
a fictitious scattering problem. The system consists of an infinite buffer region terminated on one side by the absorbing zone.
For a given channel $\alpha$ with momentum $k$, the presence of the imaginary potential creates a reflection $r_{\alpha k}$ so that the
scattering states propagating in the buffer zone can be written as
\begin{equation}
\psi^{\rm fict}_{\alpha E}(a) = \psi_{\alpha E} e^{i k_\alpha a} + \sum_\beta r_{\beta\alpha E} \psi_{\beta E}e^{-i k_\beta a}.
 \label{eq:prop_wave}
\end{equation}
The spurious reflection $r_{\beta\alpha E}$ can be calculated numerically with \textsc{kwant} or estimated analytically.
We define the total spurious reflection as
\begin{equation}
 r = \max_{\alpha\beta, E} \, |r_{\beta\alpha E}|.
 \label{eq:r_total}
\end{equation}
Only the channels $\alpha$ that have a large enough velocity $v_\alpha > L_b/t_{\rm max}$ (where $t_{\rm max}$ is the maximum time of the simulation) are taken into account into the calculation of $r$.
Indeed, slower channels may contribute to reflection, but due to the presence of the buffer region, the reflected wave will not have time to reach the scattering region and spoil the results.  Given a targeted accuracy $r_{\rm max}$ the problem reduces to optimize the parameters $A$, $n$, $L_b$ and $L_\Sigma$ such that $r \le r_{\rm max}$ while $L = L_b + L_\Sigma$ is as small as possible.

A trivial possibility is to have no imaginary
potential at all and choose $L_b$ large enough so that even the fastest channels cannot reach the scattering region. Although non-optimum, this
boundary condition is implemented in \textsc{tkwant} and referred to as ``simple boundary condition''.\cite{weston16a} \textsc{tkwant} implements an heuristic algorithm that -- although
non-optimal in general -- considerably improves on the simple boundary condition in certain cases. We stress again that \textsc{tkwant} provides a ``safe'' algorithm that seeks a given precision whatever the dynamics in the scattering region. For a given time-dependent problem, the error will usually be much smaller than $r_{\rm max}$. For large simulations where the computing time is critical, a manual control of the imaginary potential may be significantly more efficient.

\subsection{Heuristic for optimization}
\label{sec:prefactor_a}

In our heuristic, we consider only two extreme values of $(\alpha, k)$: The fastest modes that will quickly go through the buffer region but will be absorbed efficiently by the imaginary potential (with very little reflection) and the slowest modes that will take a long time to cross the buffer region but will create significantly more reflection. \textsc{KwantSpectrum} provides the necessarily tools for finding the maximum
velocity $v_{\rm fast}$ in the leads (fast modes)  as well as the points of maximum curvature $\gamma_{\rm slow} = \partial^2 E_{\alpha}/\partial k^2$ where the velocity vanishes (slow modes).

To estimate the reflection $r_{\Sigma}$ of a given mode of dispersion relation
\begin{equation}
\varepsilon(q) = \frac{1}{2}\gamma q^2 ,
\end{equation}
where $q= k-k_0$ is the momentum counted from the bottom/top of the band, we use
an analytical expression Eq.\ (34) derived in Ref.\ [\onlinecite{weston16a}].
Note the presence of a typo in  Eq.\ (34) in Ref.\ [\onlinecite{weston16a}].
The correct form has a factor $(n-1)!$ instead of $(n-1)$ in the second term
and reads,
\begin{equation}
 r_{\Sigma} = e^{- A q / \varepsilon} + \frac{A n (n + 1) (n - 1)!}{2^{n + 2} \varepsilon q^n L_\Sigma^{n + 1}} ,
 \label{eq:refl_coeff}
\end{equation}
where the first and second term respectively describe the absorption by the imaginary
potential and the reflection when it is not perfectly adiabatic.

\textbf{ Optimization of $A$}.
We first choose the optimum value of $A_*$ that minimizes Eq.\ (\ref{eq:refl_coeff})
i.e.\ that satisfies $\partial_A r_{\Sigma}(A_*) = 0$.
The value of $A_*$ strongly depends on $q$ and $\gamma$. We optimize $A_*$ with respect to the {\it fastest} mode. Indeed  the first term of Eq.\ (\ref{eq:refl_coeff}) scales as $e^{-2A/v}$ while the second scales as $A/v^{n+1}$. Hence fast modes are limited by the
first term while slow modes are limited by the second one. Since the first term is exponential, it is computationally cheap to make it negligible for all modes.
Making the second term small enough is a matter of increasing $L_\Sigma$. We arrive at,
\begin{equation}
A_* = - \frac{\varepsilon(q_{\rm fast})}{q_{\rm fast}}
\log \left( \frac{n (n + 1) (n - 1) ! }{2 (2 q_{\rm fast} L_\Sigma)^{n + 1}}\right).
  \label{eq:a_optimal}
\end{equation}

\textbf{ Optimization of $L_\Sigma/L_b$}. The second optimization is to find the best  of way of splitting the total length $L$ into $L = L_\Sigma + L$ for a given $A_*$.
Introducing $x$ as,
\begin{equation}
 L_{\Sigma} = L (1 - x), \qquad L_{b} = L x, \qquad x \in \{ 0, 1\},
 \label{eq:len_split}
\end{equation}
the second term of Eq.\ (\ref{eq:refl_coeff}) is dominated by the slowest modes that can go through the buffer layer. The corresponding $q_{\rm slow}$ satisfies $2 L_b = \gamma_{\rm slow} q_{\rm slow}t_{\rm max} $. We get,
\begin{align}
 r_{\Sigma} =\frac{ A_* t_{\rm max} n (n + 1) (n - 1) ! }{2 x L} \left(\frac{\gamma_{\rm slow} t_{\rm max}}{4 L^2 x(1 - x)}\right)^{n + 1}.
  \label{eq:refl_gamma}
\end{align}

Optimizing with respect to $x$,
$\partial_x r_{\Sigma} = 0$ in the above equation leads to the optimum splitting fraction $x_*$,
\begin{equation}
x_* = \frac{n+2}{2n +3} .
\label{eq:x_optimal}
\end{equation}
independently of the value of $\gamma_{\rm slow}$.

\textbf{Overall iterative optimization sequence.} Our overall estimate of the error reads,
\begin{equation}
 r_{\Sigma} = e^{-2A_*/v_{\rm fast}} + \frac{A_* n (n + 1) (n - 1)!}{2^{n + 1} \gamma_{\rm slow} q_{\rm slow}^{n+2} L^{n + 1} (1-x_*)^{n+1}} .
 \label{eq:reflection_total}
\end{equation}
Our overall algorithm for setting the values of $L_b$, $L_\Sigma$ and $A$ reads as follows:
\begin{enumerate}
\item We start with an initial value of $L_{0} = v_{\rm fast} t_{\rm max}/2$ that corresponds to the ``simple boundary condition'' with no imaginary potential. The choice of this length is guaranteed to induce no spurious reflection.

\item We set $A_*$ using Eq.\ (\ref{eq:a_optimal}) with $L_\Sigma = (1-x_*) L_0$ and
$x_*$ given by Eq.\ (\ref{eq:x_optimal}).

\item We use Eq.\ (\ref{eq:reflection_total}) to find the value of $L_*$ that satisfies
$r_\Sigma < r_{\rm max}$.

\item If the new value $L_* < L_0$ then it is computationally advantageous to use $L_*$ instead of $L_0$ in the simulations. We update $L_0 \rightarrow L_*$ and go back to step 2 to see if $L$ can be further decreased. If $L_* > L_0$ we terminate the optimization and keep $L_0$ as our value of $L$.

\end{enumerate}
Note that we did not perform a systematic optimization over the order of the polynomial $n$, but we have found empirically that $n= 6$ is a good compromise.

\subsection{Illustration}
\label{sec:boundary_benchmark}

\begin{figure}
	\centering
	\includegraphics[width=80mm]{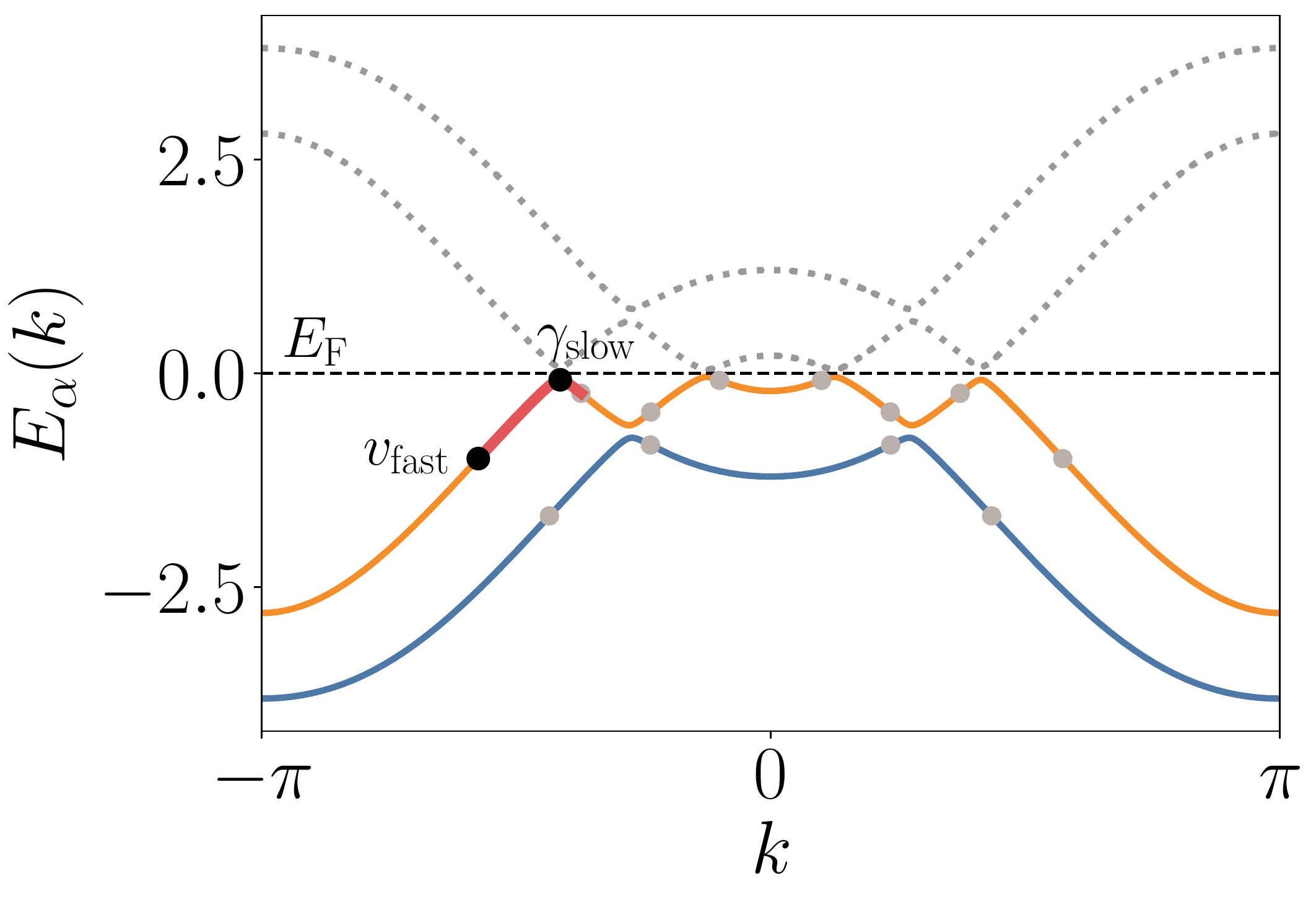}
	\vspace{-2ex}
  \caption{%
Spectrum of the model system used for the boundary benchmark.
Two bands $\alpha = 0$ (blue) and $\alpha = 1$ (orange) with inflection points (gray) are below the Fermi energy $E_{\text F}$ (dashed horizontal line). The highest velocity is at point $v_{\rm fast}$ and the local extremum with highest curvature is at point $\gamma_{\rm slow}$ (both in black).
Note that the small gaps are resulting in local extrema with high curvature values which are strongly reflected at the absorbing boundaries.
}
\label{fig:boundary_spectrum}
\end{figure}

\begin{figure}
	\centering
	\includegraphics[width=40mm]{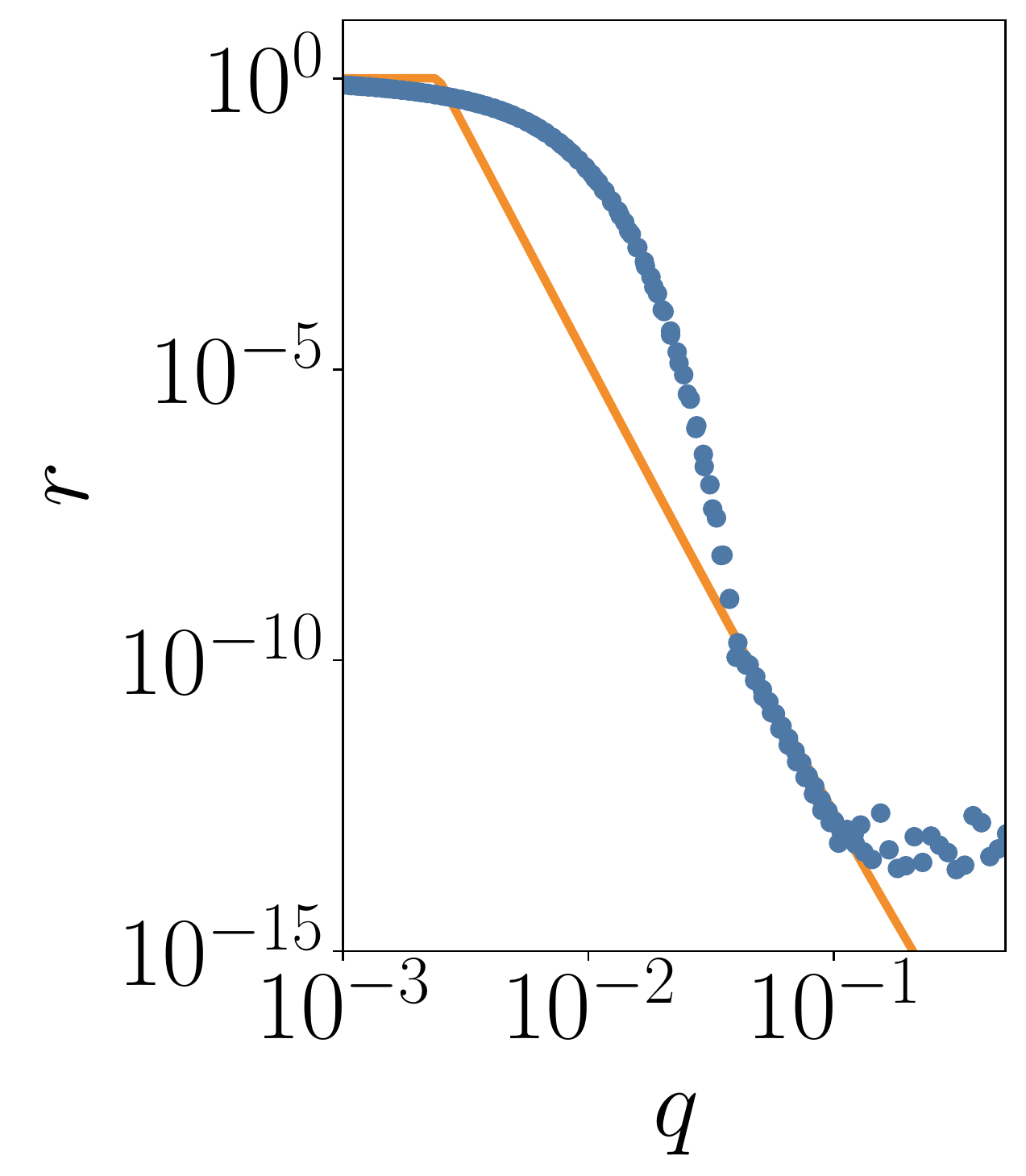}
	\includegraphics[width=40mm]{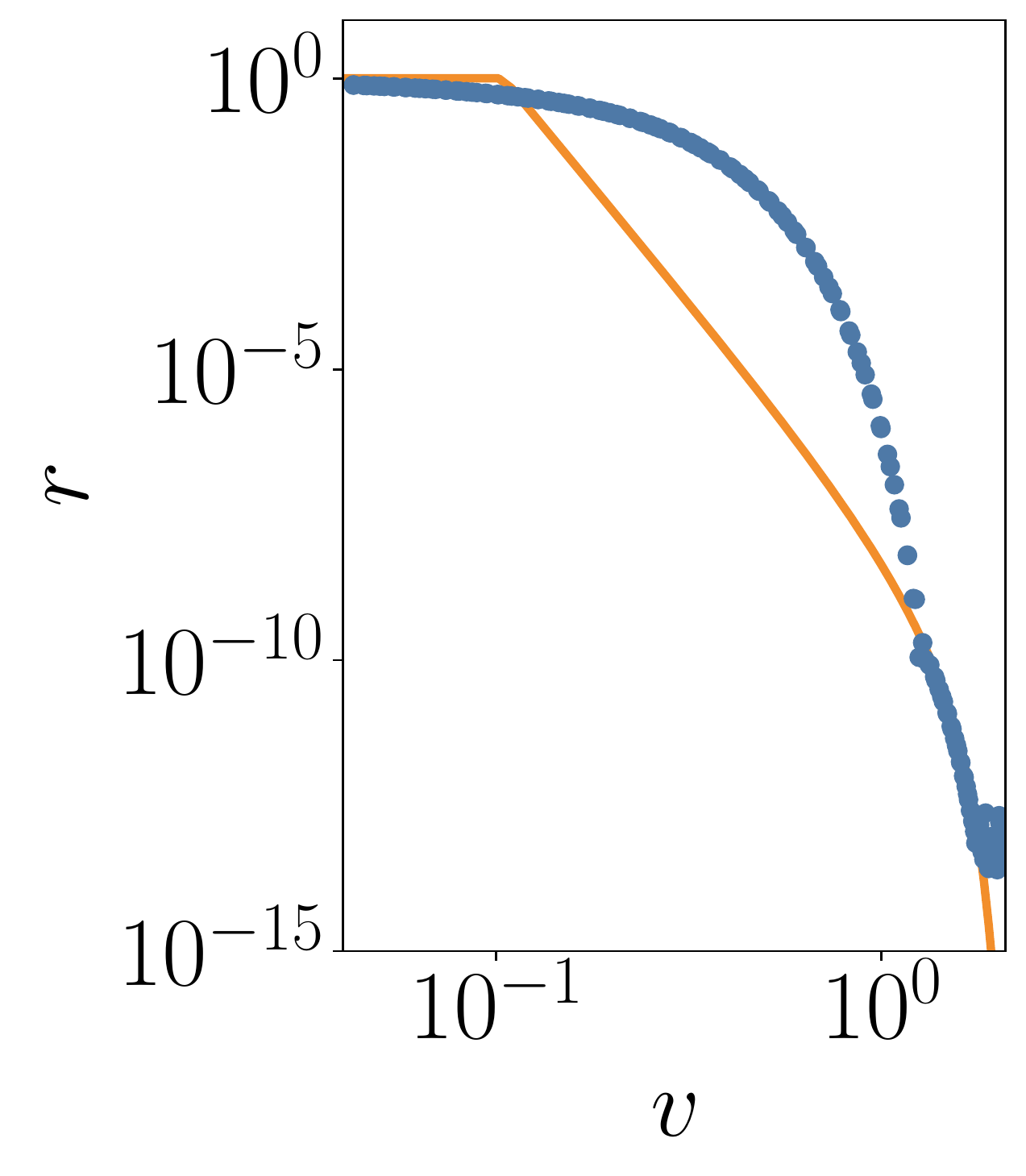}
	\vspace{-2ex}
  \caption{Comparison of the reflection coefficient $r$ estimated with Eq.\ (\ref{eq:refl_coeff}) (orange line) vs.\ exact numerical calculations from the scattering matrix (blue points). The reflection is plotted for the modes around
$\gamma_{\rm slow}$ (red highlighted part in Fig.\ \ref{fig:boundary_spectrum})
and $q$ is the relative momentum measured to this point. Parameters: $n = 6$, $t_{\text max} = 10^{4}$.
  }
\label{fig:reflection_vs_qv}
\end{figure}

To illustrate the procedure, we apply the optimizing algorithm to a real world system with a complex energy dispersion as shown in Fig.\ \ref{fig:boundary_spectrum}.
This example is difficult due to the presence of tiny gaps at the avoided crossings (high curvature/very low effective mass) which leads to a potentially large spurious reflection.  Fig.\ \ref{fig:reflection_vs_qv} compares our analytical estimate of $r$ to an exact numerical calculation performed with \textsc{kwant}.
We observe a deviation from the analytical estimate for high values of $r$ but the estimate is rather accurate for small $r$. Since it is in the latter parameter range that it is actually needed, the estimate is quite reliable. See, e.g.\ Fig.\ 3 of Ref.\ [\onlinecite{weston16a}] for a more detailed study.

\subsection{Computational complexity}

The overall computational complexity (CPU time of a simulation) of \textsc{tkwant} scales as $(N_s + N L) t_{\rm max}$ where $L$ also scales with $t_{\rm max}$. For the ``simple boundary condition'', $L \propto t_{\rm max}$, such that the overall complexity is $\propto t_{\rm max}$ for large scattering regions/short simulation times but $\propto t_{\rm max}^2$ for small scattering regions/long simulation times.

\begin{figure}[h]
	\centering
	\includegraphics[width=80mm]{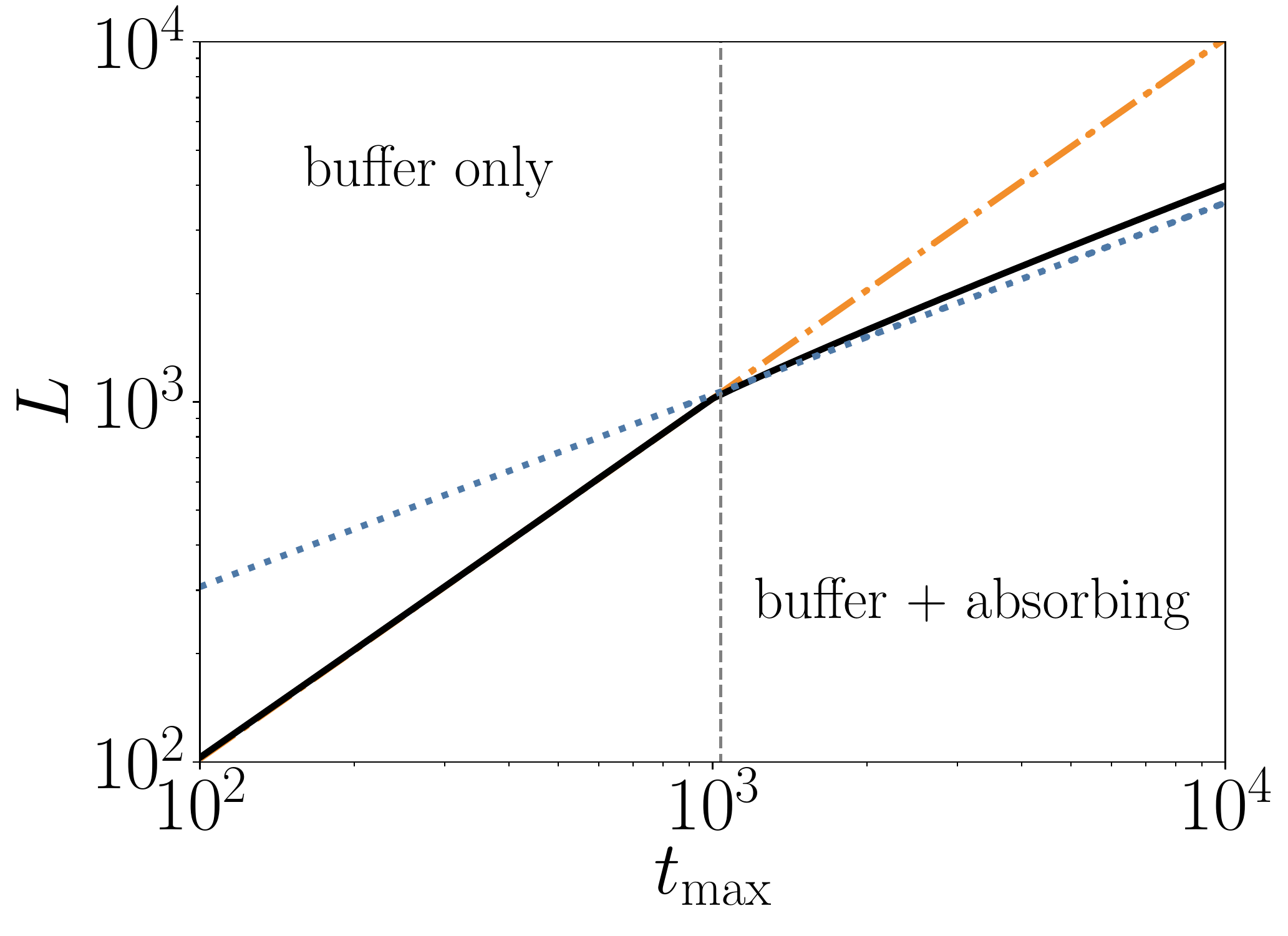}
	\vspace{-2ex}
  \caption{%
Scaling of the total length of the boundary cells $L = L_b + L_{\Sigma}$ vs.\ the maximal simulation time $t_{\text max}$ (black line).
Up to around $t_{\text max} \approx 10^3$, the ``simple boundary conditions'' with only buffer cells (linear scaling $L \sim t_{\rm max}$, orange dash-dotted line) are preferred. For larger $t_{\text max}$, the combination of buffer and absorbing cells is more effective Blue dotted line: theoretical scaling $t_{\rm max}^{x_*}$ with $x_* = 8/15$.
Same parameters as in Fig.\ \ref{fig:boundary_spectrum} : $r_{\text max} = 10^{-5}, n = 6$.
}
\label{fig:boundary_scaling}
\end{figure}

The heuristic algorithm described in this appendix has a complexity $L \propto t_{\rm max}^{x_*}$ [as can be seen from Eq.\ (\ref{eq:refl_gamma}) neglecting logarithmic corrections] which translates into a more favorable overall complexity
$\propto t_{\rm max}^{1+x_*}\approx  t_{\rm max}^{1.5}$ for large simulation times.
The crossover between the short and large time behavior is illustrated in
 Fig.\ \ref{fig:boundary_scaling}.

The scaling $t_{\rm max}^{1.5}$ corresponds to a ``safe'' usage of \textsc{tkwant} that does not make any assumptions about the actual dynamics that is taking place in the scattering region or additional symmetries in the leads.
In most cases, it is possible to obtain the optimum overall scaling $\propto t_{\rm max}$.
One can take advantage of the structure of the leads. For instance, if the lead is in the quantum Hall regime, inducing back reflection with the imaginary potential involves back scattering an chiral edge state on one edge of the lead to the other side.
As this process is exponentially suppressed with the width of the lead, extremely accurate results can be obtained with an absorbing zone that contains only a handful of sites.
Another example is graphene: since the imaginary potential does not break the symmetry between A and B sites, it conserves the corresponding pseudo-spin hence do not induce back scattering in the region close to the Dirac points. Last, in many practical situations, the time-dependent perturbation is actually slow and small with respect to $\hbar/E_F$ and $E_F$ respectively.
It follows that only the modes close to $E_F$ will actually play a role in the simulation.
Experienced users can manually set the imaginary potential $\Sigma^l(a)$ and check the convergence of the results by monitoring how they converge with $L_\Sigma$ and/or $L_b$.

\end{appendix}

\clearpage
\bibliography{meso}% no space between several names

\begin{thebibliography}{77}%
\makeatletter
\providecommand \@ifxundefined [1]{%
 \@ifx{#1\undefined}
}%
\providecommand \@ifnum [1]{%
 \ifnum #1\expandafter \@firstoftwo
 \else \expandafter \@secondoftwo
 \fi
}%
\providecommand \@ifx [1]{%
 \ifx #1\expandafter \@firstoftwo
 \else \expandafter \@secondoftwo
 \fi
}%
\providecommand \natexlab [1]{#1}%
\providecommand \enquote  [1]{``#1''}%
\providecommand \bibnamefont  [1]{#1}%
\providecommand \bibfnamefont [1]{#1}%
\providecommand \citenamefont [1]{#1}%
\providecommand \href@noop [0]{\@secondoftwo}%
\providecommand \href [0]{\begingroup \@sanitize@url \@href}%
\providecommand \@href[1]{\@@startlink{#1}\@@href}%
\providecommand \@@href[1]{\endgroup#1\@@endlink}%
\providecommand \@sanitize@url [0]{\catcode `\\12\catcode `\$12\catcode
  `\&12\catcode `\#12\catcode `\^12\catcode `\_12\catcode `\%12\relax}%
\providecommand \@@startlink[1]{}%
\providecommand \@@endlink[0]{}%
\providecommand \url  [0]{\begingroup\@sanitize@url \@url }%
\providecommand \@url [1]{\endgroup\@href {#1}{\urlprefix }}%
\providecommand \urlprefix  [0]{URL }%
\providecommand \Eprint [0]{\href }%
\providecommand \doibase [0]{http://dx.doi.org/}%
\providecommand \selectlanguage [0]{\@gobble}%
\providecommand \bibinfo  [0]{\@secondoftwo}%
\providecommand \bibfield  [0]{\@secondoftwo}%
\providecommand \translation [1]{[#1]}%
\providecommand \BibitemOpen [0]{}%
\providecommand \bibitemStop [0]{}%
\providecommand \bibitemNoStop [0]{.\EOS\space}%
\providecommand \EOS [0]{\spacefactor3000\relax}%
\providecommand \BibitemShut  [1]{\csname bibitem#1\endcsname}%
\let\auto@bib@innerbib\@empty
%</preamble>
\bibitem [{\citenamefont {Batelaan}\ and\ \citenamefont
  {Tonomura}(2009)}]{Batelaan09}%
  \BibitemOpen
  \bibfield  {author} {\bibinfo {author} {\bibfnamefont {H.}~\bibnamefont
  {Batelaan}}\ and\ \bibinfo {author} {\bibfnamefont {A.}~\bibnamefont
  {Tonomura}},\ }\bibfield  {title} {\emph {\enquote {\bibinfo {title} {{The
  Aharonov-Bohm Effects: Variations on a Subtle Theme}},}\ }}\href {\doibase
  10.1063/1.3226854} {\bibfield  {journal} {\bibinfo  {journal} {Phys. Today}\
  }\textbf {\bibinfo {volume} {62}},\ \bibinfo {pages} {38} (\bibinfo {year}
  {2009})}\BibitemShut {NoStop}%
\bibitem [{\citenamefont {Ji}\ \emph {et~al.}(2003)\citenamefont {Ji},
  \citenamefont {Chung}, \citenamefont {Sprinzak}, \citenamefont {Heiblum},
  \citenamefont {Mahalu},\ and\ \citenamefont {Shtrikman}}]{Ji03}%
  \BibitemOpen
  \bibfield  {author} {\bibinfo {author} {\bibfnamefont {Y.}~\bibnamefont
  {Ji}}, \bibinfo {author} {\bibfnamefont {Y.}~\bibnamefont {Chung}}, \bibinfo
  {author} {\bibfnamefont {D.}~\bibnamefont {Sprinzak}}, \bibinfo {author}
  {\bibfnamefont {M.}~\bibnamefont {Heiblum}}, \bibinfo {author} {\bibfnamefont
  {D.}~\bibnamefont {Mahalu}}, \ and\ \bibinfo {author} {\bibfnamefont
  {H.}~\bibnamefont {Shtrikman}},\ }\bibfield  {title} {\emph {\enquote
  {\bibinfo {title} {{An Electronic Mach-Zehnder Interferometer}},}\ }}\href
  {\doibase 10.1038/nature01503} {\bibfield  {journal} {\bibinfo  {journal}
  {Nature}\ }\textbf {\bibinfo {volume} {422}},\ \bibinfo {pages} {415}
  (\bibinfo {year} {2003})}\BibitemShut {NoStop}%
\bibitem [{\citenamefont {Roulleau}\ \emph {et~al.}(2008)\citenamefont
  {Roulleau}, \citenamefont {Portier}, \citenamefont {Roche}, \citenamefont
  {Cavanna}, \citenamefont {Faini}, \citenamefont {Gennser},\ and\
  \citenamefont {Mailly}}]{Roulleau08}%
  \BibitemOpen
  \bibfield  {author} {\bibinfo {author} {\bibfnamefont {P.}~\bibnamefont
  {Roulleau}}, \bibinfo {author} {\bibfnamefont {F.}~\bibnamefont {Portier}},
  \bibinfo {author} {\bibfnamefont {P.}~\bibnamefont {Roche}}, \bibinfo
  {author} {\bibfnamefont {A.}~\bibnamefont {Cavanna}}, \bibinfo {author}
  {\bibfnamefont {G.}~\bibnamefont {Faini}}, \bibinfo {author} {\bibfnamefont
  {U.}~\bibnamefont {Gennser}}, \ and\ \bibinfo {author} {\bibfnamefont
  {D.}~\bibnamefont {Mailly}},\ }\bibfield  {title} {\emph {\enquote {\bibinfo
  {title} {{Direct Measurement of the Coherence Length of Edge States in the
  Integer Quantum Hall Regime}},}\ }}\href {\doibase
  10.1103/PhysRevLett.100.126802} {\bibfield  {journal} {\bibinfo  {journal}
  {Phys. Rev. Lett.}\ }\textbf {\bibinfo {volume} {100}},\ \bibinfo {pages}
  {126802} (\bibinfo {year} {2008})}\BibitemShut {NoStop}%
\bibitem [{\citenamefont {Matveev}\ and\ \citenamefont
  {Glazman}(1993{\natexlab{a}})}]{matveev93}%
  \BibitemOpen
  \bibfield  {author} {\bibinfo {author} {\bibfnamefont {K.~A.}\ \bibnamefont
  {Matveev}}\ and\ \bibinfo {author} {\bibfnamefont {L.~I.}\ \bibnamefont
  {Glazman}},\ }\bibfield  {title} {\emph {\enquote {\bibinfo {title} {Coulomb
  blockade of tunneling into a quasi-one-dimensional wire},}\ }}\href {\doibase
  10.1103/PhysRevLett.70.990} {\bibfield  {journal} {\bibinfo  {journal} {Phys.
  Rev. Lett.}\ }\textbf {\bibinfo {volume} {70}},\ \bibinfo {pages} {990}
  (\bibinfo {year} {1993}{\natexlab{a}})}\BibitemShut {NoStop}%
\bibitem [{\citenamefont {Matveev}\ and\ \citenamefont
  {Glazman}(1993{\natexlab{b}})}]{matveev93a}%
  \BibitemOpen
  \bibfield  {author} {\bibinfo {author} {\bibfnamefont {K.~A.}\ \bibnamefont
  {Matveev}}\ and\ \bibinfo {author} {\bibfnamefont {L.~I.}\ \bibnamefont
  {Glazman}},\ }\bibfield  {title} {\emph {\enquote {\bibinfo {title}
  {Conductance and coulomb blockade in a multi-mode quantum wire},}\ }}\href
  {\doibase 10.1016/0921-4526(93)90169-7} {\bibfield  {journal} {\bibinfo
  {journal} {Physica B}\ }\textbf {\bibinfo {volume} {189}},\ \bibinfo {pages}
  {266 } (\bibinfo {year} {1993}{\natexlab{b}})}\BibitemShut {NoStop}%
\bibitem [{\citenamefont {Inoshita}(1998)}]{Inoshita98}%
  \BibitemOpen
  \bibfield  {author} {\bibinfo {author} {\bibfnamefont {T.}~\bibnamefont
  {Inoshita}},\ }\bibfield  {title} {\emph {\enquote {\bibinfo {title} {{Kondo
  Effect in Quantum Dots}},}\ }}\href {\doibase 10.1126/science.281.5376.526}
  {\bibfield  {journal} {\bibinfo  {journal} {Science}\ }\textbf {\bibinfo
  {volume} {281}},\ \bibinfo {pages} {526} (\bibinfo {year}
  {1998})}\BibitemShut {NoStop}%
\bibitem [{\citenamefont {Cronenwett}\ \emph {et~al.}(1998)\citenamefont
  {Cronenwett}, \citenamefont {Oosterkamp},\ and\ \citenamefont
  {Kouwenhoven}}]{Cronenwett98}%
  \BibitemOpen
  \bibfield  {author} {\bibinfo {author} {\bibfnamefont {S.~M.}\ \bibnamefont
  {Cronenwett}}, \bibinfo {author} {\bibfnamefont {T.~H.}\ \bibnamefont
  {Oosterkamp}}, \ and\ \bibinfo {author} {\bibfnamefont {L.~P.}\ \bibnamefont
  {Kouwenhoven}},\ }\bibfield  {title} {\emph {\enquote {\bibinfo {title} {{A
  Tunable Kondo Effect in Quantum Dots}},}\ }}\href {\doibase
  10.1126/science.281.5376.540} {\bibfield  {journal} {\bibinfo  {journal}
  {Science}\ }\textbf {\bibinfo {volume} {281}},\ \bibinfo {pages} {540}
  (\bibinfo {year} {1998})}\BibitemShut {NoStop}%
\bibitem [{\citenamefont {Andreev}(1964)}]{Andreev64}%
  \BibitemOpen
  \bibfield  {author} {\bibinfo {author} {\bibfnamefont {A.~F.}\ \bibnamefont
  {Andreev}},\ }\bibfield  {title} {\emph {\enquote {\bibinfo {title} {{The
  Thermal Conductivity of the Intermediate State in Superconductors}},}\
  }}\href@noop {} {\bibfield  {journal} {\bibinfo  {journal} {Zh. Eksp. Teor.
  Fiz.}\ }\textbf {\bibinfo {volume} {46}},\ \bibinfo {pages} {1823} (\bibinfo
  {year} {1964})},\ \translation{Sov. Phys. JETP {\bf{19}}, 1228,
  (1964)}\BibitemShut {NoStop}%
\bibitem [{\citenamefont {Katsnelson}\ \emph {et~al.}(2006)\citenamefont
  {Katsnelson}, \citenamefont {Novoselov},\ and\ \citenamefont
  {Geim}}]{Katsnelson06}%
  \BibitemOpen
  \bibfield  {author} {\bibinfo {author} {\bibfnamefont {M.}~\bibnamefont
  {Katsnelson}}, \bibinfo {author} {\bibfnamefont {K.}~\bibnamefont
  {Novoselov}}, \ and\ \bibinfo {author} {\bibfnamefont {A.}~\bibnamefont
  {Geim}},\ }\bibfield  {title} {\emph {\enquote {\bibinfo {title} {{Chiral
  tunneling and the Klein paradox in graphene}},}\ }}\href {\doibase
  10.1038/nphys384} {\bibfield  {journal} {\bibinfo  {journal} {Nat. Phys.}\
  }\textbf {\bibinfo {volume} {2}},\ \bibinfo {pages} {620} (\bibinfo {year}
  {2006})}\BibitemShut {NoStop}%
\bibitem [{\citenamefont {Stander}\ \emph {et~al.}(2009)\citenamefont
  {Stander}, \citenamefont {Huard},\ and\ \citenamefont
  {Goldhaber-Gordon}}]{Stander09}%
  \BibitemOpen
  \bibfield  {author} {\bibinfo {author} {\bibfnamefont {N.}~\bibnamefont
  {Stander}}, \bibinfo {author} {\bibfnamefont {B.}~\bibnamefont {Huard}}, \
  and\ \bibinfo {author} {\bibfnamefont {D.}~\bibnamefont {Goldhaber-Gordon}},\
  }\bibfield  {title} {\emph {\enquote {\bibinfo {title} {{Evidence for Klein
  Tunneling in Graphene $p\mathrm{\text{\ensuremath{-}}}n$ Junctions}},}\
  }}\href {\doibase 10.1103/PhysRevLett.102.026807} {\bibfield  {journal}
  {\bibinfo  {journal} {Phys. Rev. Lett.}\ }\textbf {\bibinfo {volume} {102}},\
  \bibinfo {pages} {026807} (\bibinfo {year} {2009})}\BibitemShut {NoStop}%
\bibitem [{\citenamefont {Blanter}\ and\ \citenamefont
  {B{\"u}ttiker}(2000)}]{Blanter00}%
  \BibitemOpen
  \bibfield  {author} {\bibinfo {author} {\bibfnamefont {Y.}~\bibnamefont
  {Blanter}}\ and\ \bibinfo {author} {\bibfnamefont {M.}~\bibnamefont
  {B{\"u}ttiker}},\ }\bibfield  {title} {\emph {\enquote {\bibinfo {title}
  {Shot noise in mesoscopic conductors},}\ }}\href {\doibase
  https://doi.org/10.1016/S0370-1573(99)00123-4} {\bibfield  {journal}
  {\bibinfo  {journal} {Phys. Rep.}\ }\textbf {\bibinfo {volume} {336}},\
  \bibinfo {pages} {1 } (\bibinfo {year} {2000})}\BibitemShut {NoStop}%
\bibitem [{\citenamefont {F{\`e}ve}\ \emph {et~al.}(2007)\citenamefont
  {F{\`e}ve}, \citenamefont {Mah{\'e}}, \citenamefont {Berroir}, \citenamefont
  {Kontos}, \citenamefont {Pla{\c c}ais}, \citenamefont {Glattli},
  \citenamefont {Cavanna}, \citenamefont {Etienne},\ and\ \citenamefont
  {Jin}}]{Feve08}%
  \BibitemOpen
  \bibfield  {author} {\bibinfo {author} {\bibfnamefont {G.}~\bibnamefont
  {F{\`e}ve}}, \bibinfo {author} {\bibfnamefont {A.}~\bibnamefont {Mah{\'e}}},
  \bibinfo {author} {\bibfnamefont {J.-M.}\ \bibnamefont {Berroir}}, \bibinfo
  {author} {\bibfnamefont {T.}~\bibnamefont {Kontos}}, \bibinfo {author}
  {\bibfnamefont {B.}~\bibnamefont {Pla{\c c}ais}}, \bibinfo {author}
  {\bibfnamefont {D.~C.}\ \bibnamefont {Glattli}}, \bibinfo {author}
  {\bibfnamefont {A.}~\bibnamefont {Cavanna}}, \bibinfo {author} {\bibfnamefont
  {B.}~\bibnamefont {Etienne}}, \ and\ \bibinfo {author} {\bibfnamefont
  {Y.}~\bibnamefont {Jin}},\ }\bibfield  {title} {\emph {\enquote {\bibinfo
  {title} {{An On-Demand Coherent Single-Electron Source}},}\ }}\href {\doibase
  10.1126/science.1141243} {\bibfield  {journal} {\bibinfo  {journal}
  {Science}\ }\textbf {\bibinfo {volume} {316}},\ \bibinfo {pages} {1169}
  (\bibinfo {year} {2007})}\BibitemShut {NoStop}%
\bibitem [{\citenamefont {Dubois}\ \emph {et~al.}(2013)\citenamefont {Dubois},
  \citenamefont {Jullien}, \citenamefont {Portier}, \citenamefont {Roche},
  \citenamefont {Cavanna}, \citenamefont {Jin}, \citenamefont {Wegscheider},
  \citenamefont {Roulleau},\ and\ \citenamefont {Glattli}}]{Dubois13}%
  \BibitemOpen
  \bibfield  {author} {\bibinfo {author} {\bibfnamefont {J.}~\bibnamefont
  {Dubois}}, \bibinfo {author} {\bibfnamefont {T.}~\bibnamefont {Jullien}},
  \bibinfo {author} {\bibfnamefont {F.}~\bibnamefont {Portier}}, \bibinfo
  {author} {\bibfnamefont {P.}~\bibnamefont {Roche}}, \bibinfo {author}
  {\bibfnamefont {A.}~\bibnamefont {Cavanna}}, \bibinfo {author} {\bibfnamefont
  {Y.}~\bibnamefont {Jin}}, \bibinfo {author} {\bibfnamefont {W.}~\bibnamefont
  {Wegscheider}}, \bibinfo {author} {\bibfnamefont {P.}~\bibnamefont
  {Roulleau}}, \ and\ \bibinfo {author} {\bibfnamefont {D.}~\bibnamefont
  {Glattli}},\ }\bibfield  {title} {\emph {\enquote {\bibinfo {title}
  {Minimal-excitation states for electron quantum optics using levitons},}\
  }}\href {\doibase 10.1038/nature12713} {\bibfield  {journal} {\bibinfo
  {journal} {Nature}\ }\textbf {\bibinfo {volume} {502}},\ \bibinfo {pages}
  {659} (\bibinfo {year} {2013})}\BibitemShut {NoStop}%
\bibitem [{\citenamefont {McNeil}\ \emph {et~al.}(2011)\citenamefont {McNeil},
  \citenamefont {Kataoka}, \citenamefont {Ford}, \citenamefont {Barnes},
  \citenamefont {Anderson}, \citenamefont {Jones}, \citenamefont {Farrer},\
  and\ \citenamefont {Ritchie}}]{McNeil11}%
  \BibitemOpen
  \bibfield  {author} {\bibinfo {author} {\bibfnamefont {R.}~\bibnamefont
  {McNeil}}, \bibinfo {author} {\bibfnamefont {M.}~\bibnamefont {Kataoka}},
  \bibinfo {author} {\bibfnamefont {C.}~\bibnamefont {Ford}}, \bibinfo {author}
  {\bibfnamefont {C.}~\bibnamefont {Barnes}}, \bibinfo {author} {\bibfnamefont
  {D.}~\bibnamefont {Anderson}}, \bibinfo {author} {\bibfnamefont
  {G.}~\bibnamefont {Jones}}, \bibinfo {author} {\bibfnamefont
  {I.}~\bibnamefont {Farrer}}, \ and\ \bibinfo {author} {\bibfnamefont
  {D.}~\bibnamefont {Ritchie}},\ }\bibfield  {title} {\emph {\enquote {\bibinfo
  {title} {On-demand single-electron transfer between distant quantum dots},}\
  }}\href {\doibase 10.1038/nature10444} {\bibfield  {journal} {\bibinfo
  {journal} {Nature}\ }\textbf {\bibinfo {volume} {477}},\ \bibinfo {pages}
  {439} (\bibinfo {year} {2011})}\BibitemShut {NoStop}%
\bibitem [{\citenamefont {Fletcher}\ \emph {et~al.}(2013)\citenamefont
  {Fletcher}, \citenamefont {See}, \citenamefont {Howe}, \citenamefont
  {Pepper}, \citenamefont {Giblin}, \citenamefont {Griffiths}, \citenamefont
  {Jones}, \citenamefont {Farrer}, \citenamefont {Ritchie}, \citenamefont
  {Janssen},\ and\ \citenamefont {Kataoka}}]{Fletcher13}%
  \BibitemOpen
  \bibfield  {author} {\bibinfo {author} {\bibfnamefont {J.~D.}\ \bibnamefont
  {Fletcher}}, \bibinfo {author} {\bibfnamefont {P.}~\bibnamefont {See}},
  \bibinfo {author} {\bibfnamefont {H.}~\bibnamefont {Howe}}, \bibinfo {author}
  {\bibfnamefont {M.}~\bibnamefont {Pepper}}, \bibinfo {author} {\bibfnamefont
  {S.~P.}\ \bibnamefont {Giblin}}, \bibinfo {author} {\bibfnamefont {J.~P.}\
  \bibnamefont {Griffiths}}, \bibinfo {author} {\bibfnamefont {G.~A.~C.}\
  \bibnamefont {Jones}}, \bibinfo {author} {\bibfnamefont {I.}~\bibnamefont
  {Farrer}}, \bibinfo {author} {\bibfnamefont {D.~A.}\ \bibnamefont {Ritchie}},
  \bibinfo {author} {\bibfnamefont {T.~J. B.~M.}\ \bibnamefont {Janssen}}, \
  and\ \bibinfo {author} {\bibfnamefont {M.}~\bibnamefont {Kataoka}},\
  }\bibfield  {title} {\emph {\enquote {\bibinfo {title} {Clock-controlled
  emission of single-electron wave packets in a solid-state circuit},}\ }}\href
  {\doibase 10.1103/PhysRevLett.111.216807} {\bibfield  {journal} {\bibinfo
  {journal} {Phys. Rev. Lett.}\ }\textbf {\bibinfo {volume} {111}},\ \bibinfo
  {pages} {216807} (\bibinfo {year} {2013})}\BibitemShut {NoStop}%
\bibitem [{\citenamefont {Roussely}\ \emph {et~al.}(2018)\citenamefont
  {Roussely}, \citenamefont {Arrighi}, \citenamefont {Georgiou}, \citenamefont
  {Takada}, \citenamefont {Schalk}, \citenamefont {Urdampilleta}, \citenamefont
  {Ludwig}, \citenamefont {Wieck}, \citenamefont {Armagnat}, \citenamefont
  {Kloss}, \citenamefont {Waintal}, \citenamefont {Meunier},\ and\
  \citenamefont {B{\"a}uerle}}]{Roussely2018}%
  \BibitemOpen
  \bibfield  {author} {\bibinfo {author} {\bibfnamefont {G.}~\bibnamefont
  {Roussely}}, \bibinfo {author} {\bibfnamefont {E.}~\bibnamefont {Arrighi}},
  \bibinfo {author} {\bibfnamefont {G.}~\bibnamefont {Georgiou}}, \bibinfo
  {author} {\bibfnamefont {S.}~\bibnamefont {Takada}}, \bibinfo {author}
  {\bibfnamefont {M.}~\bibnamefont {Schalk}}, \bibinfo {author} {\bibfnamefont
  {M.}~\bibnamefont {Urdampilleta}}, \bibinfo {author} {\bibfnamefont
  {A.}~\bibnamefont {Ludwig}}, \bibinfo {author} {\bibfnamefont {A.~D.}\
  \bibnamefont {Wieck}}, \bibinfo {author} {\bibfnamefont {P.}~\bibnamefont
  {Armagnat}}, \bibinfo {author} {\bibfnamefont {T.}~\bibnamefont {Kloss}},
  \bibinfo {author} {\bibfnamefont {X.}~\bibnamefont {Waintal}}, \bibinfo
  {author} {\bibfnamefont {T.}~\bibnamefont {Meunier}}, \ and\ \bibinfo
  {author} {\bibfnamefont {C.}~\bibnamefont {B{\"a}uerle}},\ }\bibfield
  {title} {\emph {\enquote {\bibinfo {title} {Unveiling the bosonic nature of
  an ultrashort few-electron pulse},}\ }}\href {\doibase
  10.1038/s41467-018-05203-7} {\bibfield  {journal} {\bibinfo  {journal} {Nat.
  Commun.}\ }\textbf {\bibinfo {volume} {9}},\ \bibinfo {pages} {2811}
  (\bibinfo {year} {2018})}\BibitemShut {NoStop}%
\bibitem [{\citenamefont {Hashisaka}\ \emph {et~al.}(2017)\citenamefont
  {Hashisaka}, \citenamefont {Hiyama}, \citenamefont {Akiho}, \citenamefont
  {Muraki},\ and\ \citenamefont {Fujisawa}}]{Hashisaka2017}%
  \BibitemOpen
  \bibfield  {author} {\bibinfo {author} {\bibfnamefont {M.}~\bibnamefont
  {Hashisaka}}, \bibinfo {author} {\bibfnamefont {N.}~\bibnamefont {Hiyama}},
  \bibinfo {author} {\bibfnamefont {T.}~\bibnamefont {Akiho}}, \bibinfo
  {author} {\bibfnamefont {K.}~\bibnamefont {Muraki}}, \ and\ \bibinfo {author}
  {\bibfnamefont {T.}~\bibnamefont {Fujisawa}},\ }\bibfield  {title} {\emph
  {\enquote {\bibinfo {title} {{Waveform measurement of charge- and
  spin-density wavepackets in a chiral Tomonaga--Luttinger liquid}},}\ }}\href
  {\doibase 10.1038/nphys4062} {\bibfield  {journal} {\bibinfo  {journal} {Nat.
  Phys.}\ }\textbf {\bibinfo {volume} {13}},\ \bibinfo {pages} {559} (\bibinfo
  {year} {2017})}\BibitemShut {NoStop}%
\bibitem [{\citenamefont {Vannucci}\ \emph {et~al.}(2018)\citenamefont
  {Vannucci}, \citenamefont {Ronetti}, \citenamefont {Ferraro}, \citenamefont
  {Rech}, \citenamefont {Jonckheere}, \citenamefont {Martin},\ and\
  \citenamefont {Sassetti}}]{Vannucci18}%
  \BibitemOpen
  \bibfield  {author} {\bibinfo {author} {\bibfnamefont {L.}~\bibnamefont
  {Vannucci}}, \bibinfo {author} {\bibfnamefont {F.}~\bibnamefont {Ronetti}},
  \bibinfo {author} {\bibfnamefont {D.}~\bibnamefont {Ferraro}}, \bibinfo
  {author} {\bibfnamefont {J.}~\bibnamefont {Rech}}, \bibinfo {author}
  {\bibfnamefont {T.}~\bibnamefont {Jonckheere}}, \bibinfo {author}
  {\bibfnamefont {T.}~\bibnamefont {Martin}}, \ and\ \bibinfo {author}
  {\bibfnamefont {M.}~\bibnamefont {Sassetti}},\ }\bibfield  {title} {\emph
  {\enquote {\bibinfo {title} {Photoassisted shot noise spectroscopy at
  fractional filling factor},}\ }}\href {\doibase
  10.1088/1742-6596/969/1/012143} {\bibfield  {journal} {\bibinfo  {journal}
  {J. Phys. Conf. Ser.}\ }\textbf {\bibinfo {volume} {969}},\ \bibinfo {pages}
  {012143} (\bibinfo {year} {2018})}\BibitemShut {NoStop}%
\bibitem [{\citenamefont {Klapwijk}\ \emph {et~al.}(1982)\citenamefont
  {Klapwijk}, \citenamefont {Blonder},\ and\ \citenamefont
  {Tinkham}}]{Klapwijk82}%
  \BibitemOpen
  \bibfield  {author} {\bibinfo {author} {\bibfnamefont {T.~M.}\ \bibnamefont
  {Klapwijk}}, \bibinfo {author} {\bibfnamefont {G.~E.}\ \bibnamefont
  {Blonder}}, \ and\ \bibinfo {author} {\bibfnamefont {M.}~\bibnamefont
  {Tinkham}},\ }\bibfield  {title} {\emph {\enquote {\bibinfo {title}
  {Explanation of subharmonic energy gap structure in superconducting
  contacts},}\ }}\href {\doibase https://doi.org/10.1016/0378-4363(82)90189-9}
  {\bibfield  {journal} {\bibinfo  {journal} {Physica B+C}\ }\textbf {\bibinfo
  {volume} {109-110}},\ \bibinfo {pages} {1657 } (\bibinfo {year}
  {1982})}\BibitemShut {NoStop}%
\bibitem [{\citenamefont {Averin}\ and\ \citenamefont
  {Bardas}(1995)}]{Averin95}%
  \BibitemOpen
  \bibfield  {author} {\bibinfo {author} {\bibfnamefont {D.}~\bibnamefont
  {Averin}}\ and\ \bibinfo {author} {\bibfnamefont {A.}~\bibnamefont
  {Bardas}},\ }\bibfield  {title} {\emph {\enquote {\bibinfo {title} {{ac
  Josephson Effect in a Single Quantum Channel}},}\ }}\href {\doibase
  10.1103/PhysRevLett.75.1831} {\bibfield  {journal} {\bibinfo  {journal}
  {Phys. Rev. Lett.}\ }\textbf {\bibinfo {volume} {75}},\ \bibinfo {pages}
  {1831} (\bibinfo {year} {1995})}\BibitemShut {NoStop}%
\bibitem [{\citenamefont {Rokhinson}\ \emph {et~al.}(2012)\citenamefont
  {Rokhinson}, \citenamefont {Liu},\ and\ \citenamefont
  {Furdyna}}]{Rokhinson12}%
  \BibitemOpen
  \bibfield  {author} {\bibinfo {author} {\bibfnamefont {L.}~\bibnamefont
  {Rokhinson}}, \bibinfo {author} {\bibfnamefont {X.}~\bibnamefont {Liu}}, \
  and\ \bibinfo {author} {\bibfnamefont {J.}~\bibnamefont {Furdyna}},\
  }\bibfield  {title} {\emph {\enquote {\bibinfo {title} {{The fractional A.C.
  Josephson effect in a semiconductor-superconductor nanowire as a signature of
  Majorana particles}},}\ }}\href {\doibase 10.1038/nphys2429} {\bibfield
  {journal} {\bibinfo  {journal} {Nat. Phys.}\ }\textbf {\bibinfo {volume}
  {8}},\ \bibinfo {pages} {795} (\bibinfo {year} {2012})}\BibitemShut {NoStop}%
\bibitem [{\citenamefont {San-Jose}\ \emph {et~al.}(2013)\citenamefont
  {San-Jose}, \citenamefont {Cayao}, \citenamefont {Prada},\ and\ \citenamefont
  {Aguado}}]{Sanjose13}%
  \BibitemOpen
  \bibfield  {author} {\bibinfo {author} {\bibfnamefont {P.}~\bibnamefont
  {San-Jose}}, \bibinfo {author} {\bibfnamefont {J.}~\bibnamefont {Cayao}},
  \bibinfo {author} {\bibfnamefont {E.}~\bibnamefont {Prada}}, \ and\ \bibinfo
  {author} {\bibfnamefont {R.}~\bibnamefont {Aguado}},\ }\bibfield  {title}
  {\emph {\enquote {\bibinfo {title} {{Multiple Andreev reflection and critical
  current in topological superconducting nanowire junctions}},}\ }}\href
  {\doibase 10.1088/1367-2630/15/7/075019} {\bibfield  {journal} {\bibinfo
  {journal} {New J. Physics}\ }\textbf {\bibinfo {volume} {15}},\ \bibinfo
  {pages} {075019} (\bibinfo {year} {2013})}\BibitemShut {NoStop}%
\bibitem [{\citenamefont {Bertoni}\ \emph {et~al.}(2000)\citenamefont
  {Bertoni}, \citenamefont {Bordone}, \citenamefont {Brunetti}, \citenamefont
  {Jacoboni},\ and\ \citenamefont {Reggiani}}]{Bertoni00}%
  \BibitemOpen
  \bibfield  {author} {\bibinfo {author} {\bibfnamefont {A.}~\bibnamefont
  {Bertoni}}, \bibinfo {author} {\bibfnamefont {P.}~\bibnamefont {Bordone}},
  \bibinfo {author} {\bibfnamefont {R.}~\bibnamefont {Brunetti}}, \bibinfo
  {author} {\bibfnamefont {C.}~\bibnamefont {Jacoboni}}, \ and\ \bibinfo
  {author} {\bibfnamefont {S.}~\bibnamefont {Reggiani}},\ }\bibfield  {title}
  {\emph {\enquote {\bibinfo {title} {Quantum logic gates based on coherent
  electron transport in quantum wires},}\ }}\href {\doibase
  10.1103/PhysRevLett.84.5912} {\bibfield  {journal} {\bibinfo  {journal}
  {Phys. Rev. Lett.}\ }\textbf {\bibinfo {volume} {84}},\ \bibinfo {pages}
  {5912} (\bibinfo {year} {2000})}\BibitemShut {NoStop}%
\bibitem [{\citenamefont {Ionicioiu}\ \emph {et~al.}(2001)\citenamefont
  {Ionicioiu}, \citenamefont {Amaratunga},\ and\ \citenamefont
  {Udrea}}]{Ionicioiu01}%
  \BibitemOpen
  \bibfield  {author} {\bibinfo {author} {\bibfnamefont {R.}~\bibnamefont
  {Ionicioiu}}, \bibinfo {author} {\bibfnamefont {G.}~\bibnamefont
  {Amaratunga}}, \ and\ \bibinfo {author} {\bibfnamefont {F.}~\bibnamefont
  {Udrea}},\ }\bibfield  {title} {\emph {\enquote {\bibinfo {title} {Quantum
  computation with ballistic electrons},}\ }}\href {\doibase
  10.1142/S0217979201003521} {\bibfield  {journal} {\bibinfo  {journal}
  {International Journal of Modern Physics B}\ }\textbf {\bibinfo {volume}
  {15}},\ \bibinfo {pages} {125} (\bibinfo {year} {2001})}\BibitemShut
  {NoStop}%
\bibitem [{\citenamefont {Bautze}\ \emph {et~al.}(2014)\citenamefont {Bautze},
  \citenamefont {S\"ussmeier}, \citenamefont {Takada}, \citenamefont {Groth},
  \citenamefont {Meunier}, \citenamefont {Yamamoto}, \citenamefont {Tarucha},
  \citenamefont {Waintal},\ and\ \citenamefont {B\"auerle}}]{Bautze14}%
  \BibitemOpen
  \bibfield  {author} {\bibinfo {author} {\bibfnamefont {T.}~\bibnamefont
  {Bautze}}, \bibinfo {author} {\bibfnamefont {C.}~\bibnamefont {S\"ussmeier}},
  \bibinfo {author} {\bibfnamefont {S.}~\bibnamefont {Takada}}, \bibinfo
  {author} {\bibfnamefont {C.}~\bibnamefont {Groth}}, \bibinfo {author}
  {\bibfnamefont {T.}~\bibnamefont {Meunier}}, \bibinfo {author} {\bibfnamefont
  {M.}~\bibnamefont {Yamamoto}}, \bibinfo {author} {\bibfnamefont
  {S.}~\bibnamefont {Tarucha}}, \bibinfo {author} {\bibfnamefont
  {X.}~\bibnamefont {Waintal}}, \ and\ \bibinfo {author} {\bibfnamefont
  {C.}~\bibnamefont {B\"auerle}},\ }\bibfield  {title} {\emph {\enquote
  {\bibinfo {title} {Theoretical, numerical, and experimental study of a flying
  qubit electronic interferometer},}\ }}\href {\doibase
  10.1103/PhysRevB.89.125432} {\bibfield  {journal} {\bibinfo  {journal} {Phys.
  Rev. B}\ }\textbf {\bibinfo {volume} {89}},\ \bibinfo {pages} {125432}
  (\bibinfo {year} {2014})}\BibitemShut {NoStop}%
\bibitem [{\citenamefont {B{\"a}uerle}\ \emph {et~al.}(2018)\citenamefont
  {B{\"a}uerle}, \citenamefont {Glattli}, \citenamefont {Meunier},
  \citenamefont {Portier}, \citenamefont {Roche}, \citenamefont {Roulleau},
  \citenamefont {Takada},\ and\ \citenamefont {Waintal}}]{Bauerle18}%
  \BibitemOpen
  \bibfield  {author} {\bibinfo {author} {\bibfnamefont {C.}~\bibnamefont
  {B{\"a}uerle}}, \bibinfo {author} {\bibfnamefont {D.~C.}\ \bibnamefont
  {Glattli}}, \bibinfo {author} {\bibfnamefont {T.}~\bibnamefont {Meunier}},
  \bibinfo {author} {\bibfnamefont {F.}~\bibnamefont {Portier}}, \bibinfo
  {author} {\bibfnamefont {P.}~\bibnamefont {Roche}}, \bibinfo {author}
  {\bibfnamefont {P.}~\bibnamefont {Roulleau}}, \bibinfo {author}
  {\bibfnamefont {S.}~\bibnamefont {Takada}}, \ and\ \bibinfo {author}
  {\bibfnamefont {X.}~\bibnamefont {Waintal}},\ }\bibfield  {title} {\emph
  {\enquote {\bibinfo {title} {Coherent control of single electrons: a review
  of current progress},}\ }}\href {\doibase 10.1088/1361-6633/aaa98a}
  {\bibfield  {journal} {\bibinfo  {journal} {Rep. Prog. Phys.}\ }\textbf
  {\bibinfo {volume} {81}},\ \bibinfo {pages} {056503} (\bibinfo {year}
  {2018})}\BibitemShut {NoStop}%
\bibitem [{\citenamefont {Glattli}\ \emph {et~al.}(2020)\citenamefont
  {Glattli}, \citenamefont {Nath}, \citenamefont {Taktak}, \citenamefont
  {Roulleau}, \citenamefont {Bauerle},\ and\ \citenamefont
  {Waintal}}]{Glattli20}%
  \BibitemOpen
  \bibfield  {author} {\bibinfo {author} {\bibfnamefont {D.~C.}\ \bibnamefont
  {Glattli}}, \bibinfo {author} {\bibfnamefont {J.}~\bibnamefont {Nath}},
  \bibinfo {author} {\bibfnamefont {I.}~\bibnamefont {Taktak}}, \bibinfo
  {author} {\bibfnamefont {P.}~\bibnamefont {Roulleau}}, \bibinfo {author}
  {\bibfnamefont {C.}~\bibnamefont {Bauerle}}, \ and\ \bibinfo {author}
  {\bibfnamefont {X.}~\bibnamefont {Waintal}},\ }\href@noop {} {\enquote
  {\bibinfo {title} {{Design of a Single-Shot Electron detector with
  sub-electron sensitivity for electron flying qubit operation}},}\ } (\bibinfo
  {year} {2020}),\ \Eprint {http://arxiv.org/abs/2002.03947} {arXiv:2002.03947
  [cond-mat.mes-hall]} \BibitemShut {NoStop}%
\bibitem [{\citenamefont {Caroli}\ \emph {et~al.}(1971)\citenamefont {Caroli},
  \citenamefont {Combescot}, \citenamefont {Nozieres},\ and\ \citenamefont
  {Saint-James}}]{Caroli71}%
  \BibitemOpen
  \bibfield  {author} {\bibinfo {author} {\bibfnamefont {C.}~\bibnamefont
  {Caroli}}, \bibinfo {author} {\bibfnamefont {R.}~\bibnamefont {Combescot}},
  \bibinfo {author} {\bibfnamefont {P.}~\bibnamefont {Nozieres}}, \ and\
  \bibinfo {author} {\bibfnamefont {D.}~\bibnamefont {Saint-James}},\
  }\bibfield  {title} {\emph {\enquote {\bibinfo {title} {Direct calculation of
  the tunneling current},}\ }}\href {\doibase 10.1088/0022-3719/4/8/018}
  {\bibfield  {journal} {\bibinfo  {journal} {J. Physics C}\ }\textbf {\bibinfo
  {volume} {4}},\ \bibinfo {pages} {916} (\bibinfo {year} {1971})}\BibitemShut
  {NoStop}%
\bibitem [{\citenamefont {Croy}\ and\ \citenamefont {Saalmann}(2009)}]{Croy09}%
  \BibitemOpen
  \bibfield  {author} {\bibinfo {author} {\bibfnamefont {A.}~\bibnamefont
  {Croy}}\ and\ \bibinfo {author} {\bibfnamefont {U.}~\bibnamefont
  {Saalmann}},\ }\bibfield  {title} {\emph {\enquote {\bibinfo {title}
  {Propagation scheme for nonequilibrium dynamics of electron transport in
  nanoscale devices},}\ }}\href {\doibase 10.1103/PhysRevB.80.245311}
  {\bibfield  {journal} {\bibinfo  {journal} {Phys. Rev. B}\ }\textbf {\bibinfo
  {volume} {80}},\ \bibinfo {pages} {245311} (\bibinfo {year}
  {2009})}\BibitemShut {NoStop}%
\bibitem [{\citenamefont {Moskalets}(2011)}]{Moskalets11}%
  \BibitemOpen
  \bibfield  {author} {\bibinfo {author} {\bibfnamefont {M.~V.}\ \bibnamefont
  {Moskalets}},\ }\href {\doibase 10.1142/p822} {\emph {\bibinfo {title}
  {Scattering Matrix Approach to Non-Stationary Quantum Transport}}}\ (\bibinfo
   {publisher} {Imperial College Press},\ \bibinfo {year} {2011})\BibitemShut
  {NoStop}%
\bibitem [{tkw()}]{tkwant}%
  \BibitemOpen
  \href@noop {} {}\bibinfo {note} {\textsc{Tkwant} is free software and can be
  found at
  \hyperlink{https://tkwant.kwant-project.org}{https://tkwant.kwant-project.org}}\BibitemShut
  {NoStop}%
\bibitem [{\citenamefont {Groth}\ \emph {et~al.}(2014)\citenamefont {Groth},
  \citenamefont {Wimmer}, \citenamefont {Akhmerov},\ and\ \citenamefont
  {Waintal}}]{groth14}%
  \BibitemOpen
  \bibfield  {author} {\bibinfo {author} {\bibfnamefont {C.~W.}\ \bibnamefont
  {Groth}}, \bibinfo {author} {\bibfnamefont {M.}~\bibnamefont {Wimmer}},
  \bibinfo {author} {\bibfnamefont {A.~R.}\ \bibnamefont {Akhmerov}}, \ and\
  \bibinfo {author} {\bibfnamefont {X.}~\bibnamefont {Waintal}},\ }\bibfield
  {title} {\emph {\enquote {\bibinfo {title} {Kwant: a software package for
  quantum transport},}\ }}\href
  {http://stacks.iop.org/1367-2630/16/i=6/a=063065} {\bibfield  {journal}
  {\bibinfo  {journal} {New J. Phys.}\ }\textbf {\bibinfo {volume} {16}},\
  \bibinfo {pages} {063065} (\bibinfo {year} {2014})}\BibitemShut {NoStop}%
\bibitem [{\citenamefont {Gaury}\ \emph
  {et~al.}(2014{\natexlab{a}})\citenamefont {Gaury}, \citenamefont {Weston},
  \citenamefont {Santin}, \citenamefont {Houzet}, \citenamefont {Groth},\ and\
  \citenamefont {Waintal}}]{gaury14}%
  \BibitemOpen
  \bibfield  {author} {\bibinfo {author} {\bibfnamefont {B.}~\bibnamefont
  {Gaury}}, \bibinfo {author} {\bibfnamefont {J.}~\bibnamefont {Weston}},
  \bibinfo {author} {\bibfnamefont {M.}~\bibnamefont {Santin}}, \bibinfo
  {author} {\bibfnamefont {M.}~\bibnamefont {Houzet}}, \bibinfo {author}
  {\bibfnamefont {C.}~\bibnamefont {Groth}}, \ and\ \bibinfo {author}
  {\bibfnamefont {X.}~\bibnamefont {Waintal}},\ }\bibfield  {title} {\emph
  {\enquote {\bibinfo {title} {Numerical simulations of time-resolved quantum
  electronics},}\ }}\href {\doibase
  http://dx.doi.org/10.1016/j.physrep.2013.09.001} {\bibfield  {journal}
  {\bibinfo  {journal} {Phys. Rep.}\ }\textbf {\bibinfo {volume} {534}},\
  \bibinfo {pages} {1 } (\bibinfo {year} {2014}{\natexlab{a}})}\BibitemShut
  {NoStop}%
\bibitem [{\citenamefont {Johansson}\ \emph {et~al.}(2012)\citenamefont
  {Johansson}, \citenamefont {Nation},\ and\ \citenamefont
  {Nori}}]{Johansson12a}%
  \BibitemOpen
  \bibfield  {author} {\bibinfo {author} {\bibfnamefont {J.}~\bibnamefont
  {Johansson}}, \bibinfo {author} {\bibfnamefont {P.}~\bibnamefont {Nation}}, \
  and\ \bibinfo {author} {\bibfnamefont {F.}~\bibnamefont {Nori}},\ }\bibfield
  {title} {\emph {\enquote {\bibinfo {title} {{QuTiP: An open-source Python
  framework for the dynamics of open quantum systems}},}\ }}\href {\doibase
  https://doi.org/10.1016/j.cpc.2012.02.021} {\bibfield  {journal} {\bibinfo
  {journal} {Comput. Phys. Commun.}\ }\textbf {\bibinfo {volume} {183}},\
  \bibinfo {pages} {1760 } (\bibinfo {year} {2012})}\BibitemShut {NoStop}%
\bibitem [{\citenamefont {Johansson}\ \emph {et~al.}(2013)\citenamefont
  {Johansson}, \citenamefont {Nation},\ and\ \citenamefont
  {Nori}}]{Johansson12b}%
  \BibitemOpen
  \bibfield  {author} {\bibinfo {author} {\bibfnamefont {J.}~\bibnamefont
  {Johansson}}, \bibinfo {author} {\bibfnamefont {P.}~\bibnamefont {Nation}}, \
  and\ \bibinfo {author} {\bibfnamefont {F.}~\bibnamefont {Nori}},\ }\bibfield
  {title} {\emph {\enquote {\bibinfo {title} {{QuTiP 2: A Python framework for
  the dynamics of open quantum systems}},}\ }}\href {\doibase
  https://doi.org/10.1016/j.cpc.2012.11.019} {\bibfield  {journal} {\bibinfo
  {journal} {Comput. Phys. Commun.}\ }\textbf {\bibinfo {volume} {184}},\
  \bibinfo {pages} {1234 } (\bibinfo {year} {2013})}\BibitemShut {NoStop}%
\bibitem [{\citenamefont {Gaury}\ and\ \citenamefont
  {Waintal}(2016)}]{gaury16}%
  \BibitemOpen
  \bibfield  {author} {\bibinfo {author} {\bibfnamefont {B.}~\bibnamefont
  {Gaury}}\ and\ \bibinfo {author} {\bibfnamefont {X.}~\bibnamefont
  {Waintal}},\ }\bibfield  {title} {\emph {\enquote {\bibinfo {title} {A
  computational approach to quantum noise in time-dependent nanoelectronic
  devices},}\ }}\href {\doibase https://doi.org/10.1016/j.physe.2015.09.009}
  {\bibfield  {journal} {\bibinfo  {journal} {Physica E}\ }\textbf {\bibinfo
  {volume} {75}},\ \bibinfo {pages} {72 } (\bibinfo {year} {2016})}\BibitemShut
  {NoStop}%
\bibitem [{\citenamefont {Gaury}\ \emph {et~al.}(2015)\citenamefont {Gaury},
  \citenamefont {Weston},\ and\ \citenamefont {Waintal}}]{gaury15}%
  \BibitemOpen
  \bibfield  {author} {\bibinfo {author} {\bibfnamefont {B.}~\bibnamefont
  {Gaury}}, \bibinfo {author} {\bibfnamefont {J.}~\bibnamefont {Weston}}, \
  and\ \bibinfo {author} {\bibfnamefont {X.}~\bibnamefont {Waintal}},\
  }\bibfield  {title} {\emph {\enquote {\bibinfo {title} {{The a.c. Josephson
  effect without superconductivity}},}\ }}\href {\doibase 10.1038/ncomms7524}
  {\bibfield  {journal} {\bibinfo  {journal} {Nat. Commun.}\ }\textbf {\bibinfo
  {volume} {6}},\ \bibinfo {pages} {6524} (\bibinfo {year} {2015})}\BibitemShut
  {NoStop}%
\bibitem [{\citenamefont {{van der Walt}}\ \emph {et~al.}(2011)\citenamefont
  {{van der Walt}}, \citenamefont {{Colbert}},\ and\ \citenamefont
  {{Varoquaux}}}]{numpy}%
  \BibitemOpen
  \bibfield  {author} {\bibinfo {author} {\bibfnamefont {S.}~\bibnamefont {{van
  der Walt}}}, \bibinfo {author} {\bibfnamefont {S.~C.}\ \bibnamefont
  {{Colbert}}}, \ and\ \bibinfo {author} {\bibfnamefont {G.}~\bibnamefont
  {{Varoquaux}}},\ }\bibfield  {title} {\emph {\enquote {\bibinfo {title} {{The
  NumPy Array: A Structure for Efficient Numerical Computation}},}\ }}\href
  {\doibase 10.1109/MCSE.2011.37} {\bibfield  {journal} {\bibinfo  {journal}
  {Comput. Sci. Eng.}\ }\textbf {\bibinfo {volume} {13}},\ \bibinfo {pages}
  {22} (\bibinfo {year} {2011})}\BibitemShut {NoStop}%
\bibitem [{\citenamefont {Weston}\ and\ \citenamefont
  {Waintal}(2016{\natexlab{a}})}]{weston16b}%
  \BibitemOpen
  \bibfield  {author} {\bibinfo {author} {\bibfnamefont {J.}~\bibnamefont
  {Weston}}\ and\ \bibinfo {author} {\bibfnamefont {X.}~\bibnamefont
  {Waintal}},\ }\bibfield  {title} {\emph {\enquote {\bibinfo {title} {Towards
  realistic time-resolved simulations of quantum devices},}\ }}\href {\doibase
  10.1007/s10825-016-0855-9} {\bibfield  {journal} {\bibinfo  {journal} {J.
  Comput. Electron.}\ }\textbf {\bibinfo {volume} {15}},\ \bibinfo {pages}
  {1148} (\bibinfo {year} {2016}{\natexlab{a}})}\BibitemShut {NoStop}%
\bibitem [{\citenamefont {Rossignol}\ \emph {et~al.}(2018)\citenamefont
  {Rossignol}, \citenamefont {Kloss}, \citenamefont {Armagnat},\ and\
  \citenamefont {Waintal}}]{Rossignol18}%
  \BibitemOpen
  \bibfield  {author} {\bibinfo {author} {\bibfnamefont {B.}~\bibnamefont
  {Rossignol}}, \bibinfo {author} {\bibfnamefont {T.}~\bibnamefont {Kloss}},
  \bibinfo {author} {\bibfnamefont {P.}~\bibnamefont {Armagnat}}, \ and\
  \bibinfo {author} {\bibfnamefont {X.}~\bibnamefont {Waintal}},\ }\bibfield
  {title} {\emph {\enquote {\bibinfo {title} {Toward flying qubit
  spectroscopy},}\ }}\href {\doibase 10.1103/PhysRevB.98.205302} {\bibfield
  {journal} {\bibinfo  {journal} {Phys. Rev. B}\ }\textbf {\bibinfo {volume}
  {98}},\ \bibinfo {pages} {205302} (\bibinfo {year} {2018})}\BibitemShut
  {NoStop}%
\bibitem [{\citenamefont {Weston}\ and\ \citenamefont
  {Waintal}(2016{\natexlab{b}})}]{weston16a}%
  \BibitemOpen
  \bibfield  {author} {\bibinfo {author} {\bibfnamefont {J.}~\bibnamefont
  {Weston}}\ and\ \bibinfo {author} {\bibfnamefont {X.}~\bibnamefont
  {Waintal}},\ }\bibfield  {title} {\emph {\enquote {\bibinfo {title}
  {Linear-scaling source-sink algorithm for simulating time-resolved quantum
  transport and superconductivity},}\ }}\href {\doibase
  10.1103/PhysRevB.93.134506} {\bibfield  {journal} {\bibinfo  {journal} {Phys.
  Rev. B}\ }\textbf {\bibinfo {volume} {93}},\ \bibinfo {pages} {134506}
  (\bibinfo {year} {2016}{\natexlab{b}})}\BibitemShut {NoStop}%
\bibitem [{\citenamefont {Weston}\ \emph {et~al.}(2015)\citenamefont {Weston},
  \citenamefont {Gaury},\ and\ \citenamefont {Waintal}}]{weston15}%
  \BibitemOpen
  \bibfield  {author} {\bibinfo {author} {\bibfnamefont {J.}~\bibnamefont
  {Weston}}, \bibinfo {author} {\bibfnamefont {B.}~\bibnamefont {Gaury}}, \
  and\ \bibinfo {author} {\bibfnamefont {X.}~\bibnamefont {Waintal}},\
  }\bibfield  {title} {\emph {\enquote {\bibinfo {title} {{Manipulating Andreev
  and Majorana bound states with microwaves}},}\ }}\href {\doibase
  10.1103/PhysRevB.92.020513} {\bibfield  {journal} {\bibinfo  {journal} {Phys.
  Rev. B}\ }\textbf {\bibinfo {volume} {92}},\ \bibinfo {pages} {020513}
  (\bibinfo {year} {2015})}\BibitemShut {NoStop}%
\bibitem [{\citenamefont {Rossignol}\ \emph {et~al.}(2019)\citenamefont
  {Rossignol}, \citenamefont {Kloss},\ and\ \citenamefont
  {Waintal}}]{rossignol19}%
  \BibitemOpen
  \bibfield  {author} {\bibinfo {author} {\bibfnamefont {B.}~\bibnamefont
  {Rossignol}}, \bibinfo {author} {\bibfnamefont {T.}~\bibnamefont {Kloss}}, \
  and\ \bibinfo {author} {\bibfnamefont {X.}~\bibnamefont {Waintal}},\
  }\bibfield  {title} {\emph {\enquote {\bibinfo {title} {{Role of
  Quasiparticles in an Electric Circuit with Josephson Junctions}},}\ }}\href
  {\doibase 10.1103/PhysRevLett.122.207702} {\bibfield  {journal} {\bibinfo
  {journal} {Phys. Rev. Lett.}\ }\textbf {\bibinfo {volume} {122}},\ \bibinfo
  {pages} {207702} (\bibinfo {year} {2019})}\BibitemShut {NoStop}%
\bibitem [{\citenamefont {Gaury}\ and\ \citenamefont
  {Waintal}(2014)}]{gaury14a}%
  \BibitemOpen
  \bibfield  {author} {\bibinfo {author} {\bibfnamefont {B.}~\bibnamefont
  {Gaury}}\ and\ \bibinfo {author} {\bibfnamefont {X.}~\bibnamefont
  {Waintal}},\ }\bibfield  {title} {\emph {\enquote {\bibinfo {title}
  {Dynamical control of interference using voltage pulses in the quantum
  regime},}\ }}\href {\doibase 10.1038/ncomms4844} {\bibfield  {journal}
  {\bibinfo  {journal} {Nat. Commun.}\ }\textbf {\bibinfo {volume} {5}},\
  \bibinfo {pages} {3844} (\bibinfo {year} {2014})}\BibitemShut {NoStop}%
\bibitem [{\citenamefont {Gaury}\ \emph
  {et~al.}(2014{\natexlab{b}})\citenamefont {Gaury}, \citenamefont {Weston},\
  and\ \citenamefont {Waintal}}]{gaury14b}%
  \BibitemOpen
  \bibfield  {author} {\bibinfo {author} {\bibfnamefont {B.}~\bibnamefont
  {Gaury}}, \bibinfo {author} {\bibfnamefont {J.}~\bibnamefont {Weston}}, \
  and\ \bibinfo {author} {\bibfnamefont {X.}~\bibnamefont {Waintal}},\
  }\bibfield  {title} {\emph {\enquote {\bibinfo {title} {{Stopping electrons
  with radio-frequency pulses in the quantum Hall regime}},}\ }}\href {\doibase
  10.1103/PhysRevB.90.161305} {\bibfield  {journal} {\bibinfo  {journal} {Phys.
  Rev. B}\ }\textbf {\bibinfo {volume} {90}},\ \bibinfo {pages} {161305}
  (\bibinfo {year} {2014}{\natexlab{b}})}\BibitemShut {NoStop}%
\bibitem [{\citenamefont {Fruchart}\ \emph {et~al.}(2016)\citenamefont
  {Fruchart}, \citenamefont {Delplace}, \citenamefont {Weston}, \citenamefont
  {Waintal},\ and\ \citenamefont {Carpentier}}]{fruchart16}%
  \BibitemOpen
  \bibfield  {author} {\bibinfo {author} {\bibfnamefont {M.}~\bibnamefont
  {Fruchart}}, \bibinfo {author} {\bibfnamefont {P.}~\bibnamefont {Delplace}},
  \bibinfo {author} {\bibfnamefont {J.}~\bibnamefont {Weston}}, \bibinfo
  {author} {\bibfnamefont {X.}~\bibnamefont {Waintal}}, \ and\ \bibinfo
  {author} {\bibfnamefont {D.}~\bibnamefont {Carpentier}},\ }\bibfield  {title}
  {\emph {\enquote {\bibinfo {title} {{Probing (topological) Floquet states
  through DC transport}},}\ }}\href {\doibase
  https://doi.org/10.1016/j.physe.2015.09.035} {\bibfield  {journal} {\bibinfo
  {journal} {Physica E}\ }\textbf {\bibinfo {volume} {75}},\ \bibinfo {pages}
  {287} (\bibinfo {year} {2016})}\BibitemShut {NoStop}%
\bibitem [{\citenamefont {Ivanov}\ \emph {et~al.}(1997)\citenamefont {Ivanov},
  \citenamefont {Lee},\ and\ \citenamefont {Levitov}}]{levitov97}%
  \BibitemOpen
  \bibfield  {author} {\bibinfo {author} {\bibfnamefont {D.~A.}\ \bibnamefont
  {Ivanov}}, \bibinfo {author} {\bibfnamefont {H.~W.}\ \bibnamefont {Lee}}, \
  and\ \bibinfo {author} {\bibfnamefont {L.~S.}\ \bibnamefont {Levitov}},\
  }\bibfield  {title} {\emph {\enquote {\bibinfo {title} {Coherent states of
  alternating current},}\ }}\href {\doibase 10.1103/PhysRevB.56.6839}
  {\bibfield  {journal} {\bibinfo  {journal} {Phys. Rev. B}\ }\textbf {\bibinfo
  {volume} {56}},\ \bibinfo {pages} {6839} (\bibinfo {year}
  {1997})}\BibitemShut {NoStop}%
\bibitem [{\citenamefont {Keeling}\ \emph {et~al.}(2006)\citenamefont
  {Keeling}, \citenamefont {Klich},\ and\ \citenamefont {Levitov}}]{keeling06}%
  \BibitemOpen
  \bibfield  {author} {\bibinfo {author} {\bibfnamefont {J.}~\bibnamefont
  {Keeling}}, \bibinfo {author} {\bibfnamefont {I.}~\bibnamefont {Klich}}, \
  and\ \bibinfo {author} {\bibfnamefont {L.~S.}\ \bibnamefont {Levitov}},\
  }\bibfield  {title} {\emph {\enquote {\bibinfo {title} {Minimal excitation
  states of electrons in one-dimensional wires},}\ }}\href {\doibase
  10.1103/PhysRevLett.97.116403} {\bibfield  {journal} {\bibinfo  {journal}
  {Phys. Rev. Lett.}\ }\textbf {\bibinfo {volume} {97}},\ \bibinfo {pages}
  {116403} (\bibinfo {year} {2006})}\BibitemShut {NoStop}%
\bibitem [{\citenamefont {Levitov}\ \emph {et~al.}(1996)\citenamefont
  {Levitov}, \citenamefont {Lee},\ and\ \citenamefont {Lesovik}}]{levitov96}%
  \BibitemOpen
  \bibfield  {author} {\bibinfo {author} {\bibfnamefont {L.~S.}\ \bibnamefont
  {Levitov}}, \bibinfo {author} {\bibfnamefont {H.}~\bibnamefont {Lee}}, \ and\
  \bibinfo {author} {\bibfnamefont {G.~B.}\ \bibnamefont {Lesovik}},\
  }\bibfield  {title} {\emph {\enquote {\bibinfo {title} {Electron counting
  statistics and coherent states of electric current},}\ }}\href {\doibase
  10.1063/1.531672} {\bibfield  {journal} {\bibinfo  {journal} {J. Math.
  Phys.}\ }\textbf {\bibinfo {volume} {37}},\ \bibinfo {pages} {4845} (\bibinfo
  {year} {1996})}\BibitemShut {NoStop}%
\bibitem [{\citenamefont {Abbout}\ \emph {et~al.}(2018)\citenamefont {Abbout},
  \citenamefont {Weston}, \citenamefont {Waintal},\ and\ \citenamefont
  {Manchon}}]{Abbout18}%
  \BibitemOpen
  \bibfield  {author} {\bibinfo {author} {\bibfnamefont {A.}~\bibnamefont
  {Abbout}}, \bibinfo {author} {\bibfnamefont {J.}~\bibnamefont {Weston}},
  \bibinfo {author} {\bibfnamefont {X.}~\bibnamefont {Waintal}}, \ and\
  \bibinfo {author} {\bibfnamefont {A.}~\bibnamefont {Manchon}},\ }\bibfield
  {title} {\emph {\enquote {\bibinfo {title} {{Cooperative Charge Pumping and
  Enhanced Skyrmion Mobility}},}\ }}\href {\doibase
  10.1103/PhysRevLett.121.257203} {\bibfield  {journal} {\bibinfo  {journal}
  {Phys. Rev. Lett.}\ }\textbf {\bibinfo {volume} {121}},\ \bibinfo {pages}
  {257203} (\bibinfo {year} {2018})}\BibitemShut {NoStop}%
\bibitem [{\citenamefont {Kara~Slimane}\ \emph {et~al.}(2020)\citenamefont
  {Kara~Slimane}, \citenamefont {Reck},\ and\ \citenamefont
  {Fleury}}]{slimane20}%
  \BibitemOpen
  \bibfield  {author} {\bibinfo {author} {\bibfnamefont {A.}~\bibnamefont
  {Kara~Slimane}}, \bibinfo {author} {\bibfnamefont {P.}~\bibnamefont {Reck}},
  \ and\ \bibinfo {author} {\bibfnamefont {G.}~\bibnamefont {Fleury}},\
  }\bibfield  {title} {\emph {\enquote {\bibinfo {title} {Simulating
  time-dependent thermoelectric transport in quantum systems},}\ }}\href
  {\doibase 10.1103/PhysRevB.101.235413} {\bibfield  {journal} {\bibinfo
  {journal} {Phys. Rev. B}\ }\textbf {\bibinfo {volume} {101}},\ \bibinfo
  {pages} {235413} (\bibinfo {year} {2020})}\BibitemShut {NoStop}%
\bibitem [{\citenamefont {Kloss}\ \emph {et~al.}(2018)\citenamefont {Kloss},
  \citenamefont {Weston},\ and\ \citenamefont {Waintal}}]{kloss18}%
  \BibitemOpen
  \bibfield  {author} {\bibinfo {author} {\bibfnamefont {T.}~\bibnamefont
  {Kloss}}, \bibinfo {author} {\bibfnamefont {J.}~\bibnamefont {Weston}}, \
  and\ \bibinfo {author} {\bibfnamefont {X.}~\bibnamefont {Waintal}},\
  }\bibfield  {title} {\emph {\enquote {\bibinfo {title} {{Transient and
  Sharvin resistances of Luttinger liquids}},}\ }}\href {\doibase
  10.1103/PhysRevB.97.165134} {\bibfield  {journal} {\bibinfo  {journal} {Phys.
  Rev. B}\ }\textbf {\bibinfo {volume} {97}},\ \bibinfo {pages} {165134}
  (\bibinfo {year} {2018})}\BibitemShut {NoStop}%
\bibitem [{\citenamefont {Keldysh}(1964)}]{Keldysh64}%
  \BibitemOpen
  \bibfield  {author} {\bibinfo {author} {\bibfnamefont {L.~V.}\ \bibnamefont
  {Keldysh}},\ }\bibfield  {title} {\emph {\enquote {\bibinfo {title} {Diagram
  technique for non-equilibrium processes},}\ }}\href@noop {} {\bibfield
  {journal} {\bibinfo  {journal} {Zh. Eksp. Teor. Fiz.}\ }\textbf {\bibinfo
  {volume} {47}},\ \bibinfo {pages} {1515} (\bibinfo {year} {1964})},\
  \translation{Sov. Phys. JETP {\bf{20}}, 1018, (1965)}\BibitemShut {NoStop}%
\bibitem [{\citenamefont {Rammer}\ and\ \citenamefont
  {Smith}(1986)}]{rammer86}%
  \BibitemOpen
  \bibfield  {author} {\bibinfo {author} {\bibfnamefont {J.}~\bibnamefont
  {Rammer}}\ and\ \bibinfo {author} {\bibfnamefont {H.}~\bibnamefont {Smith}},\
  }\bibfield  {title} {\emph {\enquote {\bibinfo {title} {Quantum
  field-theoretical methods in transport theory of metals},}\ }}\href {\doibase
  10.1103/RevModPhys.58.323} {\bibfield  {journal} {\bibinfo  {journal} {Rev.
  Mod. Phys.}\ }\textbf {\bibinfo {volume} {58}},\ \bibinfo {pages} {323}
  (\bibinfo {year} {1986})}\BibitemShut {NoStop}%
\bibitem [{\citenamefont {Rammer}(2007)}]{Rammer07}%
  \BibitemOpen
  \bibfield  {author} {\bibinfo {author} {\bibfnamefont {J.}~\bibnamefont
  {Rammer}},\ }\href {\doibase 10.1017/CBO9780511618956} {\emph {\bibinfo
  {title} {{Quantum Field Theory of Non-equilibrium States}}}}\ (\bibinfo
  {publisher} {Cambridge University Press, Cambridge},\ \bibinfo {year}
  {2007})\BibitemShut {NoStop}%
\bibitem [{\citenamefont {Meir}\ and\ \citenamefont {Wingreen}(1992)}]{Meir92}%
  \BibitemOpen
  \bibfield  {author} {\bibinfo {author} {\bibfnamefont {Y.}~\bibnamefont
  {Meir}}\ and\ \bibinfo {author} {\bibfnamefont {N.~S.}\ \bibnamefont
  {Wingreen}},\ }\bibfield  {title} {\emph {\enquote {\bibinfo {title}
  {Landauer formula for the current through an interacting electron region},}\
  }}\href {\doibase 10.1103/PhysRevLett.68.2512} {\bibfield  {journal}
  {\bibinfo  {journal} {Phys. Rev. Lett.}\ }\textbf {\bibinfo {volume} {68}},\
  \bibinfo {pages} {2512} (\bibinfo {year} {1992})}\BibitemShut {NoStop}%
\bibitem [{\citenamefont {Wingreen}\ \emph {et~al.}(1993)\citenamefont
  {Wingreen}, \citenamefont {Jauho},\ and\ \citenamefont {Meir}}]{Wingreen93}%
  \BibitemOpen
  \bibfield  {author} {\bibinfo {author} {\bibfnamefont {N.~S.}\ \bibnamefont
  {Wingreen}}, \bibinfo {author} {\bibfnamefont {A.-P.}\ \bibnamefont {Jauho}},
  \ and\ \bibinfo {author} {\bibfnamefont {Y.}~\bibnamefont {Meir}},\
  }\bibfield  {title} {\emph {\enquote {\bibinfo {title} {Time-dependent
  transport through a mesoscopic structure},}\ }}\href {\doibase
  10.1103/PhysRevB.48.8487} {\bibfield  {journal} {\bibinfo  {journal} {Phys.
  Rev. B}\ }\textbf {\bibinfo {volume} {48}},\ \bibinfo {pages} {8487}
  (\bibinfo {year} {1993})}\BibitemShut {NoStop}%
\bibitem [{\citenamefont {Jauho}\ \emph {et~al.}(1994)\citenamefont {Jauho},
  \citenamefont {Wingreen},\ and\ \citenamefont {Meir}}]{Jauho94}%
  \BibitemOpen
  \bibfield  {author} {\bibinfo {author} {\bibfnamefont {A.-P.}\ \bibnamefont
  {Jauho}}, \bibinfo {author} {\bibfnamefont {N.~S.}\ \bibnamefont {Wingreen}},
  \ and\ \bibinfo {author} {\bibfnamefont {Y.}~\bibnamefont {Meir}},\
  }\bibfield  {title} {\emph {\enquote {\bibinfo {title} {Time-dependent
  transport in interacting and noninteracting resonant-tunneling systems},}\
  }}\href {\doibase 10.1103/PhysRevB.50.5528} {\bibfield  {journal} {\bibinfo
  {journal} {Phys. Rev. B}\ }\textbf {\bibinfo {volume} {50}},\ \bibinfo
  {pages} {5528} (\bibinfo {year} {1994})}\BibitemShut {NoStop}%
\bibitem [{\citenamefont {Li}\ \emph {et~al.}(2007)\citenamefont {Li},
  \citenamefont {Zhang}, \citenamefont {Hou}, \citenamefont {Qian},
  \citenamefont {Shen}, \citenamefont {Zhao},\ and\ \citenamefont
  {Xue}}]{Li07}%
  \BibitemOpen
  \bibfield  {author} {\bibinfo {author} {\bibfnamefont {R.}~\bibnamefont
  {Li}}, \bibinfo {author} {\bibfnamefont {J.}~\bibnamefont {Zhang}}, \bibinfo
  {author} {\bibfnamefont {S.}~\bibnamefont {Hou}}, \bibinfo {author}
  {\bibfnamefont {Z.}~\bibnamefont {Qian}}, \bibinfo {author} {\bibfnamefont
  {Z.}~\bibnamefont {Shen}}, \bibinfo {author} {\bibfnamefont {X.}~\bibnamefont
  {Zhao}}, \ and\ \bibinfo {author} {\bibfnamefont {Z.}~\bibnamefont {Xue}},\
  }\bibfield  {title} {\emph {\enquote {\bibinfo {title} {{A corrected NEGF+DFT
  approach for calculating electronic transport through molecular devices:
  Filling bound states and patching the non-equilibrium integration}},}\
  }}\href {\doibase https://doi.org/10.1016/j.chemphys.2007.06.011} {\bibfield
  {journal} {\bibinfo  {journal} {Chem. Phys.}\ }\textbf {\bibinfo {volume}
  {336}},\ \bibinfo {pages} {127 } (\bibinfo {year} {2007})}\BibitemShut
  {NoStop}%
\bibitem [{\citenamefont {Dhar}\ and\ \citenamefont {Sen}(2006)}]{Dhar06}%
  \BibitemOpen
  \bibfield  {author} {\bibinfo {author} {\bibfnamefont {A.}~\bibnamefont
  {Dhar}}\ and\ \bibinfo {author} {\bibfnamefont {D.}~\bibnamefont {Sen}},\
  }\bibfield  {title} {\emph {\enquote {\bibinfo {title} {{Nonequilibrium
  Green's function formalism and the problem of bound states}},}\ }}\href
  {\doibase 10.1103/PhysRevB.73.085119} {\bibfield  {journal} {\bibinfo
  {journal} {Phys. Rev. B}\ }\textbf {\bibinfo {volume} {73}},\ \bibinfo
  {pages} {085119} (\bibinfo {year} {2006})}\BibitemShut {NoStop}%
\bibitem [{\citenamefont {Khosravi}\ \emph {et~al.}(2009)\citenamefont
  {Khosravi}, \citenamefont {Stefanucci}, \citenamefont {Kurth},\ and\
  \citenamefont {Gross}}]{Khosravi09}%
  \BibitemOpen
  \bibfield  {author} {\bibinfo {author} {\bibfnamefont {E.}~\bibnamefont
  {Khosravi}}, \bibinfo {author} {\bibfnamefont {G.}~\bibnamefont
  {Stefanucci}}, \bibinfo {author} {\bibfnamefont {S.}~\bibnamefont {Kurth}}, \
  and\ \bibinfo {author} {\bibfnamefont {E.}~\bibnamefont {Gross}},\ }\bibfield
   {title} {\emph {\enquote {\bibinfo {title} {Bound states in time-dependent
  quantum transport: oscillations and memory effects in current and density},}\
  }}\href {\doibase 10.1039/B906528H} {\bibfield  {journal} {\bibinfo
  {journal} {Phys. Chem. Chem. Phys.}\ }\textbf {\bibinfo {volume} {11}},\
  \bibinfo {pages} {4535} (\bibinfo {year} {2009})}\BibitemShut {NoStop}%
\bibitem [{\citenamefont {Stefanucci}(2007)}]{Stefanucci07}%
  \BibitemOpen
  \bibfield  {author} {\bibinfo {author} {\bibfnamefont {G.}~\bibnamefont
  {Stefanucci}},\ }\bibfield  {title} {\emph {\enquote {\bibinfo {title} {Bound
  states in ab initio approaches to quantum transport: A time-dependent
  formulation},}\ }}\href {\doibase 10.1103/PhysRevB.75.195115} {\bibfield
  {journal} {\bibinfo  {journal} {Phys. Rev. B}\ }\textbf {\bibinfo {volume}
  {75}},\ \bibinfo {pages} {195115} (\bibinfo {year} {2007})}\BibitemShut
  {NoStop}%
\bibitem [{\citenamefont {Khosravi}\ \emph {et~al.}(2008)\citenamefont
  {Khosravi}, \citenamefont {Kurth}, \citenamefont {Stefanucci},\ and\
  \citenamefont {Gross}}]{Khosravi08}%
  \BibitemOpen
  \bibfield  {author} {\bibinfo {author} {\bibfnamefont {E.}~\bibnamefont
  {Khosravi}}, \bibinfo {author} {\bibfnamefont {S.}~\bibnamefont {Kurth}},
  \bibinfo {author} {\bibfnamefont {G.}~\bibnamefont {Stefanucci}}, \ and\
  \bibinfo {author} {\bibfnamefont {E.~K.~U.}\ \bibnamefont {Gross}},\
  }\bibfield  {title} {\emph {\enquote {\bibinfo {title} {The role of bound
  states in time-dependent quantum transport},}\ }}\href {\doibase
  10.1007/s00339-008-4864-9} {\bibfield  {journal} {\bibinfo  {journal} {Appl.
  Phys. A}\ }\textbf {\bibinfo {volume} {93}},\ \bibinfo {pages} {355}
  (\bibinfo {year} {2008})}\BibitemShut {NoStop}%
\bibitem [{\citenamefont {Istas}\ \emph {et~al.}(2018)\citenamefont {Istas},
  \citenamefont {Groth}, \citenamefont {Akhmerov}, \citenamefont {Wimmer},\
  and\ \citenamefont {Waintal}}]{istas18}%
  \BibitemOpen
  \bibfield  {author} {\bibinfo {author} {\bibfnamefont {M.}~\bibnamefont
  {Istas}}, \bibinfo {author} {\bibfnamefont {C.}~\bibnamefont {Groth}},
  \bibinfo {author} {\bibfnamefont {A.~R.}\ \bibnamefont {Akhmerov}}, \bibinfo
  {author} {\bibfnamefont {M.}~\bibnamefont {Wimmer}}, \ and\ \bibinfo {author}
  {\bibfnamefont {X.}~\bibnamefont {Waintal}},\ }\bibfield  {title} {\emph
  {\enquote {\bibinfo {title} {{A general algorithm for computing bound states
  in infinite tight-binding systems}},}\ }}\href {\doibase
  10.21468/SciPostPhys.4.5.026} {\bibfield  {journal} {\bibinfo  {journal}
  {SciPost Phys.}\ }\textbf {\bibinfo {volume} {4}},\ \bibinfo {pages} {26}
  (\bibinfo {year} {2018})}\BibitemShut {NoStop}%
\bibitem [{kwa()}]{kwantspectrum}%
  \BibitemOpen
  \href@noop {} {}\bibinfo {note} {\textsc{kwantSpectrum} is a Python package
  can be found at
  \hyperlink{https://kwant-project.org/extensions/kwantspectrum/}{https://kwant-project.org/extensions/kwantspectrum/}}\BibitemShut
  {NoStop}%
\bibitem [{\citenamefont {Piessens}\ \emph {et~al.}(1983)\citenamefont
  {Piessens}, \citenamefont {de~Doncker-Kapenga}, \citenamefont {\"Uberhuber},\
  and\ \citenamefont {Kahaner}}]{quadpack}%
  \BibitemOpen
  \bibfield  {author} {\bibinfo {author} {\bibfnamefont {R.}~\bibnamefont
  {Piessens}}, \bibinfo {author} {\bibfnamefont {E.}~\bibnamefont
  {de~Doncker-Kapenga}}, \bibinfo {author} {\bibfnamefont {C.~W.}\ \bibnamefont
  {\"Uberhuber}}, \ and\ \bibinfo {author} {\bibfnamefont {D.~K.}\ \bibnamefont
  {Kahaner}},\ }\bibfield  {title} {\emph {\enquote {\bibinfo {title}
  {{QUADPACK A Subroutine Package for Automatic Integration.}}}\ }}\href
  {https://doi.org/10.1007/978-3-642-61786-7} {\bibfield  {journal} {\bibinfo
  {journal} {Springer Series in Comput. Math.}\ } (\bibinfo {year}
  {1983})}\BibitemShut {NoStop}%
\bibitem [{\citenamefont {Press}\ \emph {et~al.}(2007)\citenamefont {Press},
  \citenamefont {Teukolsky}, \citenamefont {Vetterling},\ and\ \citenamefont
  {Flannery}}]{numerical_recipes}%
  \BibitemOpen
  \bibfield  {author} {\bibinfo {author} {\bibfnamefont {W.~H.}\ \bibnamefont
  {Press}}, \bibinfo {author} {\bibfnamefont {S.~A.}\ \bibnamefont
  {Teukolsky}}, \bibinfo {author} {\bibfnamefont {W.~T.}\ \bibnamefont
  {Vetterling}}, \ and\ \bibinfo {author} {\bibfnamefont {B.~P.}\ \bibnamefont
  {Flannery}},\ }\href@noop {} {\emph {\bibinfo {title} {Numerical Recipes 3rd
  Edition: The Art of Scientific Computing}}},\ \bibinfo {edition} {3rd}\ ed.\
  (\bibinfo  {publisher} {Cambridge University Press},\ \bibinfo {address} {New
  York, NY, USA},\ \bibinfo {year} {2007})\BibitemShut {NoStop}%
\bibitem [{Note1()}]{Note1}%
  \BibitemOpen
  \bibinfo {note} {\protect \textsc {Kwant} provides a discretizer to translate
  continuum into tight-binding models.}\BibitemShut {Stop}%
\bibitem [{\citenamefont {Hairer}\ \emph {et~al.}(1993)\citenamefont {Hairer},
  \citenamefont {N{\o}rsett},\ and\ \citenamefont {Wanner}}]{hairer93}%
  \BibitemOpen
  \bibfield  {author} {\bibinfo {author} {\bibfnamefont {E.}~\bibnamefont
  {Hairer}}, \bibinfo {author} {\bibfnamefont {S.}~\bibnamefont {N{\o}rsett}},
  \ and\ \bibinfo {author} {\bibfnamefont {G.}~\bibnamefont {Wanner}},\ }\href
  {\doibase 10.1007/978-3-540-78862-1} {\emph {\bibinfo {title} {Solving
  Ordinary Differential Equations, I: Nonstiff Problems}}},\ Vol.~\bibinfo
  {volume} {8}\ (\bibinfo  {publisher} {Springer, Berlin, Heidelberg},\
  \bibinfo {year} {1993})\BibitemShut {NoStop}%
\bibitem [{\citenamefont {{Message Passing Interface
  Forum}}(2015)}]{mpi32_standard}%
  \BibitemOpen
  \bibfield  {author} {\bibinfo {author} {\bibnamefont {{Message Passing
  Interface Forum}}},\ }\href@noop {} {\enquote {\bibinfo {title}
  {\hyperlink{http://www.mpi-forum.org/}{MPI: A Message-passing Interface
  Standard, Version 3.1}},}\ } (\bibinfo {year} {2015})\BibitemShut {NoStop}%
\bibitem [{\citenamefont {Ma\ifmmode~\check{c}\else \v{c}\fi{}ek}\ \emph
  {et~al.}(2020)\citenamefont {Ma\ifmmode~\check{c}\else \v{c}\fi{}ek},
  \citenamefont {Dumitrescu}, \citenamefont {Bertrand}, \citenamefont {Triggs},
  \citenamefont {Parcollet},\ and\ \citenamefont {Waintal}}]{macek20}%
  \BibitemOpen
  \bibfield  {author} {\bibinfo {author} {\bibfnamefont {M.}~\bibnamefont
  {Ma\ifmmode~\check{c}\else \v{c}\fi{}ek}}, \bibinfo {author} {\bibfnamefont
  {P.~T.}\ \bibnamefont {Dumitrescu}}, \bibinfo {author} {\bibfnamefont
  {C.}~\bibnamefont {Bertrand}}, \bibinfo {author} {\bibfnamefont
  {B.}~\bibnamefont {Triggs}}, \bibinfo {author} {\bibfnamefont
  {O.}~\bibnamefont {Parcollet}}, \ and\ \bibinfo {author} {\bibfnamefont
  {X.}~\bibnamefont {Waintal}},\ }\bibfield  {title} {\emph {\enquote {\bibinfo
  {title} {Quantum quasi-monte carlo technique for many-body perturbative
  expansions},}\ }}\href {\doibase 10.1103/PhysRevLett.125.047702} {\bibfield
  {journal} {\bibinfo  {journal} {Phys. Rev. Lett.}\ }\textbf {\bibinfo
  {volume} {125}},\ \bibinfo {pages} {047702} (\bibinfo {year}
  {2020})}\BibitemShut {NoStop}%
\bibitem [{\citenamefont {Bertrand}\ \emph {et~al.}(2019)\citenamefont
  {Bertrand}, \citenamefont {Florens}, \citenamefont {Parcollet},\ and\
  \citenamefont {Waintal}}]{bertrand19}%
  \BibitemOpen
  \bibfield  {author} {\bibinfo {author} {\bibfnamefont {C.}~\bibnamefont
  {Bertrand}}, \bibinfo {author} {\bibfnamefont {S.}~\bibnamefont {Florens}},
  \bibinfo {author} {\bibfnamefont {O.}~\bibnamefont {Parcollet}}, \ and\
  \bibinfo {author} {\bibfnamefont {X.}~\bibnamefont {Waintal}},\ }\bibfield
  {title} {\emph {\enquote {\bibinfo {title} {Reconstructing nonequilibrium
  regimes of quantum many-body systems from the analytical structure of
  perturbative expansions},}\ }}\href {\doibase 10.1103/PhysRevX.9.041008}
  {\bibfield  {journal} {\bibinfo  {journal} {Phys. Rev. X}\ }\textbf {\bibinfo
  {volume} {9}},\ \bibinfo {pages} {041008} (\bibinfo {year}
  {2019})}\BibitemShut {NoStop}%
\bibitem [{\citenamefont {Armagnat}\ \emph {et~al.}(2019)\citenamefont
  {Armagnat}, \citenamefont {Lacerda-Santos}, \citenamefont {Rossignol},
  \citenamefont {Groth},\ and\ \citenamefont {Waintal}}]{armagnat19}%
  \BibitemOpen
  \bibfield  {author} {\bibinfo {author} {\bibfnamefont {P.}~\bibnamefont
  {Armagnat}}, \bibinfo {author} {\bibfnamefont {A.}~\bibnamefont
  {Lacerda-Santos}}, \bibinfo {author} {\bibfnamefont {B.}~\bibnamefont
  {Rossignol}}, \bibinfo {author} {\bibfnamefont {C.}~\bibnamefont {Groth}}, \
  and\ \bibinfo {author} {\bibfnamefont {X.}~\bibnamefont {Waintal}},\
  }\bibfield  {title} {\emph {\enquote {\bibinfo {title} {{The self-consistent
  quantum-electrostatic problem in strongly non-linear regime}},}\ }}\href
  {\doibase 10.21468/SciPostPhys.7.3.031} {\bibfield  {journal} {\bibinfo
  {journal} {SciPost Phys.}\ }\textbf {\bibinfo {volume} {7}},\ \bibinfo
  {pages} {31} (\bibinfo {year} {2019})}\BibitemShut {NoStop}%
\bibitem [{\citenamefont {Kuhn}(1955)}]{Kuhn55}%
  \BibitemOpen
  \bibfield  {author} {\bibinfo {author} {\bibfnamefont {H.~W.}\ \bibnamefont
  {Kuhn}},\ }\bibfield  {title} {\emph {\enquote {\bibinfo {title} {{The
  Hungarian method for the assignment problem}},}\ }}\href {\doibase
  10.1002/nav.3800020109} {\bibfield  {journal} {\bibinfo  {journal} {Naval
  Res. Logist.}\ }\textbf {\bibinfo {volume} {2}},\ \bibinfo {pages} {83}
  (\bibinfo {year} {1955})}\BibitemShut {NoStop}%
\bibitem [{\citenamefont {{Virtanen}}\ \emph {et~al.}(2020)\citenamefont
  {{Virtanen}}, \citenamefont {{Gommers}}, \citenamefont {{Oliphant}},
  \citenamefont {{Haberland}}, \citenamefont {{Reddy}},\ and\ \citenamefont
  {\textit{et al.}}}]{scipy20}%
  \BibitemOpen
  \bibfield  {author} {\bibinfo {author} {\bibfnamefont {P.}~\bibnamefont
  {{Virtanen}}}, \bibinfo {author} {\bibfnamefont {R.}~\bibnamefont
  {{Gommers}}}, \bibinfo {author} {\bibfnamefont {T.~E.}\ \bibnamefont
  {{Oliphant}}}, \bibinfo {author} {\bibfnamefont {M.}~\bibnamefont
  {{Haberland}}}, \bibinfo {author} {\bibfnamefont {T.}~\bibnamefont
  {{Reddy}}}, \ and\ \bibinfo {author} {\bibnamefont {\textit{et al.}}},\
  }\bibfield  {title} {\emph {\enquote {\bibinfo {title} {{SciPy 1.0:
  Fundamental Algorithms for Scientific Computing in Python}},}\ }}\href
  {\doibase https://doi.org/10.1038/s41592-019-0686-2} {\bibfield  {journal}
  {\bibinfo  {journal} {Nat. Methods}\ }\textbf {\bibinfo {volume} {17}},\
  \bibinfo {pages} {261} (\bibinfo {year} {2020})}\BibitemShut {NoStop}%
\bibitem [{\citenamefont {Gonnet}(2010)}]{Gonnet10}%
  \BibitemOpen
  \bibfield  {author} {\bibinfo {author} {\bibfnamefont {P.}~\bibnamefont
  {Gonnet}},\ }\bibfield  {title} {\emph {\enquote {\bibinfo {title}
  {{Increasing the Reliability of Adaptive Quadrature Using Explicit
  Interpolants}},}\ }}\href {\doibase 10.1145/1824801.1824804} {\bibfield
  {journal} {\bibinfo  {journal} {ACM Trans. Math. Softw.}\ }\textbf {\bibinfo
  {volume} {37}},\ \bibinfo {pages} {26} (\bibinfo {year} {2010})}\BibitemShut
  {NoStop}%
\bibitem [{\citenamefont {Gonnet}(2012)}]{Gonnet12}%
  \BibitemOpen
  \bibfield  {author} {\bibinfo {author} {\bibfnamefont {P.}~\bibnamefont
  {Gonnet}},\ }\bibfield  {title} {\emph {\enquote {\bibinfo {title} {{A Review
  of Error Estimation in Adaptive Quadrature}},}\ }}\href {\doibase
  10.1145/2333112.2333117} {\bibfield  {journal} {\bibinfo  {journal} {ACM
  Comput. Surv.}\ }\textbf {\bibinfo {volume} {44}},\ \bibinfo {pages} {22}
  (\bibinfo {year} {2012})}\BibitemShut {NoStop}%
\end{thebibliography}%
\bibliographystyle{apsrev4-1_own}

\end{document}